\pdfoutput=1
\documentclass[11pt,twoside,a4paper,cmspaper,final,collab]{cms-tdr}
\def\svnVersion{fa2c4bc}\def\svnDate{2022/03/17}\def\cmsCernNoTag{CERN-EP-2021-097}\def\cmsCernDate{\today}\def\cmsMessage{Published in the Journal of High Energy Physics as \href{http://dx.doi.org/10.1007/JHEP04(2022)147}{\doi{10.1007/JHEP04(2022)147}.}}
\begin{document}\cmsNoteHeader{SUS-19-012}

\newcommand{\WZ}{\ensuremath{\PW\PZ}\xspace}
\newcommand{\WH}{\ensuremath{\PW\PH}\xspace}
\newcommand{\HZ}{\ensuremath{\PH\PZ}\xspace}
\newcommand{\HH}{\ensuremath{\PH\PH}\xspace}
\newcommand{\ttX}{\ensuremath{\ttbar \PX}\xspace}
\newcommand{\ttZ}{\ensuremath{\ttbar \PZ}\xspace}
\newcommand{\ttH}{\ensuremath{\ttbar \PH}\xspace}
\newcommand{\ttW}{\ensuremath{\ttbar \PW}\xspace}
\newcommand{\ttXX}{\ensuremath{\ttbar \PX\PX}\xspace}
\newcommand{\ttZZ}{\ensuremath{\ttbar \PZ\PZ}\xspace}
\newcommand{\ttZH}{\ensuremath{\ttbar \PZ\PH}\xspace}
\newcommand{\ttWZ}{\ensuremath{\ttbar \PW\PZ}\xspace}
\newcommand{\ttWW}{\ensuremath{\ttbar \PW\PW}\xspace}
\newcommand{\ttWH}{\ensuremath{\ttbar \PW\PH}\xspace}
\newcommand{\ttHH}{\ensuremath{\ttbar \PH\PH}\xspace}
\newcommand{\tttt}{\ensuremath{\ttbar \ttbar}\xspace}
\newcommand{\tX}{\ensuremath{\PQt\PX}\xspace}
\newcommand{\tZq}{\ensuremath{\PQt\PZ \cPq}\xspace}
\newcommand{\qqZZ}{\ensuremath{\Pq\Pq\to\PZ\PZ}\xspace}
\newcommand{\ggZZ}{\ensuremath{\Pg\Pg\to\PZ\PZ}\xspace}
\newcommand{\hZZ}{\ensuremath{\PH\to\PZ\PZ}\xspace}
\newcommand{\ZZ}{\ensuremath{\PZ\PZ}\xspace}
\newcommand{\WWW}{\ensuremath{\PW\PW\PW}\xspace}
\newcommand{\WWZ}{\ensuremath{\PW\PW\PZ}\xspace}
\newcommand{\WZZ}{\ensuremath{\PW\PZ\PZ}\xspace}
\newcommand{\ZZZ}{\ensuremath{\PZ\PZ\PZ}\xspace}
\newcommand{\VH}{\ensuremath{\PV\PH}\xspace}
\newcommand{\Zg}{\ensuremath{\PZ\Pg}\xspace}
\newcommand{\ttg}{\ensuremath{\ttbar\Pg}\xspace}
\newcommand{\Xg}{\ensuremath{\PX\Pg}\xspace}
\newcommand{\Wg}{\ensuremath{\PW\Pg}\xspace}

\newcommand{\xpm}{\PSGcpmDo}
\newcommand{\xone}{\PSGczDo}
\newcommand{\xonexone}{\ensuremath{\PSGczDo\PSGczDo}\xspace}
\newcommand{\xpmxtwo}{\ensuremath{\PSGcpmDo\PSGczDt}\xspace}
\newcommand{\xtwo}{\PSGczDt}
\newcommand{\slep}{\ensuremath{\widetilde{\ell}}\xspace}
\newcommand{\snu}{\PSGn}
\newcommand{\grav}{\PXXSG}

\newcommand{\ptisr}{\ensuremath{\pt^{\text{ISR}}}\xspace}
\newcommand{\mz}{\ensuremath{m_{\PZ}}\xspace}
\newcommand{\mw}{\ensuremath{m_{\PW}}\xspace}

\newcommand{\dm}{\ensuremath{\delta m}\xspace}
\newcommand{\mlll}{\ensuremath{M_{3\ell}}\xspace}
\newcommand{\mt}{\ensuremath{M_{\text{T}}}\xspace}
\newcommand{\mzone}{\ensuremath{M_{\PZ 1}}\xspace}
\newcommand{\mztwo}{\ensuremath{M_{\PZ 2}}\xspace}
\newcommand{\mtlll}{\ensuremath{M_{\text{T}}^{3\ell}}\xspace}
\newcommand{\mtll}{\ensuremath{M_{\text{T}}^{2\ell}}\xspace}
\newcommand{\mttzz}{\ensuremath{\mtt(\PZ \PZ)}\xspace}
\newcommand{\LTmet}{\ensuremath{L_{\text{T}} + \ptmiss }\xspace}
\newcommand{\mll}{\ensuremath{M_{\ell\ell}}\xspace}
\newcommand{\mlt}{\ensuremath{M_{\ell\tauh}}\xspace}
\newcommand{\mtt}{\ensuremath{M_{\text{T2}}}\xspace}
\newcommand{\mttll}{\ensuremath{\mtt(\ell\ell)}\xspace}
\newcommand{\mttlt}{\ensuremath{\mtt (\ell, \tauh)}\xspace}

\newcommand{\miniiso}{\ensuremath{I_{\text{rel}}^{\text{mini}}}\xspace}
\newcommand{\ptll}{\ensuremath{\pt(\ell\ell)}\xspace}
\newcommand{\ip}{\ensuremath{d_{\text{3D}}}\xspace}
\newcommand{\sip}{\ensuremath{\abs{\ip} / \sigma( \ip )}\xspace}
\newcommand{\abseta}{\ensuremath{\abs{\eta}}\xspace}
\newcommand{\dxy}{\ensuremath{d_{0}}\xspace}
\newcommand{\dz}{\ensuremath{d_{z}}\xspace}
\newlength\cmsTabSkip\setlength{\cmsTabSkip}{1ex}
\newcommand\cmsTableResize[1]{\resizebox*{\textwidth}{!}{#1}}
\newcommand\tableSlug[1]{\vspace*{2\cmsTabSkip} \textbf{#1} \\ \vspace*{2\cmsTabSkip} }
\cmsNoteHeader{SUS-19-012}
\title{Search for electroweak production of charginos and neutralinos in proton-proton collisions at \texorpdfstring{$\sqrt{s} = 13\TeV$}{sqrt(s) = 13 TeV}}

\date{\today}

\abstract{A direct search for electroweak production of charginos and neutralinos is presented. Events with three or four leptons, with up to two hadronically decaying \PGt leptons, or two same-sign light leptons are analyzed. The data sample consists of 137\fbinv of proton-proton collisions with a center of mass energy of 13\TeV, recorded with the CMS detector at the LHC. The results are interpreted in terms of several simplified models. These represent a broad range of production and decay scenarios for charginos and neutralinos. A parametric neural network is used to target several of the models with large backgrounds. In addition, results using orthogonal search regions are provided for all the models, simplifying alternative theoretical interpretations of the results. Depending on the model hypotheses, charginos and neutralinos with masses up to values between 300 and 1450\GeV are excluded at 95\% confidence level.}

\hypersetup{
pdfauthor={CMS Collaboration},%
pdftitle={Search for electroweak production of charginos and neutralinos in proton-proton collisions at sqrt(s) = 13 TeV},%
pdfsubject={CMS},
pdfkeywords={CMS, supersymmetry}}

\maketitle

\section{Introduction} \label{sec:introduction}

Supersymmetry (SUSY) is a promising extension of the standard model (SM) with the potential to solve several of the outstanding problems in particle physics by introducing a new symmetry between bosons and fermions~\cite{PhysRevD.3.2415,WESS197452,WESS197439,FAYET1975104,NILLES19841}. This symmetry leads to the prediction of many new particles, called superpartners of the SM particles~\cite{Martin:1997ns}. The addition of superpartners can mend the hierarchy problem by introducing cancellations between the large loop corrections to the mass of the Higgs boson (\PH). Additionally, SUSY models in which R-parity~\cite{WESS197439} is conserved, implying pair production of superpartners, provide a suitable dark matter candidate in the form of the lightest SUSY particle (LSP).

Searches for the production of SUSY particles have already been carried out in a multitude of final states by the ATLAS and CMS Collaborations at the CERN LHC, however none resulted in evidence of the existence of new particles. Particularly stringent exclusion limits have been placed on the production of strongly interacting superpartners (squarks and gluinos) due to the relatively large production cross section of such processes~\cite{Sirunyan:2019ctn,Sirunyan:2019xwh,CMS:2019tlp,Sirunyan:2020ztc,Sirunyan:2019glc,Sirunyan:2020ztc,Aad:2020qwe,Aad:2020sgw,Aad:2020srt,Aad:2019ftg,Aad:2019pfy}. The absence of any evidence for the production of such particles could mean that colored superpartners are too heavy to be produced at the LHC. The lower production cross sections associated with electroweak production directly lead to the currently softer exclusion limits on the superpartner masses. This makes searches for electroweak SUSY production especially interesting. Such superpartners might still be observed, even if their strongly interacting counterparts are out of reach at the LHC.

In this paper, we present a search for the direct production of charginos (\xpm) and neutralinos (\xtwo), mixed states of the SUSY partners of the electroweak gauge and Higgs bosons, in final states with multiple leptons ($\ell$). Events with three or four leptons, with up to two hadronically decaying $\tau$ leptons ($\tauh$), as well as events with two light leptons (electrons or muons) of the same sign are analyzed. The multitude of final states in this analysis mirrors the complexity of chargino and neutralino decay modes. A data set of proton-proton ($\Pp\Pp$) collision events collected with the CMS detector from 2016 to 2018 is used, corresponding to an integrated luminosity of 137\fbinv. Previous searches in these final states were performed on data samples of approximately 36\fbinv by ATLAS \cite{Aad:2019vvi,Aaboud:2018jiw,Aaboud:2018zeb} and CMS~\cite{SUS-16-039,SUS-17-004}, resulting in exclusion limits on chargino masses up to 1150\GeV for particular model assumptions. The use of parametric neural networks~\cite{Baldi:2016fzo}, which is the main novelty in this paper, together with the re-optimization of the search strategy, and the increased data volume, significantly extend the reach of this search compared to previous results.

This paper is structured as follows. Section~\ref{sec:detector} contains a brief description of the CMS detector. Descriptions and diagrams of all targeted models can be found in Section~\ref{sec:signals}. Section ~\ref{sec:selection} outlines the baseline requirements imposed to select events corresponding to final states of interest in the search. Details on the simulation of the different background and signal processes that populate such selections are included in Section~\ref{sec:simulations}. Section~\ref{sec:strategy} includes a description of the search strategies developed to isolate the different signals from the background processes. The different techniques used for the estimation of the contributions of the SM backgrounds are detailed in Section~\ref{sec:background}. A summary of all sources of uncertainty affecting the interpretation of the results is included in Section~\ref{sec:systematics}. A comparison between the observed data and the expectations for the different signal extraction strategies are presented in Section~\ref{sec:results}. Section~\ref{sec:interpretations} is composed of an interpretation of such information in terms of several SUSY models. Finally, Section~\ref{sec:summary} contains a brief summary of the obtained results.

\section{The CMS detector} \label{sec:detector}

The central feature of the CMS detector is a superconducting solenoid of 6\unit{m} internal diameter, providing a magnetic field of 3.8\unit{T}. Silicon pixel and strip trackers, a lead tungstate crystal electromagnetic calorimeter, and a brass and scintillator hadron calorimeter, each composed of a barrel and two endcap sections, reside within the solenoid. Forward calorimeters extend the pseudorapidity ($\eta$) coverage provided by the barrel and endcap detectors. Muons are detected in gas-ionization detectors embedded in the steel flux-return yoke outside the solenoid. A more detailed description of the CMS detector, together with a definition of the coordinate system used and the relevant variables, can be found in Ref.~\cite{Chatrchyan:2008zzk}.

A two-tiered trigger system~\cite{Khachatryan:2016bia} is used to reduce the rate of recorded events and select those of interest. The first level, composed of custom hardware processors, uses information from the calorimeters and muon detectors to select events at a rate of 100\unit{kHz} within a time latency of less than 4\mus. The second level consists of a processor farm which runs a version of the full event reconstruction, optimized for fast processing, and decreases the event rate to around 1\unit{kHz} before data storage.

\section{Signal models} \label{sec:signals}

This search is aimed at the production of charginos and neutralinos, specifically in decay modes that lead to final states with three or more leptons. The results will be interpreted in the context of several simplified models in which the only free parameters are the superpartner masses~\cite{Alves:2011wf, Chatrchyan:2013sza}. Interpretations are performed for both \xpmxtwo production and effective \xonexone production in gauge mediated models with mass degenerate \xone, \xtwo and \xpm. In the former models \xpm and \xtwo are assumed to be wino-like, i.e. mass-degenerate mixtures of superpartners of the $SU(2)_{L}$ gauge field, while \xone is the LSP and bino-like, i.e. the superpartner of the $U(1)_{Y}$ field. The latter models consider Higgsino-like \xpm, \xtwo, and \xone that are nearly mass-degenerate with \xone being the next-to-LSP (NLSP), and a gravitino being the LSP. In all models, the other superpartners are assumed to be heavy and decoupled. The lightest of the CP even bosons inside the Higgs sector of the minimal supersymmetric SM is assumed to have SM-like properties, including the mass and branching fractions~\cite{deFlorian:2016spz}, and is referred to as the Higgs boson. The rest of the bosons inside the Higgs sector are assumed to be heavy and decoupled. An overview of all specific models used for the interpretation of the search is given below. Scenarios in which the mass splitting between any of the superpartners in the decay chain is small are referred to as ``compressed'' in this paper, and generally have one or more decay products with low \pt. Cases where the mass splittings between all superpartners are relatively large, resulting in high \pt decay products, are called ``uncompressed''.

\subsection{Production of \texorpdfstring{\xpmxtwo} ~with decays via intermediate sleptons}
Charginos and neutralinos can decay to leptons and the LSP via intermediate sleptons (\slep) and sneutrinos (\snu), the respective superpartners of charged leptons and neutrinos ($\nu$). These decays are shown in Fig.~\ref{fig:tchislepsnu}. We only target diagrams in which at least three leptons are produced. Whether the decays are more likely to result in $\tau$ leptons than the other lepton flavors depends on the combination of gauge eigenstates making up \xpm and \xtwo, and their masses~\cite{Martin:1997ns}. Three scenarios are considered:

\begin{figure}[ht]
\centering
\includegraphics[width=.32\textwidth]{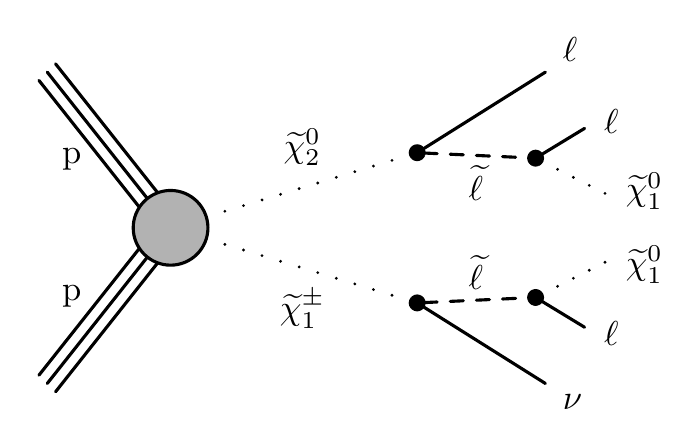}
\includegraphics[width=.32\textwidth]{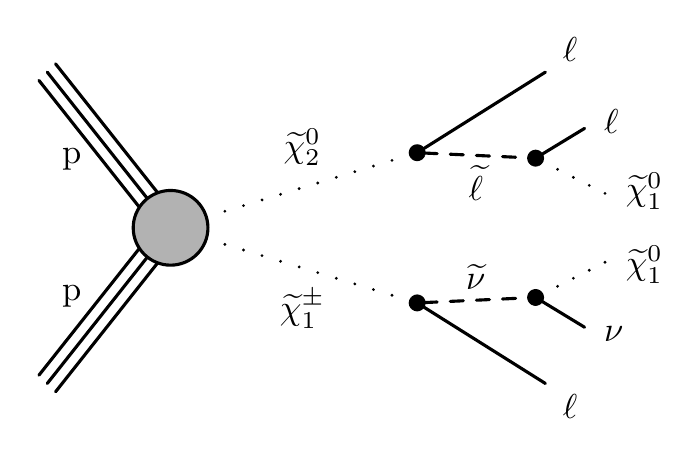}
\topcaption{Production of \xpmxtwo with subsequent decays via sleptons (left) and a slepton and a sneutrino (right).}
\label{fig:tchislepsnu}
\end{figure}

\begin{itemize}
    \item The ``flavor-democratic'' scenario, in which the \xpm and \xtwo decays are mediated by left-handed sleptons, resulting in decays to all lepton flavors with equal probability.
    \item The ``$\tau$-enriched'' scenario, where \xpm couples only to right-handed sleptons, while the decay of \xtwo still goes via left-handed sleptons. Right-handed sleptons only couple to the Higgsino component of \xpm, resulting in \xpm decays that strongly favor the heavier $\tau$ leptons. The decay of \xtwo will still result in all lepton flavors with equal probability.
    \item The ``$\tau$-dominated'' scenario with both \xpm and \xtwo decays mediated by $\tau$ sleptons because the other slepton flavors are heavy and decoupled~\cite{Martin:1997ns}. In this case both \xpm and \xtwo decay exclusively to $\tau$ leptons.
\end{itemize}

In each of these scenarios the branching fraction to leptons is assumed to be 100\%, and both charged \slep and \snu masses are assumed to lie between $m_{\xtwo}$ and $m_{\xone}$, where we take $m_{\xpm}$ to be equal to $m_{\xtwo}$. The kinematics of the leptons and LSPs vary depending on the mass difference between \xtwo and \slep. A parameter $x$ is introduced that governs the \slep mass as follows: $m_{\slep} = x m_{\xtwo} + ( 1 - x ) m_{\xone}$. The interpretation of the search is done using three different values of $x$ as benchmarks for possible manifestations of SUSY in nature:

\begin{itemize}
    \item $x = 0.5$: The slepton mass lies in the middle between $m_{\xtwo}$ and $m_{\xone}$. Each of the leptons and neutrinos emitted in the decay will carry half of the mass difference so they all have identical momentum spectra.
    \item $x = 0.95$: The slepton mass is close to $m_{\xtwo}$, resulting in softer leptons from the initial \xtwo and \xpm decays to sleptons.
    \item $x = 0.05$: The slepton mass is similar to $m_{\xone}$. The second lepton produced in the \xtwo decay will be soft.
\end{itemize}

\subsection{Production of \texorpdfstring{\xpmxtwo} ~with decays via \texorpdfstring{\PH}~, \texorpdfstring{\PW} ~or  \texorpdfstring{\PZ} ~bosons}
\begin{figure}[ht]
\centering
\includegraphics[width=.32\textwidth]{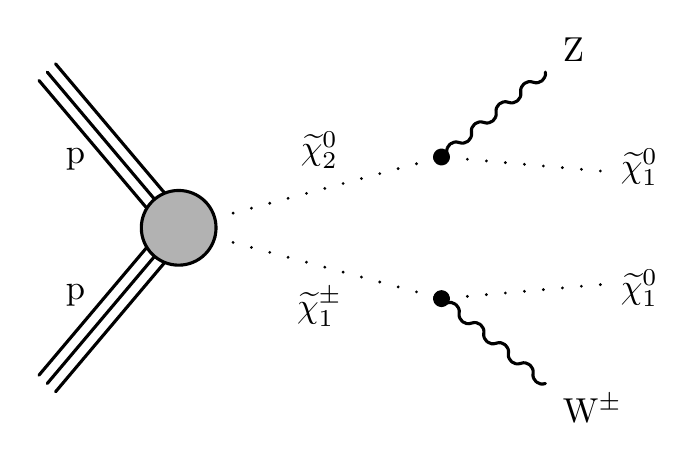}
\includegraphics[width=.32\textwidth]{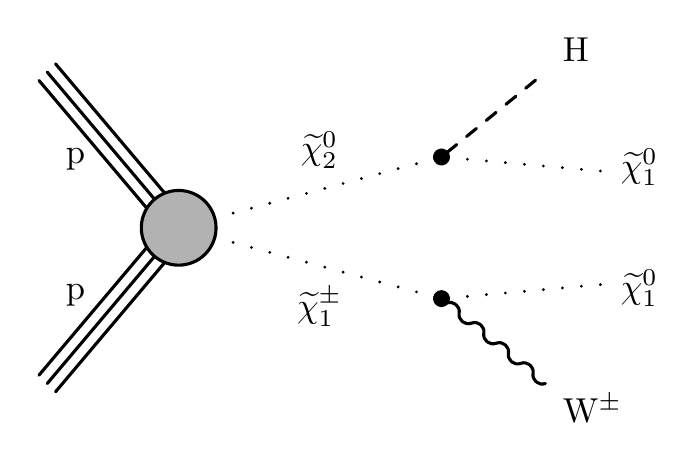}
\caption{Production of \xpmxtwo with subsequent decay of \xpm via a \PW boson and \xtwo via a \PZ boson (left) or \PH boson (right).}
\label{fig:tchiWZ}
\end{figure}

If the sleptons are too heavy, \xtwo is forced to decay to the LSP by emitting either an \PH or \PZ boson, while \xpm decays to a \PW boson and the LSP. These decays are illustrated in Fig.~\ref{fig:tchiWZ}. Final states with multiple leptons can occur by means of subsequent decays of the electroweak bosons to leptons. In the case of \WZ (\WH) mediated \xpmxtwo decays, a branching fraction of 3.3 (2.9)\% to multiple leptons is expected, much lower than what is assumed for the slepton-mediated decays.

\subsection{Production of \texorpdfstring{\xonexone} ~with decays via \texorpdfstring{\PH} ~or  \texorpdfstring{\PZ} ~bosons}
\begin{figure}[ht]
\centering
\includegraphics[width=.32\textwidth]{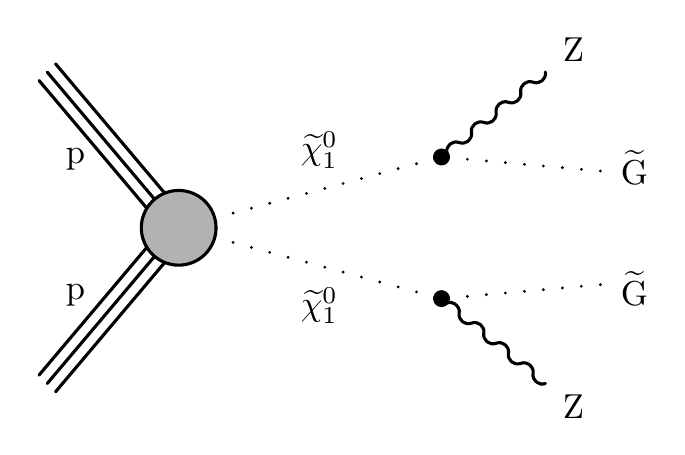}
\includegraphics[width=.32\textwidth]{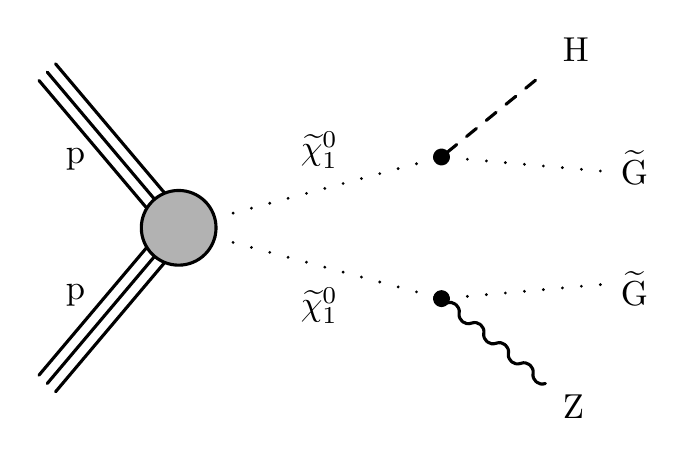}
\includegraphics[width=.32\textwidth]{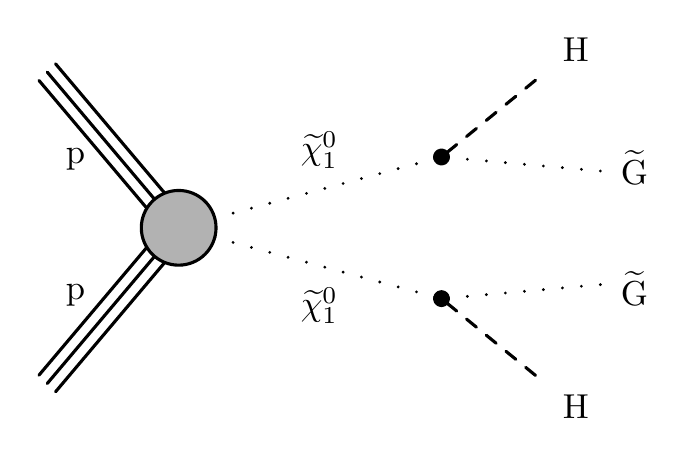}
\caption{Effective \xone pair production with decays mediated by \PZ or \PH bosons.}
\label{fig:tchiZH}
\end{figure}

Lastly, we consider \xone pair production in a gauge-mediated SUSY breaking model with Higgsino-like neutralinos and charginos, and a near massless gravitino (\grav) as the LSP~\cite{Matchev:1999ft,Ruderman:2011vv,Meade:2009qv}. The cross section for direct pair production of neutralinos is expected to be vanishingly small~\cite{Beenakker:1999xh,Fuks:2012qx,Fuks:2013vua}, so we consider a model in which \xtwo, \xone, and \xpm are almost mass-degenerate. In such a model, \xpm and \xtwo decay to \xone via soft particles that escape detection, resulting in effective \xone pair production. The \xonexone pair subsequently decays to LSPs by emitting \PH or \PZ bosons, as depicted in Fig.~\ref{fig:tchiZH}.

\section{Event selection} \label{sec:selection}

This analysis employs the particle-flow (PF) algorithm~\cite{Sirunyan:2017ulk} for event reconstruction. The algorithm aims to identify and reconstruct the individual particles in an event from an optimized combination of various elements in the CMS detector. Particles reconstructed by the PF algorithm (PF candidates) are classified into charged and neutral hadrons, photons, electrons, and muons.

After reconstruction, PF candidates are clustered into jets using the anti-$\kt$ algorithm~\cite{Cacciari:2008gp}, with a distance parameter of 0.4, implemented in the \FASTJET package~\cite{Cacciari:2005hq,Cacciari:2011ma}. Several selection criteria are applied, designed to remove jets that are likely to originate from extraneous energy deposits in the calorimeters~\cite{CMS-PAS-JME-10-003}. The missing transverse momentum vector $\ptvecmiss$ is defined as the negative vector sum of transverse momenta (\pt) of all PF candidates in the event, taking into account jet energy corrections~\cite{Chatrchyan:2011ds,Khachatryan:2016kdb}. Its magnitude is referred to as \ptmiss. The vertex with the largest squared \pt sum of all objects returned by the jet finding algorithm, with the tracks associated with this candidate vertex as inputs, as well as the $\ptvecmiss$ computed from the vector sum of the \pt of those jets, is taken to be the primary $\Pp\Pp$ interaction vertex.

Electrons are reconstructed from a combination of the tracker and the electromagnetic calorimeter measurements. They are required to satisfy $\abseta < 2.5$, ensuring they are within the volume of the tracker, and $\pt > 10 \GeV$. Additionally, requirements are placed on the shower shape, and on a multivariate discriminant based on the shower shape and track quality of the electrons~\cite{Khachatryan:2015hwa}. Electrons that are matched to a secondary vertex consistent with a photon conversion or have a missing hit in the tracker are vetoed.

Muon reconstruction uses a global fit combining information from the tracker, muon spectrometers, and calorimeters. Muons must be within the acceptance of the muon spectrometers, $\abseta < 2.4$, and have $\pt > 10 \GeV$. Selected muons further pass criteria on the geometrical matching between the track in the inner tracker and the muon spectrometers, and on the quality of the global fit~\cite{Sirunyan:2018fpa}.

Both electron and muon candidates must be consistent with originating from the primary $\Pp\Pp$ interaction vertex. This is ensured by requiring the transverse impact parameter (\dxy) to be smaller than 0.5\mm, and the longitudinal one (\dz) not to exceed $1.0 \mm$. The significance of the impact parameter must satisfy $\sip < 8$, where \ip and $\sigma(\ip)$ are, respectively, the three-dimensional impact parameter and its uncertainty.

In order to select leptons resulting from superpartner production, it is important to identify ``prompt'' leptons that originate from the decay of electroweak bosons or superpartners. Prompt leptons have to be separated from other genuine leptons produced in hadron decays, as well as particles in jets that are incorrectly reconstructed as leptons. Such lepton candidates are collectively called ``nonprompt''. As a first step in rejecting nonprompt leptons, electrons and muons must fulfill several prerequisites on their relative mini-isolation (\miniiso), defined as the scalar \pt sum of all other PF candidates in a cone of \pt dependent radius around the lepton's direction, divided by the lepton \pt. The radius of this cone is given by $\Delta R(\pt(\ell)) = \sqrt{\smash[b]{(\Delta\eta)^2+(\Delta\phi)^2}} = 10 \GeV/\min[ \max(\pt(\ell), 50 \GeV) , 200 \GeV]$ in ($\eta$, $\phi$) space, where $\phi$ is the azimuthal angle in radians, taking into account increased particle collimation at high lepton \pt values~\cite{Khachatryan:2016kod}. All electrons and muons must satisfy $\miniiso < 0.4$. The lepton selection discussed up to here is referred to as the baseline selection.

A gradient boosted decision tree (BDT) trained to distinguish prompt from nonprompt light leptons is used~\cite{Sirunyan:2018shy,Sirunyan:2018zgs}. This BDT uses the properties of the jet, as returned by the jet clustering algorithm, containing the lepton: its \textsc{DeepFlavor}~\cite{CMS-DP-2018-033} \PQb tagging score, the ratio of the lepton \pt to that of the jet, and the component of the jet momentum that is transverse to the lepton's direction. Other input variables are \pt, $\eta$, \miniiso, \dxy, \dz, and \sip of the lepton. The BDT additionally has access to the muon segment compatibility for muons and to the earlier mentioned multivariate discriminant for electrons. Two selection criteria on the BDT output are used in the analysis, one for events with three or more leptons, and a tighter one for events with two leptons of the same sign. The latter results in a smaller nonprompt background at the cost of slightly lower selection efficiencies for superpartner production. For prompt muons, the BDT-based selection results in typical efficiencies ranging 90--99\%. Misidentification rates for nonprompt muons passing the baseline selection range 5--10\%. Prompt electrons are identified with an efficiency of around 75\% in events with three or more leptons, with a corresponding misidentification rate of about 5\% for nonprompt electrons passing the baseline selection. The efficiency is typically in the range 50--60\% for the tighter same-sign dilepton selection, with a misidentification rate around 2\%.

Reconstruction of \tauh candidates is performed using the ``hadron-plus-strips'' algorithm~\cite{Sirunyan:2018pgf}. The \tauh candidates are required to be consistent with one- or three-pronged hadronic $\tau$ lepton decays, and must have $\abseta < 2.3$ and $\pt > 20 \GeV$. In order to reject a large background from hadrons misidentified as $\tau$ leptons, the \tauh candidates must pass a stringent selection on a BDT discriminant aimed at identifying prompt \tauh candidates~\cite{Sirunyan:2018pgf}. This selection has typical efficiencies around 50\% for prompt \tauh candidates in the analysis, while having a misidentification rate of 0.2\% for quantum chromodynamics (QCD) jets. Additional selection criteria based on the consistency between the measurements from the tracker, calorimeters and muon detectors are required to reduce the proportion of electrons and muons misidentified as \tauh candidates.

Leptons passing the BDT-based selection criteria mentioned above are labeled ``tight'' leptons. Electrons or muons are ``loose'' if they either pass the same BDT discriminant or pass additional requirements on the properties of the jet containing the lepton in case they fail the BDT selection. Similarly, loose \tauh candidates are those passing a looser requirement on the BDT discriminant. Tight leptons always satisfy the conditions of the loose selection, but not the other way around. The final analysis selection consists of tight leptons, while loose leptons are used to categorize events based on their lepton content and to predict the background from nonprompt leptons. The loose definition of electrons and muons is tuned to facilitate this background prediction, as explained in Section \ref{sec:background}.

In events with two same-sign light leptons, with or without an additional \tauh candidate, further requirements are placed on tight leptons to ensure that their sign is well-measured. For electrons, the sign is determined by the position of a linear extrapolation of the deposits in the pixel detector to the inner calorimeter surface relative to the calorimeter deposit, and compared to the sign determined from the full fit used for electron reconstruction. Electrons in which the two sign measurements are inconsistent are not considered tight. Tight muons are required to have $\sigma( \pt ) / \pt < 0.2$ where, \pt and $\sigma(\pt)$ are respectively the \pt as measured from a tracker-only fit and the associated uncertainty. These requirements are found to reduce the sign mismeasurement probability to under 0.0001 (0.3)\% with efficiencies for prompt well measured leptons greater than 99.9 (99)\% for muons (electrons).

Jets retained for analysis must satisfy $\pt >25 \GeV$, $\abseta < 2.4$, and have a separation of $\Delta R > 0.4$ from any loose lepton. Jets originating from the hadronization of \PQb quarks are identified with the \textsc{DeepCSV} algorithm~\cite{Sirunyan:2017ezt}. Jets satisfying the tight working point of this algorithm are referred to as \PQb-tagged jets. The chosen working point corresponds to a typical efficiency of 50\% for correctly identifying \PQb quark jets, with a misidentification probability of 2.4 (0.1)\% for \PQc quark (light-flavor) jets.

Events that have at least three loose leptons, or two loose light leptons of the same sign, are selected for further analysis. To enter the nominal analysis selection either all loose leptons, or at least four leptons, must be tight. Events in which one or more of the loose leptons fail the tight selection are used to predict the background from nonprompt leptons, following the procedure explained in Section~\ref{sec:background}. Events with one or more \PQb-tagged jets are vetoed to reduce the backgrounds from processes involving top quarks. To match the analysis selection to the online selection, events must satisfy the requirements of trigger algorithms selecting one, two, or three electrons or muons. The lepton \pt cuts mentioned in Section~\ref{sec:strategy} are designed to ensure selected events efficiently pass the trigger selection. Events with any opposite-sign and same flavor (OSSF) pair of light leptons passing the baseline selection, with a dilepton invariant mass below 12\GeV, are vetoed to reduce the background from photon conversions and low-mass resonances.

\section{Simulation} \label{sec:simulations}

Monte Carlo (MC) simulated event samples are used for the estimation of most of the backgrounds, the determination of signal efficiencies, and the training of the parametric neural networks used in the analysis. Separate samples, simulating the data--taking conditions in 2016, 2017, and 2018, are used for each process. In each instance we cite below, the generator program and parameters used for its simulation are chosen to be the most advanced available. Since the required computational resources are large, the samples produced for a given period of data taking (2016, 2017, or 2018) are retained for that data set, while for subsequent data sets newer generator configurations are used, along with updates to the detector model and running conditions. None of the differences in the configuration of the simulation between the data taking periods are found to have a significant impact on the analysis results.

Signal samples are generated with the \MGvATNLO program~\cite{Frixione:2002ik,Alwall:2014hca} at leading order (LO) in perturbative QCD, with up to two additional partons in the matrix element computations. Background samples of diboson \WZ, \qqZZ and \hZZ production are generated at next-to-LO (NLO) precision using the \POWHEG~v2~\cite{Nason:2004rx,Frixione:2007vw,Melia:2011tj,Nason:2013ydw} generator.  \MCFM v7.0 \cite{Campbell:2010ff} has been used to generate the gluon induced diboson processes \ggZZ. Other major background samples (\Zg, \Wg, \ttg, \VH, \WWW, \WWZ, \WZZ, \ZZZ, \ttZ, \ttW, \ttH, \tttt, \tZq)  are simulated using \MGvATNLO, at NLO in QCD. The small contributions from processes involving two top quarks and two massive bosons (\ttZZ, \ttZH, \ttWZ, \ttWH, \ttHH, \ttWW) are simulated using \MGvATNLO at LO in QCD. Version 2.2.2 (2.4.2) of \MGvATNLO is used for simulating 2016 (2017 and 2018) collisions in each case.

The NNPDF3.0~\cite{Ball:2014uwa} (NNPDF3.1~\cite{Ball:2017nwa}) parton distribution function (PDF) sets are used in the simulation of 2016 (2017 and 2018) collisions. The perturbative order in QCD used for the PDFs is LO for the signal samples and NLO (NNLO) for the 2016 (2017 and 2018) background samples. The simulation of hadronization, parton showering and the underlying event is performed by \PYTHIA 8.212 (8.230)~\cite{Sjostrand:2014zea} with the CUETP8M1~\cite{Skands:2014pea,Khachatryan:2015pea} (CP5~\cite{CP5PAS}) tune in samples matching 2016 (2017 and 2018) conditions. Double counting of partons generated by \PYTHIA and \MGvATNLO is eliminated with the \textsc{FxFx}~\cite{Frederix:2012ps} (MLM~\cite{Alwall:2007fs}) matching scheme in NLO (LO) simulations. For signal samples, the same setup is used but using the CP2 tune ~\cite{CP5PAS} instead for 2017 and 2018 samples.

Signal samples of \xpmxtwo production are normalized to cross sections computed at NLO plus next-to-leading logarithmic (NLL) precision using the Resummino framework~\cite{Beenakker:1999xh,Fuks:2012qx,Fuks:2013vua,Fuks:2013lya} in the limit of mass-degenerate wino-like \xpm and \xtwo, and bino-like \xone. Cross sections at the same precision are computed for effective \xonexone production, assuming mass-degenerate Higgsino-like \xpm, \xtwo and \xone. All other superpartners are always assumed heavy and decoupled.

Each event is overlaid with additional inelastic $\Pp\Pp$ collisions generated in \PYTHIA to mimic the presence of additional collisions in the same or adjacent bunch crossings (pileup). The simulated number of interactions per bunch crossing is reweighted to match the one observed in data. Simulated background events include a full \GEANTfour-based~\cite{Geant} detector simulation, while signal events use the CMS fast simulation package~\cite{Abdullin:2011zz} to simulate the detector response. All simulated events are subsequently reconstructed using the same software employed for collision data.

\section{Search strategy} \label{sec:strategy}
As explained in Section~\ref{sec:signals}, the search targets several models for the production of charginos and neutralinos in final states with multiple leptons. In each model we work under the assumption of R-parity conservation, meaning that the LSP is stable, giving a significant \ptmiss in most cases. Many final states, including events with two leptons of the same sign, three leptons, and four or more leptons are selected to target several possible SUSY signals that might be present in the collision data. In the case of same-sign dilepton events, only electrons or muons are considered, whereas up to two \tauh candidates are selected in the other final states. The choices here are dictated by the varied background levels in different kinematic regions, the lepton multiplicity, and the quality of lepton identification. For example, because of the lower purity in tau lepton reconstruction, the selection criteria differentiate between these candidates and the light leptons, electrons and muons. Events are further categorized according to the lepton flavors and signs to focus on various signal hypotheses. A summary of this categorization is presented in Table \ref{tab:categories}. In each of these categories, a set of search regions is defined based on the kinematics of the events to further separate potential signal events in data from the SM backgrounds. Because of the large background in events with three light leptons including an OSSF pair of leptons, and the difficulty of optimizing kinematic bins for sensitivity to a host of models, parametric neural networks are trained for separating signal and background in this region.

\begin{table}[ht]
\centering
\topcaption{Brief description of the categories used to classify events in the search.}
\label{tab:categories}
\cmsTableResize{
\begin{tabular}{cl}
Category & Requirements                                                                               \\ \hline
2$\ell$SS     & Two light leptons with the same sign                                                     \\
3$\ell$A      & Three light leptons including one or more OSSF pairs                                           \\
3$\ell$B      & Three light leptons including no OSSF pairs                                                      \\
3$\ell$C      & A pair of light leptons forming an OSSF pair and a \tauh candidate \\
3$\ell$D      & A pair of light leptons of different flavor and opposite sign and a \tauh candidate \\
3$\ell$E      & A pair of light leptons of same sign and a \tauh candidate \\
3$\ell$F      & A light lepton and two \tauh candidates                                                \\
4$\ell$G      & Four light leptons including two independent OSSF pairs                                                     \\
4$\ell$H      & Four light leptons including one or less OSSF pairs                                           \\
4$\ell$I      & Three light leptons and a \tauh candidate                                        \\
4$\ell$J      & Two light leptons and two \tauh candidates, including two OSSF pairs                       \\
4$\ell$K      & Two light leptons and two \tauh candidates, including one or no OSSF pair               \\
\end{tabular}}
\end{table}

\subsection{Same-sign dilepton events} \label{section:same_sign_strategy}
                                                                           
The signal models described in Section~\ref{sec:signals} yield final states with three or more leptons. In models where the mass difference between the NLSP and LSP is small, or the slepton mass is close to either the NLSP or LSP mass, one or more of the leptons in the final state could have a high probability to fail the lepton selection. The sensitivity of the analysis to such models is increased by retaining events with two leptons. Dilepton events with an opposite-sign lepton pair suffer from a very large SM background, but events with same-sign lepton pairs are relatively rare in the SM. For this reason, we select only events in which both leptons have the same sign (2$\ell$SS).

To ensure efficient triggering on these events, the leading lepton is required to have $\pt > 20 \GeV$ in $\PGm^{\pm}\PGm^{\pm}$ events, and $\pt > 25 \GeV$ in $\PGm^{\pm}\Pe^{\pm}$ and $\Pe^{\pm}\Pe^{\pm}$ events. The subleading lepton must satisfy $\pt > 15 (10) \GeV$ in case it is an electron (muon). The slightly higher \pt thresholds for electrons are mandated by the fact that the corresponding trigger requirements are more stringent. In addition the turn-on curve of electron triggers is softer than that of muon triggers. Events in which a third loose light lepton or tight \tauh candidate is present are vetoed to ensure orthogonality with the other event categories. As this category mainly targets signal events with a lepton that fails the selection or fails to be reconstructed, we do not veto events with a third lepton passing the baseline selection, as long as it fails the loose selection. When a third lepton passing the baseline selection is present, it is not allowed to form a mass within a 15\GeV window around the \PZ boson mass (\mz) with another lepton in the event. This requirement is found to reduce the SM \WZ background while keeping a signal efficiency over 99$\%$ for the most compressed signal scenarios with mass splittings between the NLSP and the LSP under 50\GeV. Events with more than one jet with $\pt > 40 \GeV$ are rejected to reduce the \ttbar background, while still allowing for some hadronic recoil in signal events. Finally, $\ptmiss > 60 \GeV$ is required.

Events are then binned according to their kinematic properties to maximally separate SUSY signal from the background. The stransverse mass \mtt, defined to have an endpoint at the parent particle's mass for events with two semi-invisible decays~\cite{Lester:1999tx}, is used, because its tails tend to be populated by signal events with high \ptmiss. Additional discriminating variables are the \pt of the dilepton system (\ptll), which tends to be high in uncompressed models, and \ptmiss. Bins with expected yields sufficient for an accurate background description are further split according to the sign of the leptons. This is motivated by the $\Pp\Pp$ nature of LHC collisions, which makes same-sign lepton pairs of positive sign more common than those of negative sign. The magnitude of this sign asymmetry depends on the initial state of the process and is generally different between signal and background. The full set of search regions is shown in Table~\ref{tab:SR2lss}.

\begin{table}
\centering
\topcaption{Definition of the search regions used for events with two same-sign light leptons (SSXX). The symbols ($++$) and ($--$) represent requirements on the sign of the leptons. The first \mttll bin contains only events where \mttll is exactly 0~\cite{Lester:2011nj}, whereas the second bin contains events where \mttll is larger than 0 and less than or equal to 80\GeV. The last \mttll bin contains events where \mttll exceeds 80\GeV.}
\label{tab:SR2lss}
\tableSlug{2$\ell$SS : two light leptons of the same sign }
\cmsTableResize{
\begin{tabular}{c|c|c|c|c}

\multicolumn{1}{c}{\mttll (\GeVns)}& \multicolumn{1}{c}{\ptll (\GeVns)} & \multicolumn{1}{c}{$ 60 < \ptmiss < 100 \GeV$} &\multicolumn{1}{c}{ $100 \leq \ptmiss < 200 \GeV$ }&\multicolumn{1}{c}{ $\ptmiss \geq 200 \GeV$} \\ \hline
    \multirow{3}{*}{ $0$} & $<$70 & \multicolumn{3}{c}{ SS01} \\ \cline{2-5}
                              & \multirow{2}{*}{ $\geq$70 } & \multirow{2}{*}{ SS02} &  SS03 ($++$) &  SS05 ($++$) \\ \cline{4-5}
                              &                          &                            &  SS04 ($--$) &  SS06 ($--$) \\ \hline

    \multirow{3}{*}{ 0--80} & \multirow{2}{*}{ $<$30 } & \multicolumn{2}{c|}{ SS07 ($++$)} & \multirow{2}{*}{ SS09} \\ \cline{3-4}
                               &                           & \multicolumn{2}{c|}{ SS08 ($--$)} & \\ \cline{2-5}
                               & $ \geq$30                   & \multicolumn{3}{c}{ SS10} \\ \hline

    \multirow{4}{*}{ $> 80$ }  & \multirow{2}{*}{ $<$200 } & \multirow{2}{*}{ SS11} &  SS12 ($++$) &  SS14 ($++$) \\ \cline{4-5}
                               &                           &                            &  SS13 ($--$) &  SS15 ($--$) \\ \cline{2-5}
                               & \multirow{2}{*}{ $\geq$200 } & \multirow{2}{*}{ SS16} &  SS17 ($++$) &  SS19 ($++$) \\\cline{4-5}
                               &                           &                            &  SS18 ($--$) &  SS20 ($--$) \\ \hline
\end{tabular}}
\end{table}

\subsection{Three-lepton events}
All signal models considered in this analysis yield at least three leptons in the final state, so the analysis retains all events with three or more leptons, with up to two \tauh candidates. This section describes the search strategy for events with exactly three leptons, while events with four or more leptons are discussed in Section~\ref{sec:four_leptons}.

In addition to the selection requirements specified in Section~\ref{sec:selection} we impose \pt thresholds on the leptons. In a similar fashion to those discussed in Section~\ref{section:same_sign_strategy} for same-sign dilepton events, these additional requirements ensure efficient triggering by at least one of the leptonic triggers used in the analysis. The leading light lepton is required to satisfy $\pt > 25\,(20) \GeV$ if it is an electron (muon). If two or more light leptons are present, the subleading light lepton must have $\pt > 15\,(10) \GeV$. In events with just a single muon, where this is also the leading light lepton, the muon must satisfy $\pt > 25 \GeV$. In events with just a single muon, where this is also the leading light lepton, the muon must satisfy $\pt > 25 \GeV$. These \pt requirements are added in addition of those specified in Section~\ref{sec:selection}, in order to ensure efficient triggering on the events by at least one of the leptonic triggers used in the analysis. As we target signals with escaping particles, we require $\ptmiss > 50 \GeV$, significantly reducing the background from processes without particles evading detection.

\subsubsection{Three light leptons with an OSSF pair}

If no $\tau$ leptons are present in the decay, the signal models in Section~\ref{sec:signals} mainly give final states with an OSSF pair of leptons. As such, these events will dominate the sensitivity to \xpmxtwo production with flavor-democratic decays through sleptons, or decays via the emission of a \PW and a \PZ boson. Meanwhile, this event category also suffers from the largest amount of background among all analysis categories, dominated by SM \WZ production. Because of the category's importance and the relatively large background, several parametric neural networks are trained to distinguish the signal models from the background in this region. Additionally, a set of search regions is also defined, which are less sensitive than the neural networks, but that facilitate alternative interpretations of the results. This event category is referred to as 3$\ell$A as shown in Table~\ref{tab:categories}.

Our signal model has several varying parameters, namely the masses of the NLSP and LSP. One could search for such a model by training a single machine learning discriminant based on reconstructed quantities, or by training one such discriminant for each value of the signal parameters. If the event kinematics depend on the signal parameters, the former approach will be suboptimal for most or all signal points, while the latter introduces a great deal of complexity. Additionally, the second approach of training separate discriminants for each signal point does not allow for the interpolation of the results to signal parameters not seen while training the discriminant. A solution to these problems is the training of a ``parametric'' machine learning discriminant~\cite{Baldi:2016fzo}. On top of a set of reconstructed quantities, such a discriminant uses one or more signal parameters as additional input features. In the training each background event is given a value randomly drawn from the parameter distribution in the signal simulation. This results in a discriminant that learns to optimally distinguish each signal hypothesis from the background, and that can be evaluated using signal parameters not seen during training.

The kinematics of the signal events are largely determined by the mass splitting $\dm = m_{\text{\xpm}} - m_{\text{\xone}}$, with relatively small kinematic differences between signal points having equal \dm, but differing $m_{\text{\xpm}}$ values. This is exploited by training a neural network parametric in $\dm$ for each of the four different signal models: \xpmxtwo production with decays through \PW and \PZ bosons, and \xpmxtwo with slepton-mediated decays at $x = 0.95$, $0.50$ and $0.05$, with $x$ being the parameter governing the mass splitting between the NLSP and the sleptons. The following reconstructed input variables are used when training the neural networks: the mass of the OSSF lepton pair that is closest to \mz out of all such pairs in the event (\mll), the transverse mass of the lepton not forming the \PZ boson candidate and \ptvecmiss (\mt), \ptmiss, the transverse mass of the trilepton system and \ptvecmiss (\mtlll), the trilepton invariant mass (\mlll), the scalar sum of lepton \pt's and \ptmiss (\LTmet), and the scalar sum of the \pt of all jets in the event (\HT). For each of these variables the distribution is compared between the fast simulation of the CMS detector used in the signal simulation and the nominal \GEANTfour-based simulation. This comparison is made for several representative signal points: compressed, noncompressed and at \dm close to \mz. No significant differences are observed between the distributions.

The neural networks are fully connected feed-forward networks with a single output node representing the probability that an event is signal. They are trained in \textsc{TensorFlow}~\cite{tensorflow2015-whitepaper} using the \textsc{Keras}~\cite{chollet2015keras} interface. To reinforce the learning of the parametrization, the signal parameter is fed as an additional input to each hidden layer of the network, and the \dm values assigned to background events are resampled from the signal distribution for each training epoch. The gradient descent of the network weights for training is done with a variant of the Adam~\cite{kingma2014adam} algorithm using Nesterov momentum~\cite{Nesterov1983AMF}. Batch normalization~\cite{DBLP:journals/corr/IoffeS15} is added between all of the hidden layers to reduce the internal covariance shift of the network, speeding up training and increasing the final performance. To regularize the network, dropout~\cite{dropout} is added to each hidden layer. At each node a parametric rectified linear unit activation function is used, except for the output node, which uses a sigmoid activation. The number of nodes in each layer of the network, the number of hidden layers, the learning rate, the learning rate decay, the dropout rate, and the used activation function, are all varied in grid scans, training the neural network each time with a different configuration. The performance of each configuration is then evaluated, in terms of the area under the receiver operating characteristic curve (AUC), on a validation set. The optimal values of these parameters are chosen for the final training of each network. The results of the grid scan optimization are cross-checked with a custom--made evolutionary algorithm designed to optimize the neural network hyperparameters. The evolutionary algorithm results in an equivalent final neural network performance, though with significantly fewer training iterations than needed in the grid scan.

\begin{figure}
\centering
\includegraphics[width=.32\textwidth]{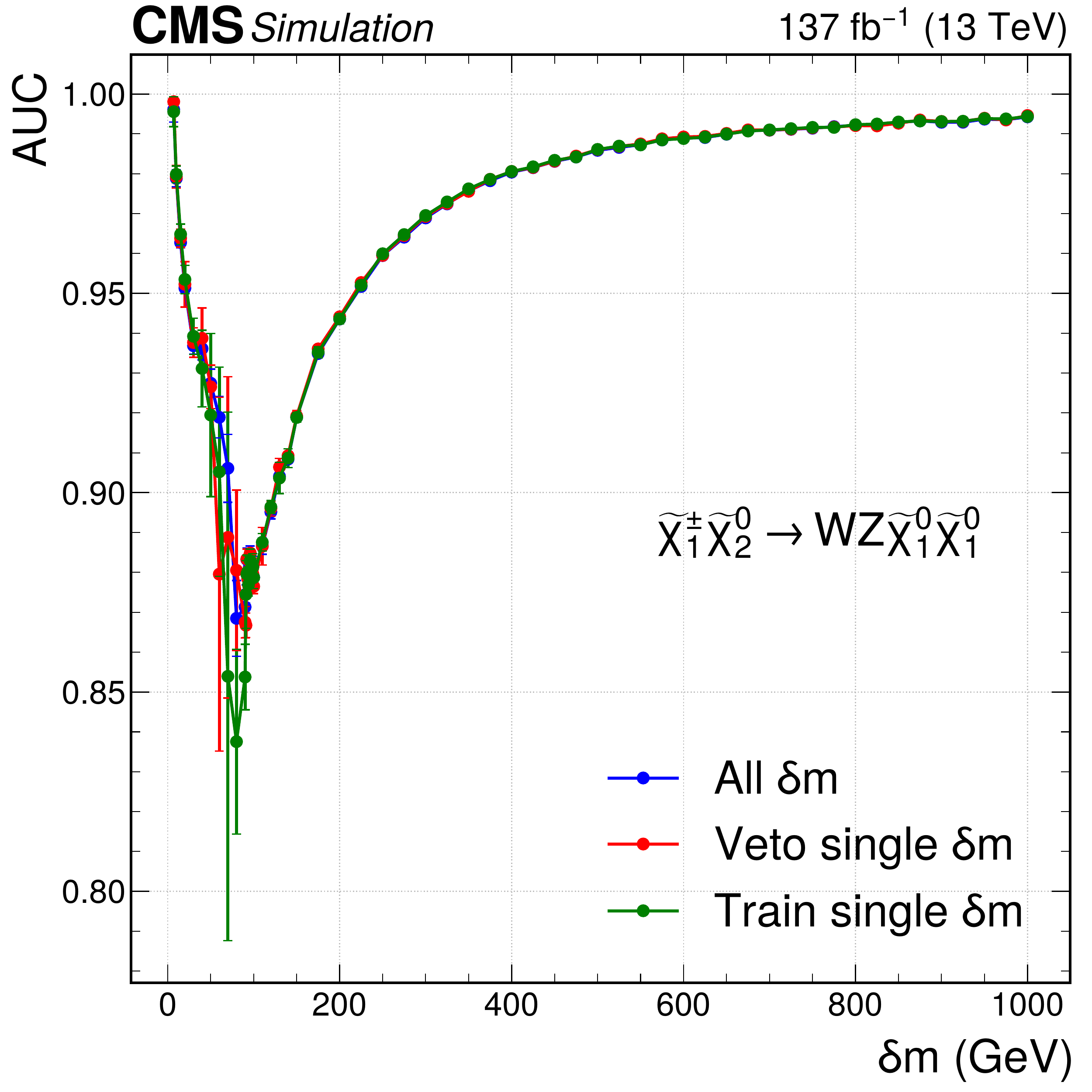}
\includegraphics[width=.32\textwidth]{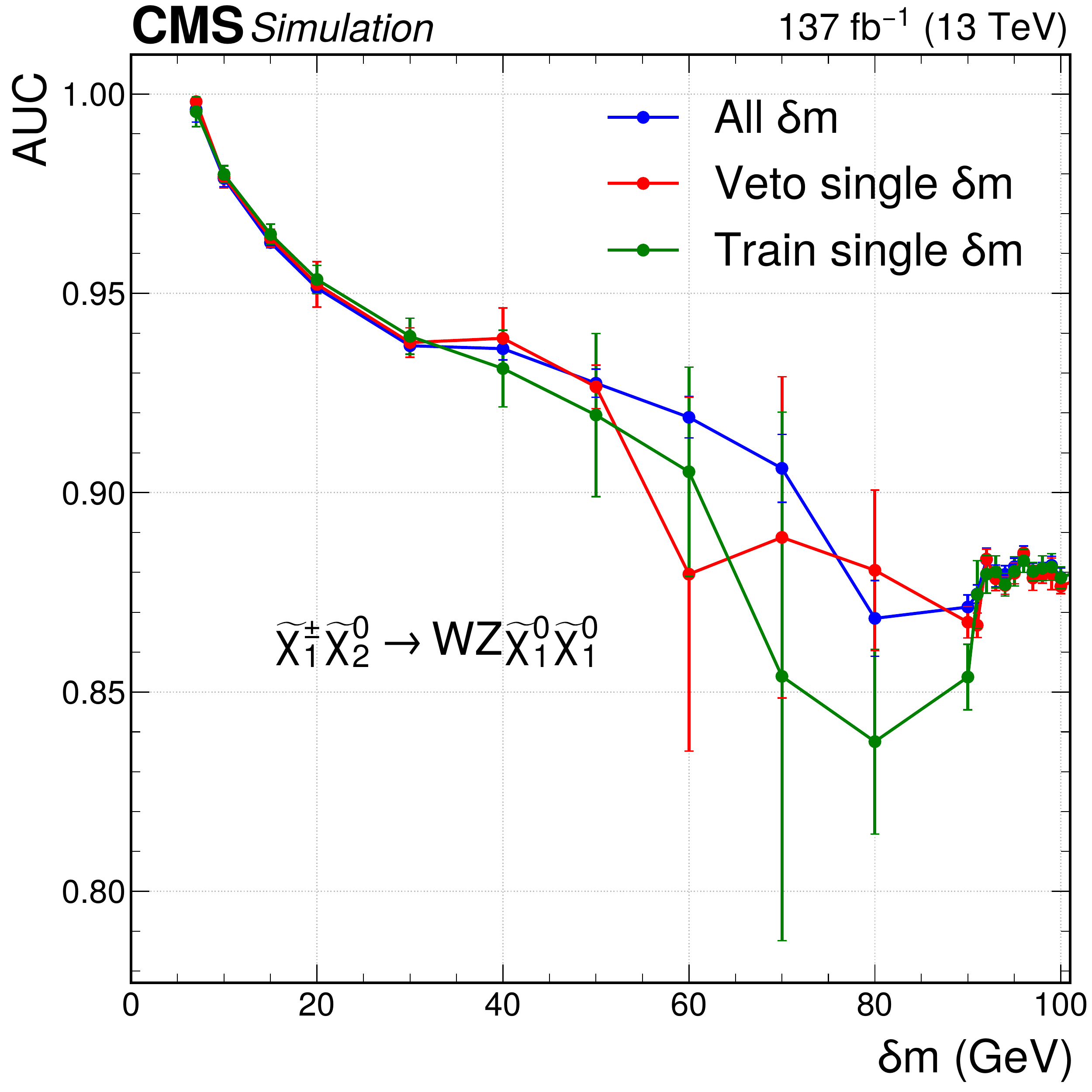}
\includegraphics[width=.32\textwidth]{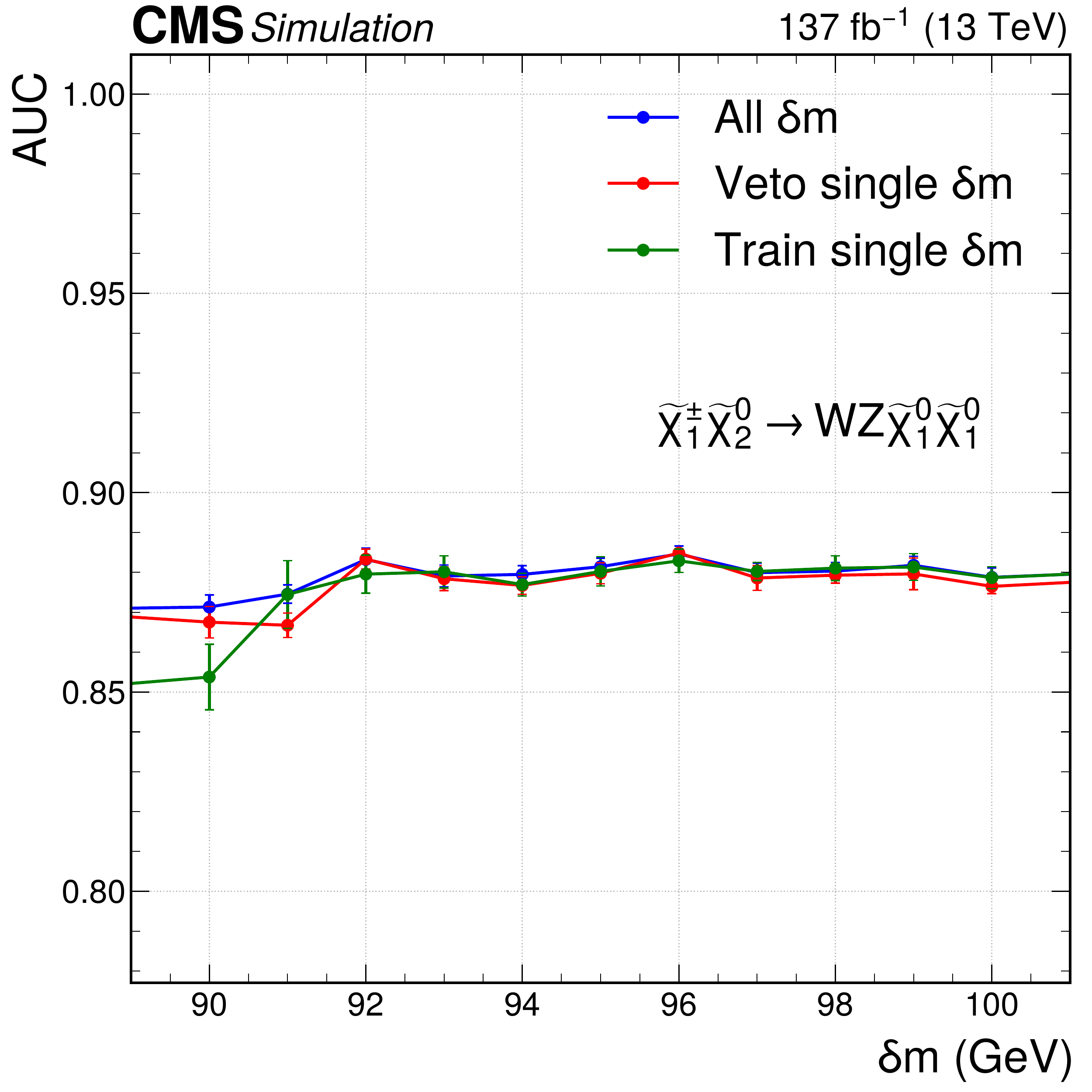}\\
\includegraphics[width=.32\textwidth]{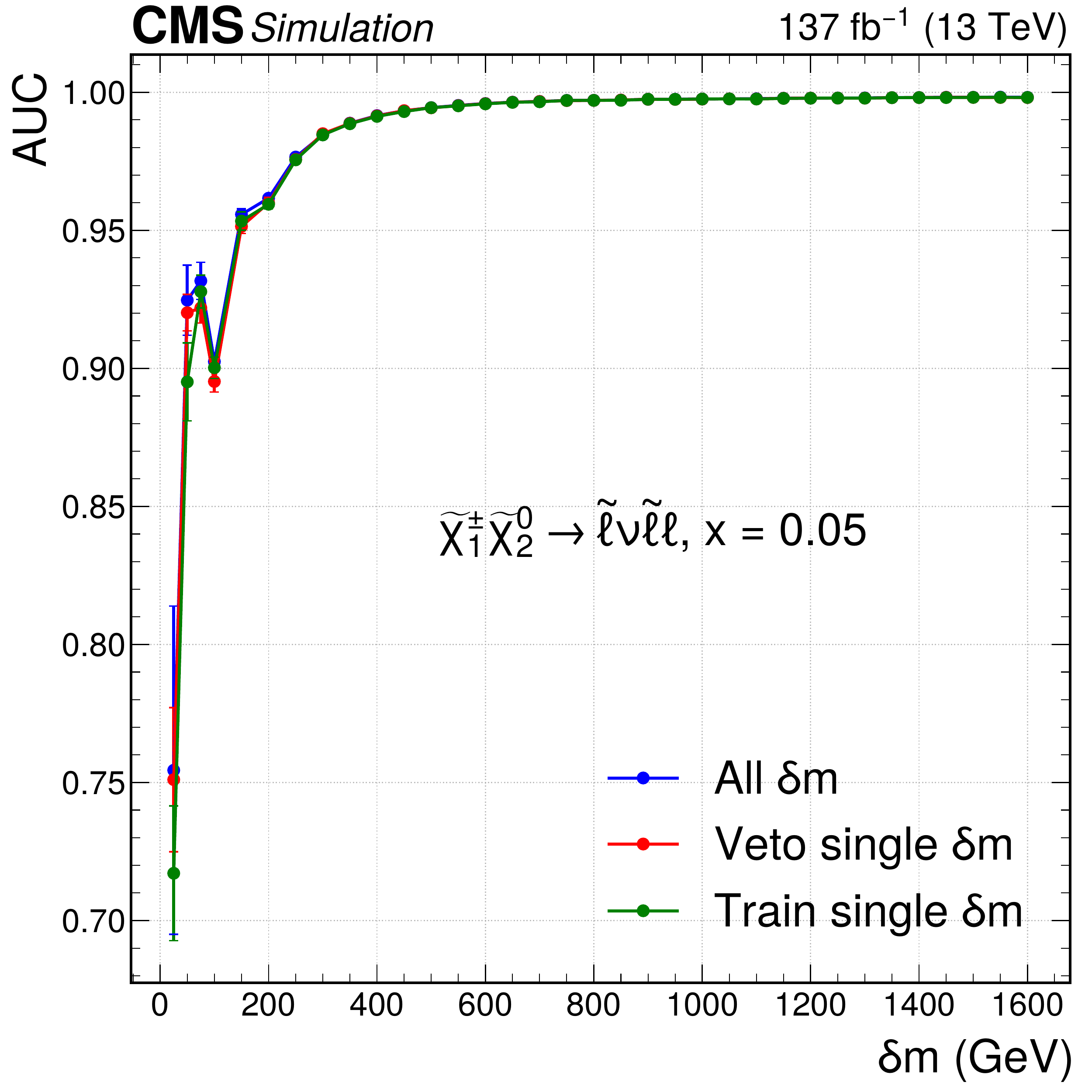}
\includegraphics[width=.32\textwidth]{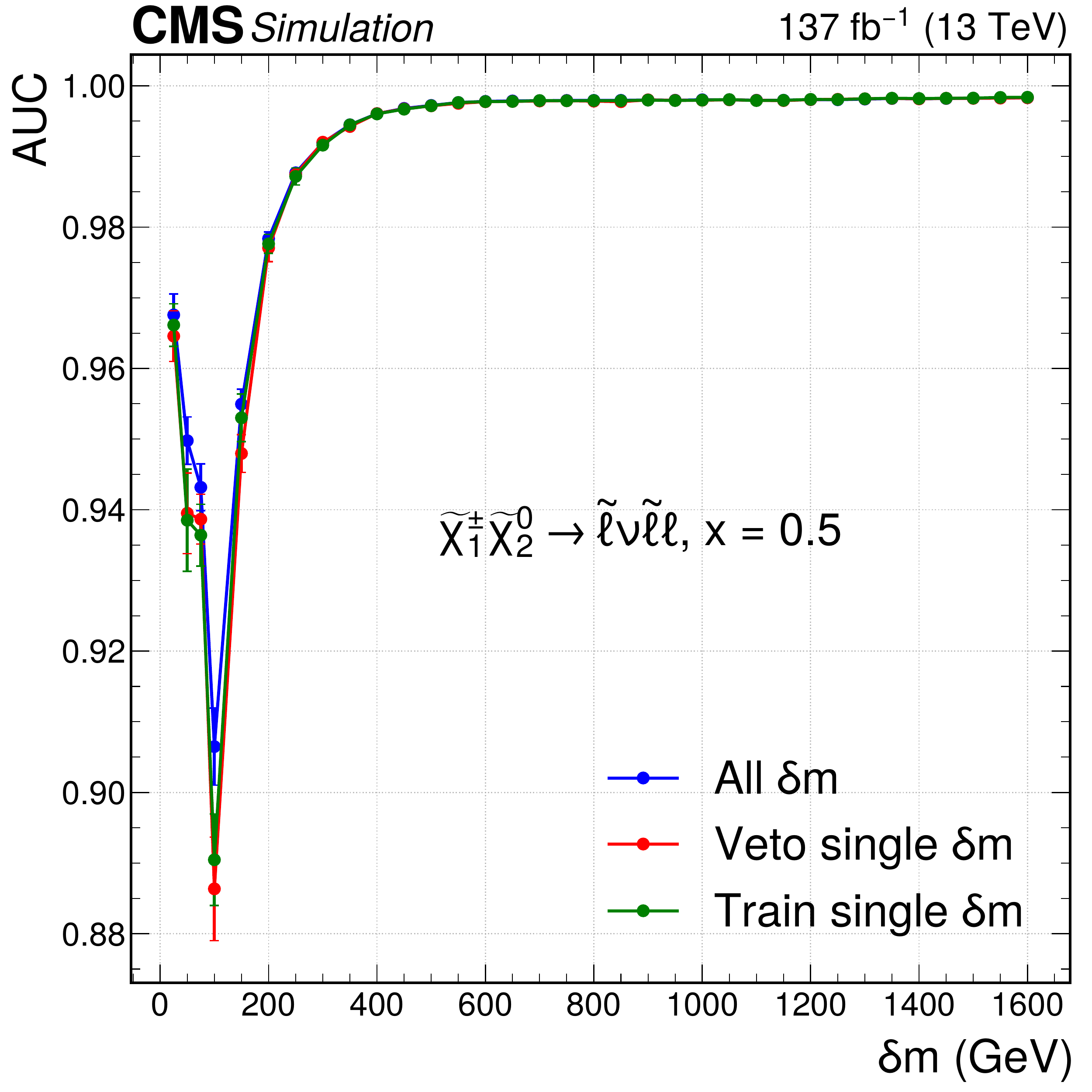}
\includegraphics[width=.32\textwidth]{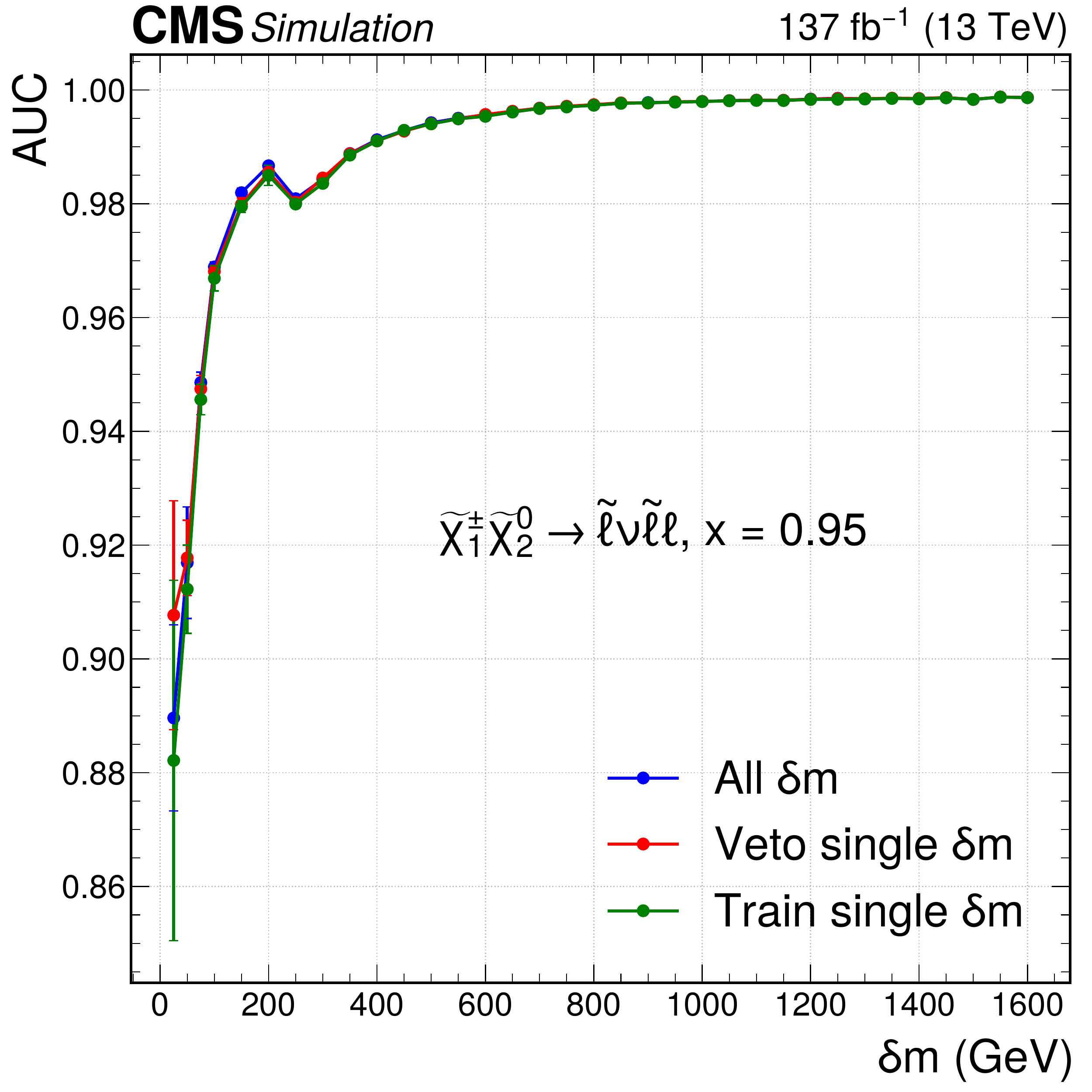}
\caption{The AUC performance of the parametric neural networks for discriminating the signal from the total background predicted in simulation, as a function of \dm for the trainings targeting different signal models. The top row corresponds to the neutral network targeting signals with \WZ-mediated decays with different mass ranges to show all points. The bottom row corresponds to the models with slepton-mediated decays at $x=0.05$ (left), $x=0.5$ (middle) and $x=0.95$ (right). Neural network models shown in blue are trained using all available \dm points, those in red are trained with all available points except the point for which the performance is shown. The models in green are not parametric and only trained to find a signal at the point where the performance is indicated. Each neural network is retrained ten times, and the mean performances are shown, with error bars indicating the standard deviation computed from ten performance values. This means that each red and green point correspond to ten neural network trainings. The entire blue curve in each figure also corresponds to ten trainings.}
\label{fig:neuralnet_interpolation}
\end{figure}

It is explicitly verified that the trained parametric networks are optimally performing at each \dm point, and able to interpolate to unseen points. The ability of the network to interpolate to a particular point is checked by training a parametric neural network excluding all events at a particular \dm value as well as a nonparametric network trained for just events at this \dm value. If both the new parametric model and our nominal one perform equally well it implies that the latter performs equally well on seen and unseen parameter points. The comparison between the nominal model and the nonparametric network tells us if the network's parametrization performs optimally or not. This check is repeated for each \dm point present in each of the signal models, training 10 neural networks of each type at each point to estimate the variations due to random weight initializations. It is found that the parametric network is able to achieve optimal performance at each \dm point present in the signal simulation even without explicitly seeing it during training, as shown in Fig.~\ref{fig:neuralnet_interpolation}.

The signal and background predictions, as well as the yield in data, are then evaluated for each of the four neural networks at every \dm value present in the signal simulation. For the interpretation of the results in a particular signal model at a given \dm only the corresponding neural network output is used. At each \dm value, the neural network output is binned in terms of the expected background yields. The last bin is defined to have a single expected background event in the 2016 data set, corresponding to 35.9\fbinv, and each preceding bin has twice the expected yield of the following bin. The shape of the outputs of the neural networks varies substantially with the \dm parameter, and this method allows for a robust binning definition across all values of \dm.

Aside from the neural network, a set of search regions is also defined to extract the signals from the background in a cut-based manner. Most of the SM \WZ background, as well as \xpmxtwo production with \PW and \PZ boson mediated decays result in \mll values close to the \PZ boson mass. For this reason, the search regions with $75 < \mll < 105 \GeV$ are optimized for finding \WZ-mediated signal decays, while the other search regions are optimized for finding slepton-mediated decays. The search region definitions are given in Tables~\ref{tab:SR3lA_1} and~\ref{tab:SR3lA_2}, using some, but not all, of the neural network input variables to define the bins. The \WZ background falls off quickly when \mt exceeds the \PW boson mass (\mw), making it a powerful tool to reduce the background. For signal events with slepton-mediated decays, \mll and \mtlll provide sensitivity to \dm, and are used to separate signal and background events. The search regions targeting \WZ-mediated superpartner decays are further binned in \ptmiss and \HT. Due to escaping LSPs in signal events, their \ptmiss spectrum tends to be harder than that of SM events, a fact which is further enhanced at large \HT.

The neural network analysis and search regions use the same data and thus can not be analyzed simultaneously. The results of both approaches are interpreted separately, and shown in Section~\ref{sec:interpretations}. The neural network approach has higher sensitivity, while the search region results are easier to reinterpret.

\begin{table}
\centering
\topcaption{Definition of the search regions used for events with three light leptons, at least two of which form an OSSF pair, excluding those with $75 < \mll < 105 \GeV$ (AXX).}
\label{tab:SR3lA_1}
\tableSlug{3$\ell$A : three light leptons with at least one OSSF pair, $\mll \leq 75 \GeV$ or $\mll \geq 105 \GeV$}
\cmsTableResize{
    \begin{tabular}{c|c|c|c|c|c}

        \multicolumn{1}{c}{$\mt$ (\GeVns)} & \multicolumn{1}{c}{$\mtlll$ (\GeVns)} & \multicolumn{1}{c}{$ \mll < 50\GeV$} & \multicolumn{1}{c}{$ 50 \leq \mll < 75\GeV $} & \multicolumn{1}{c}{$105 \leq \mll < 250 \GeV$} & \multicolumn{1}{c}{$\mll \geq 250 \GeV$} \\ \hline
        \multirow{5}{*}{0--100}   & 0--50&  A01 & \multirow{2}{*}{ A06} & \multirow{3}{*}{ A13} & \multirow{3}{*}{ A19} \\ \cline{2-3}
                                      & 50--100 &  A02 & & & \\  \cline{2-4}
                                      & 100--400 & \multirow{2}{*}{ A03} &  A07 & & \\ \cline{2-2}   \cline{4-6}
                                      & $\geq$400 & &  A08 &  A14 &  A20 \\ \hline

        \multirow{2}{*}{100--200} & 0--200 & \multirow{2}{*}{ A04} &  A09 &  A15 & \multirow{2}{*}{ A21} \\ \cline{2-2}   \cline{4-5}
                                      & $\geq$200 &  &  A10 &  A16 & \\ \hline

        \multirow{2}{*}{$\geq$200} & 0--400 & \multirow{2}{*}{ A05} &  A11 &  A17 & \multirow{2}{*}{ A22} \\   \cline{2-2} \cline{4-5}
                                      & $\geq$400 &  &  A12 &  A18 \\ \hline
\end{tabular}}
\end{table}

\begin{table}
\centering
\topcaption{Definition of the search regions used for events with three light leptons, at least two of which form an OSSF pair, and which satisfy $75 < \mll < 105 \GeV$ (AXX).}
\label{tab:SR3lA_2}
\tableSlug{3$\ell$A : three light leptons with at least one OSSF pair, $75 < \mll < 105 \GeV$ }
    \begin{tabular}{c|c|c|c|c}

        \multicolumn{1}{c}{$\mt$ (\GeVns)} & \multicolumn{1}{c}{\ptmiss (\GeVns)} & \multicolumn{1}{c}{$\HT < 100 \GeV$} & \multicolumn{1}{c}{$100 \leq \HT < 200 \GeV$} & \multicolumn{1}{c}{$\HT \geq 200 \GeV$} \\ \hline
        \multirow{6}{*}{ 0--100 }   & 50--100  &  A23 &  A36 & \multirow{2}{*}{  A49} \\ \cline{2-4}
                                        & 100--150 &  A24 &  A37 &  \\ \cline{2-5}
                                        & 150--200 &  A25 &  A38 & \multirow{2}{*}{  A50} \\ \cline{2-4}
                                        & 200--250 &  A26 &  A39 &  \\ \cline{2-5}
                                        & 250--350 & \multirow{2}{*}{  A27} & \multirow{2}{*}{  A40} &  A51 \\ \cline{2-2} \cline{5-5}
                                        & $\geq$350 & & &  A52 \\ \hline

        \multirow{6}{*}{ 100--160 } & 50--100  &  A28 &  A41 &  A53 \\ \cline{2-5}
                                        & 100--150 &  A29 &  A42 &  A54 \\ \cline{2-5}
                                        & 150--200 &  A30 &  A43 &  A55 \\ \cline{2-5}
                                        & 200--250 & \multirow{3}{*}{  A31} & \multirow{3}{*}{  A44} &  A56 \\ \cline{2-2} \cline{5-5}
                                        & 250--300 & & &  A57 \\ \cline{2-2} \cline{5-5}
                                        & $\geq$300 & & &  A58 \\ \hline

        \multirow{6}{*}{ $\geq$160 }  & 50--100  &  A32 &  A45 &  A59 \\ \cline{2-5}
                                        & 100--150 &  A33 &  A46 &  A60 \\ \cline{2-5}
                                        & 150--200 &  A34 &  A47 &  A61 \\ \cline{2-5}
                                        & 200--250 & \multirow{3}{*}{  A35} & \multirow{3}{*}{  A48} &  A62 \\ \cline{2-2} \cline{5-5}
                                        & 250--300 & & &  A63 \\ \cline{2-2} \cline{5-5}
                                        & $\geq$300 & &&  A64 \\ \hline

\end{tabular}
\end{table}

\subsubsection{Three light leptons without an OSSF pair}
Events with three light leptons that do not contain an OSSF pair (3$\ell$B) do not occur frequently in the SM, because most events with multiple leptons involve a \PZ boson decay. This category of events is particularly sensitive to signal models with nonresonant lepton production from the decay of an \PH. Since SM production of an \PH with an additional lepton is exceedingly rare, the search regions are designed to target possible $\PH \to \PW\PW$ decays in signal events. The events are binned in the minimum $\Delta R$ between any two leptons in the event ($\min(\Delta R (\ell,\ell))$), exploiting the increased collimation of leptons in $\PH \to \PW\PW$ events when compared to events with nonprompt leptons or nonresonant $\PW\PW$ production. The search region definitions are given in Table~\ref{tab:SR3lB}.

\begin{table}
\centering
\topcaption{Definition of the search regions used for events with three light leptons, none of which form an OSSF pair (BXX).}
\label{tab:SR3lB}
\tableSlug{3$\ell$B: three light leptons without an OSSF pair}
\begin{tabular}{c|c|c}
    \multicolumn{1}{c}{ $\min(\Delta R (\ell,\ell))  < 0.4$} & \multicolumn{1}{c}{$0.4 \leq \min(\Delta R(\ell,\ell))  < 1$} & \multicolumn{1}{c}{$\min( \Delta R (\ell,\ell)) \geq 1$} \\ \hline
     B01                        &  B02                            &  B03                      \\ \hline
\end{tabular}
\end{table}

\subsubsection{Three leptons with one or more \tauh candidates}

If chargino or neutralino decays are mediated by right-handed sleptons, or the first- and second- generation sleptons are heavy and decoupled, signal events will favor final states with one or more $\tau$ leptons. To retain sensitivity to such models, events with \tauh candidates are selected and split into further categories.

The first category consists of events with an OSSF pair of light leptons and a \tauh candidate (3$\ell$C). These events are mainly sensitive to $\tau$-enriched \xpmxtwo production and contain a large background from Drell--Yan and top quark pair production (\ttbar) events with a nonprompt \tauh candidate. Events are required to have $\abs{\mll - \mz} > 15 \GeV$, with \mll being the mass of the light lepton pair, to veto the bulk of the Drell--Yan background. At low and moderate \ptmiss values, \mtt is used to reduce the \ttbar background. At higher \ptmiss values, the transverse mass of the combined dilepton system and \ptvecmiss (\mtll), a proxy for the \xtwo mass, is found to be a strong discriminator. The full set of search region definitions in this category is given in Table~\ref{tab:SR3lC}.

\begin{table}
\centering
\topcaption{Definition of the search regions for events with a $\PGmp\PGmm$ or $\Pep\Pem$ pair and an additional \tauh candidate (CXX).}
\label{tab:SR3lC}
\tableSlug{3$\ell$C: $\PGmp\PGmm$ or $\Pep\Pem$ + \tauh }
\begin{tabular}{c|c|c|c|c}
    \multicolumn{1}{c}{\ptmiss (\GeVns)} & \multicolumn{1}{c}{\mtll (\GeVns)} &  \multicolumn{1}{c}{ $\mtt < 80 \GeV$} & \multicolumn{1}{c}{ $80 \leq \mtt < 120 \GeV$} & \multicolumn{1}{c}{ $\mtt \geq 120 \GeV$} \\ \hline
    50--200  & $\geq$0 &  C01 &  C02 &  C03 \\\hline

    200--300 & $\geq$0  &  C04 &  C05 &  C06 \\ \hline

    \multirow{3}{*}{ $\geq$300 } & 0--250 & \multicolumn{3}{c}{ C07} \\  \cline{2-5}
    & 250--500 & \multicolumn{3}{c}{ C08} \\ \cline{2-5}
    & $\geq$500 & \multicolumn{3}{c}{ C09} \\ \hline
    \end{tabular}
\end{table}

Events with a single \tauh in which the light leptons do not form an OSSF pair are split according to whether the light leptons have opposite sign (3$\ell$D) or not (3$\ell$E). One of the discriminating variables in such events is the opposite-sign lepton pair mass closest to what is expected for a \PZ boson decay, called \mll (\mlt) in 3$\ell$D (3$\ell$E) events. The expected reconstructed mass of a $\PZ \to \PGt\PGt$ decay is 50\GeV in $\Pe\PGm$ pairs and 60\GeV in $\Pe\tauh$ and $\PGm\tauh$ pairs. For events in 3$\ell$E where the \tauh is of the same sign as the light leptons, \mll is set to zero. Additional discrimination power is provided by the stransverse mass, computed with the two light leptons in category 3$\ell$D (\mttll), and with the leading light lepton and \tauh candidate in category 3$\ell$E (\mttlt). This variable has a sharply falling distribution beyond \mw in the SM. Definitions of the search regions in category 3$\ell$D (3$\ell$E) can be found in Table~\ref{tab:SR3lD} (\ref{tab:SR3lE}).

\begin{table}
\centering
\topcaption{Definition of the search regions for events with a $\Pe^{\pm}\PGm^{\mp}$ pair and a \tauh candidate (DXX).}
\label{tab:SR3lD}
\tableSlug{3$\ell$D: $\Pe^{\pm}\PGm^{\mp}$ + \tauh}
\begin{tabular}{c|c|c|c|c}
    \multicolumn{1}{c}{\mttll (\GeVns)} & \multicolumn{1}{c}{\ptmiss (\GeVns)} & \multicolumn{1}{c}{$\mll < 60\GeV$} &  \multicolumn{1}{c}{$60 \leq \mll < 100 \GeV$}  & \multicolumn{1}{c}{$\mll \geq 100\GeV$} \\ \hline
    \multirow{5}{*}{0--100}   & 50--100 &  D01 &  D06 &  D11 \\ \cline{2-5}
                                  & 100--150 &  D02 &  D07 &  D12 \\ \cline{2-5}
                                  & 150--200 &  D03 &  D08 &  D13 \\ \cline{2-5}
                                  & 200--250 &  D04 &  D09 & \multirow{2}{*}{ D14} \\ \cline{2-4}
                                  & $\geq$250 &  D05 &  D10 &  \\ \hline
    \multirow{2}{*}{$\geq$100}   & 50--200 & \multicolumn{3}{c}{ D15} \\ \cline{2-5}
                                  & $\geq$200 & \multicolumn{3}{c}{ D16} \\ \hline
\end{tabular}
\end{table}

\begin{table}
\centering
\topcaption{Definition of the search regions for events with a pair of light leptons of the same sign and a \tauh candidate (EXX).}
\label{tab:SR3lE}
\tableSlug{3$\ell$E: same-sign light lepton pair + \tauh}
\cmsTableResize{\begin{tabular}{c|c|c|c|c}

    \multicolumn{1}{c}{$\mttlt (\GeVns)$} & \multicolumn{1}{c}{\ptmiss (\GeVns)} & \multicolumn{1}{c}{ $\mlt \leq 50 \GeV$ } & \multicolumn{1}{c}{ $50 < \mlt \leq 100 \GeV$} & \multicolumn{1}{c}{ $\mlt > 100 \GeV$ } \\ \hline
    \multirow{3}{*}{ 0--80 } & 50--100 &  E01 & \multicolumn{2}{c}{ E04} \\ \cline{2-5}
                                          & 100--250 &  E02 & \multicolumn{2}{c}{\multirow{2}{*}{ E05}} \\ \cline{2-3}
                                          & $\geq$250 &  E03 & \multicolumn{2}{c}{} \\ \hline

    \multirow{3}{*}{ $\geq$80 } & 50--150 & \multicolumn{2}{c|}{ E06} & \multirow{2}{*}{ E08} \\ \cline{2-4}
                                         & 150--200 & \multicolumn{2}{c|}{\multirow{2}{*}{ E07}} & \\ \cline{2-2} \cline{5-5}
                                         & $\geq$200 & \multicolumn{2}{c|}{} &  E09 \\ \hline
\end{tabular}}
\end{table}

Events with two \tauh candidates provide additional sensitivity to models with $\tau$ dominated slepton decays. Events in this category are binned in the invariant mass of the leading \tauh and light lepton (\mlt), which tends to be high for uncompressed signal events. The same lepton pair and the \ptvecmiss enter the computation of the stransverse mass (\mttlt), which is used to further suppress the SM background. The complete set of bins is shown in Table~\ref{tab:SR3lF}.

\begin{table}
\centering
\topcaption{Definition of the search regions for events with 2 \tauh candidates and one light lepton (FXX).}
\label{tab:SR3lF}
\tableSlug{3$\ell$F: 2\tauh + light lepton}
\begin{tabular}{c|c|c|c}
    \multicolumn{1}{c}{\mttlt (\GeVns)} & \multicolumn{1}{c}{\ptmiss (\GeVns)} & \multicolumn{1}{c}{$\mlt < 100 \GeV$} & \multicolumn{1}{c}{$\mlt \geq 100 \GeV$} \\ \hline
    \multirow{6}{*}{0--100}   & 50--100 &  F01 &  F07 \\ \cline{2-4}
                                  & 100--150 &  F02 &  F08 \\ \cline{2-4}
                                  & 150--200 &  F03 &  F09 \\ \cline{2-4}
                                  & 200--250 &  F04 & \multirow{3}{*}{ F10} \\ \cline{2-3}
                                  & 250--300 &  F05 &  \\ \cline{2-3}
                                  & $\geq$300 &  F06 &  \\ \hline
    \multirow{2}{*}{$\geq$100}   & 50--200 & \multicolumn{2}{c}{ F11} \\ \cline{2-4}
                                  & $\geq$200 & \multicolumn{2}{c}{ F12} \\ \hline
\end{tabular}
\end{table}

\subsection{Four or more lepton events} \label{sec:four_leptons}
Events with four leptons provide sensitivity to effective \xonexone production with subsequent decays via \PH or \PZ bosons. Further categorization of the events is done depending on the number of OSSF pairs and light leptons. If more than four loose leptons are present in an event, the event categorization and computation of analysis variables uses the four highest \pt leptons.

\begin{table}
\centering
\topcaption{Definition of the search regions for events with 4 light leptons, including 2 separate OSSF pairs (GXX).}
\label{tab:SR4lG}
\tableSlug{4$\ell$G: 4 light leptons with 2 separate OSSF pairs}
\begin{tabular}{c|c|c}
    \multicolumn{1}{c}{\mttzz (\GeVns)} & \multicolumn{1}{c}{$\mztwo \geq 60 \GeV$} & \multicolumn{1}{c}{$\mztwo < 60 \GeV$} \\ \hline
    0--150 & \multicolumn{2}{c}{ G01} \\ \hline
    150--250 &  G02  &  G03 \\ \hline
    250--400 & \multicolumn{2}{c}{ G04} \\ \hline
    $\geq$400 & \multicolumn{2}{c}{ G05} \\ \hline
\end{tabular}
\end{table}

Decays of \xonexone via two \PZ bosons tend to give two OSSF pairs. For this reason, the first category consists of events with four light leptons forming two separate OSSF pairs (4$\ell$G). The OSSF dilepton pair with the closest invariant mass to \mz forms the first \PZ boson candidate ($\PZ_{1}$), while the remaining OSSF pair is taken to be the second \PZ boson candidate ($\PZ_{2}$). The \mtt computed with both \PZ boson candidates (\mttzz) is expected to have a sharply falling distribution beyond $m_{\text{NLSP}}$, providing a handle to separate different signal points and to discriminate signal from the background. Events are further binned in the mass of the $\PZ_{2}$ candidate (\mztwo) to enhance the sensitivity to signal models without two \PZ bosons in the \xonexone decay. The search region definitions are listed in Table~\ref{tab:SR4lG}.

The remaining events are further split up as follows: four light leptons forming one or no OSSF pairs (4$\ell$H), one \tauh candidate and three light leptons (4$\ell$I), two \tauh candidates and two light leptons forming two OSSF pairs (4$\ell$J), and two \tauh and two light leptons forming one or fewer OSSF pairs (4$\ell$K). The same binning is used in each of these categories, as they are sensitive to the same signal models, and it is shown in Table~\ref{tab:SR4lHIJK}. If at least one OSSF pair is present, the OSSF pair of mass closest to \mz is taken to reconstruct a $\PZ$ boson candidate ($\PZ1$). If no OSSF pair is present, other opposite sign lepton combinations are considered when finding the \PZ boson candidate. The \PZ boson candidate mass \mzone is used to discriminate between processes with and without a true on-shell \PZ boson involved. The remaining two leptons in the event are assigned to be the decay of an \PH candidate. Events are further subdivided according to the $\Delta R$ between those two remaining leptons ($\Delta R^{\PH}$), as these are expected to be collimated if they are from genuine \PH decay products.

\begin{table}
\centering
\topcaption{Definition of the search regions for events with 4 leptons with one or more \tauh, or without two light-lepton OSSF pairs (XYY).}
\label{tab:SR4lHIJK}
\tableSlug{4$\ell$H-K : 4 leptons with one or more \tauh or without two light-lepton OSSF pairs}
\begin{tabular}{c|c|c}
    \multicolumn{1}{c}{\mzone (\GeVns)}  & \multicolumn{1}{c}{$\Delta R^{\PH} < 0.8$} & \multicolumn{1}{c}{$\Delta R^{\PH} \geq 0.8$} \\ \hline
    \multicolumn{1}{c|}{ 0--60}  & \multirow{2}{*}{ X03} &  X02 \\ \cline{1-1} \cline{3-3}
    \multicolumn{1}{c|}{ $>$60 } & &  X01 \\ \hline
\end{tabular}
\end{table}

\section{Background estimation} \label{sec:background}

The background contributions in each of the search categories can be subdivided into four distinct categories. Firstly, there are SM events with three or more prompt leptons, or two prompt leptons of same sign. Secondly, external and internal conversions of photons also result in events entering our search region. Backgrounds from both conversions and prompt leptons are estimated using simulated samples. Thirdly, backgrounds involving one or more nonprompt leptons are directly predicted from data. Lastly, events that enter a particular event category due to the mismeasurement of a lepton sign are estimated from data for events in the 2$\ell$SS and 3$\ell$E categories, while its importance is minute in other event categories and is estimated from simulation.

The dominant background contribution to events in the 3$\ell$A category comes from \WZ production. With leptonic decays of the \PZ and \PW bosons, respectively, \WZ production results in events with three prompt leptons and a neutrino giving sizable \ptmiss, thus mimicking the topology of \xpmxtwo production. The background is estimated from simulation and is validated in a control region that is contained within the search regions but nearly
depleted of signal events. The control region has the same selection criteria as 3$\ell$A events, with the following additional requirements: $\abs{\mll - \mz} < 15 \GeV$, $50 < \ptmiss < 100 \GeV$, $50 < \mt < 100 \GeV$ and $\abs{\mlll - \mz} > 15 \GeV$. A fit is performed to the data in the \WZ CR, in which the \WZ normalization is free to float. This fit takes into account all relevant analysis uncertainties and assumes that no signal is present. The result is a normalization factor of $r_{\WZ} = 1.17 \pm 0.05$ over the \POWHEG prediction, which is applied in the analysis. The 3$\ell$A events are interpreted twice, once using the neural network and once using the search regions. The \WZ control region is included into the fit for the signal region interpretation and, to avoid double counting the data, the partially overlapping search regions A23, A36, and A49 are removed from it. When the neural network results are used for the interpretation, the \WZ control region is excluded from the fits used for the interpretation of the results in terms of superpartner production.

One of the most important discriminating variables used in both the parametric neural networks and the 3$\ell$A search regions is \mt, the transverse mass of the lepton not forming the \PZ boson candidate. Simulation studies indicate that the tails of the \mt distribution mainly originate from the mispairing of leptons when forming the \PZ boson candidate, leading to \mt being computed with one of the leptons from the \PZ boson decay. The prediction of such mispaired events is validated by selecting $\Pe\Pe\PGm$ and $\Pe\PGm\PGm$ events in the aforementioned \WZ control region. In these events, where there is no ambiguity when assigning leptons to \PW and \PZ boson decays, the leptons are intentionally mispaired, and the simulated predictions are validated by comparison to data. A second, though smaller, source of events in the \mt tails comes from \ptvecmiss mismeasurements. This effect is studied in $\PGm\gamma$ events enriched in the \Wg process where no possible lepton ambiguity may arise. The muon is required to have $\pt > 25\,(28) \GeV$ in 2016 (2017 and 2018) data. To reduce the contribution from final-state radiation (FSR) photons, which affect the \mt distribution, the photon is required to have $\pt > 40 \GeV$, and be separated by $\Delta R > 0.3$ from the muon. Additional requirements are $\mt > 40 \GeV$ and $\ptmiss > 50 \GeV$. After this selection, the simulated \Wg prediction is compared to the data in bins of \mt and \ptmiss to derive uncertainties based on the agreement, which are applied to the \WZ prediction. The uncertainties vary from around 5\% at low \ptmiss and \mt to slightly below 30\% at high \ptmiss and \mt. The relative uncertainties derived in the $\PGm\gamma$ region thus provide an upper bound on the uncertainty in the \WZ prediction, as a function of \mt and \ptmiss.

Production of \ZZ with subsequent leptonic \PZ boson decays leads to final states with four leptons. This process is the largest background in the four-lepton categories, while also entering the three lepton categories in case one of the leptons fails to be identified. The normalization of \ZZ is constrained using a four-lepton control region with identical selection to the 4$\ell$G region, but an inverted requirement $\ptmiss < 50 \GeV$. The predictions as a function of the analysis variables are also checked in this control region. We perform a fit to the control region, in which the \ZZ normalization is free to float, all analysis uncertainties are included and the signal presence is suppressed. The fit results in a normalization factor of $r_{\ZZ} = 1.02 \pm 0.10$ over the \POWHEG simulation prediction. Contributions from gluon induced (\ggZZ), quark induced (\qqZZ), and resonant (\hZZ) production are split for plotting purposes in the four lepton categories where they are the most relevant background sources. For categories with three or fewer leptons, all contributions to \ZZ production are grouped together under the \ZZ/\PH label in figures.

Processes involving one or more top quarks and electroweak bosons can produce many prompt leptons and contribute to all of the analysis final states. The main contributions come from \tttt, \ttH, \ttW, and \ttZ, which are collectively labeled \ttX. Smaller contributions originate from processes with a single top quark or with top quarks and multiple electroweak bosons and are labeled \tX. These are minor backgrounds because of their small cross sections, and they are further reduced by the \PQb veto applied in the event selection. Even smaller contributions appear due to processes in which two top quarks and two massive bosons are produced, labeled \ttXX, which have cross sections so small that barely one event in total appears at the analysis level. The predictions for the \ttZ background are verified in a \ttZ-enriched control region with the selection of the 3$\ell$A category, but requiring at least one \PQb jet, and $\abs{\mll - \mz} < 15 \GeV$. All other contributions are estimated from simulation. For plotting purposes, the \ttX and \tX contributions are grouped together.

Rare processes involving the production of three or more electroweak bosons (\WWW, \WWZ, \WZZ, \ZZZ) can also lead to events with enough prompt leptons to enter the search regions. These processes have extremely small cross sections, and thus constitute only a tiny fraction of the background. Their contributions are estimated from simulation. For plotting purposes this contributions are labeled as multiboson.

Internal and external photon conversions can lead to additional leptons in an event. Such events typically enter the search regions when the conversion is asymmetric and one of the leptons coming from the conversion has a very low \pt and fails to be reconstructed. This background is dominated by \Zg for events with three or more leptons, while \Wg provides the dominant source for 2$\ell$SS events. The conversion background is estimated from simulation, which is validated and normalized in a \Zg-enriched control region in data. This region is obtained by applying the 3$\ell$A selection with inverted requirements $\ptmiss < 50 \GeV$, and $\mll < 75 \GeV$. The last requirement is applied because asymmetric conversions from \Zg tend to have \mlll rather than \mll values close to the \PZ boson mass. Processes involving a final state photon (\Zg, \Wg, \ttg) which undergoes an asymmetric conversion are included into the \Xg background group for plotting and fitting purposes. A fit is performed in the control region, in which the photon conversion process normalization is free to float, all analysis uncertainties are included, and the signal presence is suppressed. This leads to a normalization factor of $r_{\Xg} = 1.12 \pm 1.10$ over the \MGvATNLO simulation prediction.

Events with nonprompt leptons entering the search regions come mostly from \ttbar and Drell--Yan production with an additional nonprompt lepton. It is a dominant background source in categories 3$\ell$B, 2$\ell$SS, and all of the categories involving one or more \tauh candidates. This background contribution is estimated from data using the ``tight-to-loose'' ratio method, as described in Ref.~\cite{Khachatryan:2016kod}. The probability for a loose nonprompt lepton to also pass the tight lepton selection, the ``nonprompt rate'', is measured as a function of \pt and \abseta. For light leptons, this is done in a QCD-enriched sample of single lepton events. The nonprompt rate of \tauh candidates is measured in both Drell--Yan- and \ttbar-enriched events. These nonprompt rates differ due to the flavor content of the jets in the event. In the 3$\ell$D and 3$\ell$E categories the background from nonprompt $\tau$ leptons is expected to be dominated by \ttbar, so the \ttbar based nonprompt rate measurement is used. For 3$\ell$C and 3$\ell$F events, Drell--Yan is the dominant background source, and the nonprompt rates measured in Drell--Yan enriched data are used. The measured nonprompt rates are applied to events passing the search region selection but with one or more leptons failing the tight selection while still passing the loose selection. Both simulated events and a data sample enriched in nonprompt leptons are used to validate the method.

Electron sign mismeasurements are an important source of background in 2$\ell$SS and 3$\ell$E events. This background is reduced by the application of additional requirements on the leptons designed to ensure a well-determined sign, as discussed in Section~\ref{sec:selection}. The remaining background for electron sign mismeasurement is predicted from data in 2$\ell$SS and 3$\ell$E events. The probability for an electron sign mismeasurement is computed as a function of \pt and \abseta in a large sample of simulated events from Drell--Yan, \ttbar, and diboson production. The resulting background contribution in the search regions is then determined by applying this probability to data events with two light leptons of opposite sign. A sample of same-sign dilepton events dominated by Drell--Yan is selected by requiring $\abs{\mll - \mz} < 10 \GeV$, in which the predictions are validated, and an integral normalization factor is measured for each data-taking year by which the predictions are scaled. The scale factors are 0.89, 1.26 and 1.17 in the 2016, 2017 and 2018 data sets, respectively. Studies in simulated events indicate that the probability of sign misidentification for muons is negligible, and the minuscule background contribution that results in the search regions is estimated using simulation.

\section{Systematic uncertainties} \label{sec:systematics}

Several sources of systematic uncertainties affect both the background and signal predictions, changing both the total yields and the contribution of each process to the different analysis bins. The experimental sources of uncertainty that affect the simulated samples are pileup modeling, jet energy scale and resolution, \PQb tagging, lepton identification and trigger efficiencies, \ptmiss resolution, and the measurement of the integrated luminosity. Additional sources of systematic uncertainty come from the uncertainties in the theoretical calculations used to generate samples of simulated events. The effects of each of these uncertainties, aside from those associated with the integrated luminosity and the trigger efficiency, vary across the analysis bins.

Light lepton identification efficiencies are measured in a \PZ boson enriched data sample using the ``tag-and-probe'' technique~\cite{Khachatryan:2015hwa,Sirunyan:2018fpa}. The corresponding corrections are applied to simulated events. Uncertainties on these measured corrections, as well as on the validity of their extrapolation to the search regions, are applied to simulated events. Signal events are expected to contain relatively high \pt leptons because of the potentially large superpartner masses. For this reason the lepton efficiencies are measured separately for events with a reconstructed \PZ boson \pt above and below 80\GeV. The difference between the corresponding corrections at high and low \pt of the \PZ boson is taken as the uncertainty in the application of these corrections, and is around 0.5\% for most of the leptons, but ranges up to 3\% for very high and low \pt leptons.

Similarly, identification efficiencies for \tauh candidates are measured in $\PGm\tauh$ events enriched in \PZ bosons for \pt values up to approximately 60\GeV. For \tauh candidates with intermediate \pt values, up to 100\GeV, \ttbar enriched $\PGm\tauh$ events are used. At even higher \pt values the efficiencies are measured using single \tauh events enriched in highly virtual \PW bosons. The associated uncertainties in the measured efficiencies applied in the analysis range from 1 to 3\%.

The uncertainty in the correction of the number of events per bunch crossing applied to simulated events is estimated by varying the total $\Pp\Pp$ inelastic cross section up and down by $4.6\%$~\cite{Sirunyan:2018nqx}. The uncertainty in the measurement of the integrated luminosity, used to normalize all simulated yields, is 2.3 (2.5)\% for the data set collected in 2017~\cite{CMS-PAS-LUM-17-004} (2016~\cite{CMS-PAS-LUM-17-001} and 2018~\cite{CMS-PAS-LUM-18-002}). The integrated luminosity of the total data set has an uncertainty of 1.8\%, where the improved understanding comes from the independence of some parts of the uncertainty between data-taking years. The correlated (uncorrelated) components correspond to 1.2, 1.1 and 2.0 (2.2, 2.0 and 1.5\%) for the 2016, 2017 and 2018 data-taking years.  The trigger efficiency is measured in an unbiased sample of events, triggered on the \ptmiss and total hadronic momentum in the event. The uncertainties in the trigger efficiency range from 1.4\% for events with three or more light leptons to 3\% for events in the 2$\ell$SS and 5\% for category 3$\ell$F events have less redundancy to pass the leptonic trigger requirements. The trigger uncertainties are split into a component correlated across years accounting for possible biases in the method used to measure trigger efficiencies and an uncorrelated component per year because of the limited data available in the data sideband used for such measurements. The latter accounts for effects of roughly $1\%$ for each category and year with the former accounting for the rest of the uncertainty size.

The uncertainties due to the jet energy scale are computed by varying the scale for all jets up and down within its uncertainty. Similarly, the uncertainties from the jet energy resolution are estimated by smearing the jets according to the resolution uncertainty~\cite{Chatrchyan:2011ds,Khachatryan:2016kdb}. Both effects are subsequently propagated to all steps of the analysis, affecting \ptvecmiss, the \PQb veto and all analysis variables calculated using jets or \ptvecmiss~\cite{Khachatryan:2016kdb}. The \ptvecmiss is affected by additional resolution uncertainties due to objects not clustered into jets, which are also propagated to all affected analysis variables. Corrections are applied to account for differences between data and simulation in the \PQb tagging efficiency and misidentification rate. Uncertainties in this correction affect the \PQb veto used in the analysis, and the effects are propagated to all analysis bins. These effects are partially correlated across data-taking years accounting for the possible year dependency of the origin of each source of uncertainty included into these variations. The overall effect of the correlated and uncorrelated components is approximately equal for all data-taking years.

Uncertainties stemming from a limited knowledge of the proton PDFs are estimated using a set of NNPDF3.0 (NNPDF3.1) replicas in simulations corresponding to 2016 (2017 and 2018) data-taking conditions. Uncertainties stemming from missing higher order corrections are estimated by varying the renormalization and factorization scales up and down simultaneously by a factor two and evaluating the effect on simulated events. Both of these theoretical uncertainty sources lead to changes in the predicted cross sections of simulated processes, as well as additional kinematic variations across analysis bins. The shape variations are taken into account for all simulated events, whereas for several processes the cross section uncertainties are replaced by a prior uncertainty which is constrained by a fit to data. This is the case for \WZ, \ZZ and \Zg processes.

The experimental and theoretical uncertainties listed earlier in this section affect all simulated processes, both signal and background, and the effects are considered correlated across processes. A number of additional process-specific systematic uncertainty sources are taken into account.

The modeling of QCD ISR in signal events is done by \MGvATNLO and affects the total ISR transverse momentum of the events (\ptisr). The \ptisr distribution in 2016 signal events is reweighted based on the \PZ boson \pt spectrum observed in data~\cite{Chatrchyan:2013xna}. Differences between corrected and uncorrected signal events are taken into account as systematic uncertainties. For 2017 and 2018 data, the \ptisr distribution was found to be well modeled, but a small correction based on the distribution of the number of reconstructed jets in a \PZ boson enriched data sample is applied. The size of the corrections are also considered as uncertainties in this case.

As discussed in Section~\ref{sec:background}, \Wg data are used to validate the modeling of events with \ptvecmiss mismeasurements entering the \mt tail in \WZ events. The deviations from unity observed in \Wg as a function of \mt and \ptmiss serve as an upper bound on the potential uncertainties in \WZ as the \mt distribution of \Wg is more strongly affected by mismeasurements. The resulting uncertainties typically range from 10\% at low-\mt and \ptmiss to 20\% at high-\ptmiss values. These uncertainties are mainly driven by the size of the \Wg-enriched data sample.

A prior normalization uncertainty of 10\% is assigned to \WZ events, which is constrained implicitly by the fit to the data in 3$\ell$A events. Similarly prior normalization uncertainties of 10\% are assigned to both the \ZZ and \Zg processes, which are further constrained by including their respective control regions in the analysis fit. A normalization uncertainty of 30\% is assigned to the nonprompt lepton background prediction, covering any biases found in simulated studies of the method. Three separate uncorrelated nuisance parameters with priors of the same size are used for nonprompt light leptons, nonprompt \tauh candidates from \ttbar and nonprompt \tauh candidates from Drell--Yan production. A 20\% uncertainty is assigned to the normalization of the sign misidentification background to cover deviations observed in the \PZ boson enriched control region mentioned in Section~\ref{sec:background}.

A summary of the systematic uncertainties applied in the analysis, and their effects on the predicted event yields across analysis bins is shown in Table~\ref{tab:systematics}.

\begin{table}[ht]
\topcaption{Systematic uncertainty sources affecting the analysis, with their typical size across signal regions, and the treatment of the correlations across data-taking years. Uncertainties in the jet energy corrections and \PQb tagging efficiencies are considered separately for signal events which use CMS fast simulation, as explained in Section~\ref{sec:simulations}, and for the other simulated processes. Both the overall integrated luminosity and all normalization uncertainties have effects on the predicted yields of the corresponding processes that are of the same size across all signal regions. Their quoted typical size corresponds to the size of such variations. All other uncertainties can have different effects on the predicted yields for each process and signal region. Their typical uncertainty corresponds to the range of sizes that such effects take across the analysis search regions.}

\label{tab:systematics}
\cmsTableResize{\begin{tabular}{lcc}
        Source                                                & Typical uncertainty (\%)       & \begin{tabular}[x]{@{}c@{}}Correlation across\\data-taking years\end{tabular} \\ \hline
\Pe/\PGm efficiency                                   & 1--2 per lepton              & Correlated                            \\
\tauh efficiency                                      & 1--3 per lepton              & Correlated                            \\
Pileup                                                & 1--2                         & Correlated                            \\
Integrated luminosity                                 & $1.8$                          & Partially correlated                  \\
Trigger efficiency                                    & 1.4--5                       & Partially correlated                  \\
Jet energy corrections                                & $1$                            & Partially correlated                  \\
Jet energy corrections (fast simulation)              & $1$                            & Correlated                            \\
\PQb tagging efficiency                               & 1--3                         & Correlated                            \\
\PQb tagging efficiency (fast simulation)             & 1--3                            & Correlated                            \\
PDF                                                   & 1--10                           & Correlated                            \\
Renormalization and factorization scales              & 1--10                           & Correlated                            \\
Signal ISR                                            & 1--5                           & Correlated                            \\
Signal \ptmiss                                        & 1--2                           & Correlated                            \\
\WZ shape                                             & 5--30                           & Uncorrelated                          \\
\WZ normalization                                     & $10$                             & Correlated                            \\
\ZZ normalization                                     & $10$                             & Correlated                            \\
Conversion normalization                              & $10$                             & Correlated                            \\
Nonprompt normalization (\Pe/\PGm/\tauh)              & $30$                             & Correlated                            \\
Charge misidentification normalization                & $20$                             & Correlated                            \\
\ttX normalization                                    & $15$                             & Correlated                            \\
Multiboson normalization                              & $50$                             & Correlated                            \\
\end{tabular}}
\end{table}

\clearpage

\section{Results} \label{sec:results}

The observed and expected SM yields in each of the search regions introduced in Section~\ref{sec:strategy} are shown in this section. The expected yields are obtained using the background estimation procedures elucidated in Section~\ref{sec:background}, with systematic uncertainties as explained in Section~\ref{sec:systematics}.

\begin{figure}
\includegraphics[width=.33\textwidth]{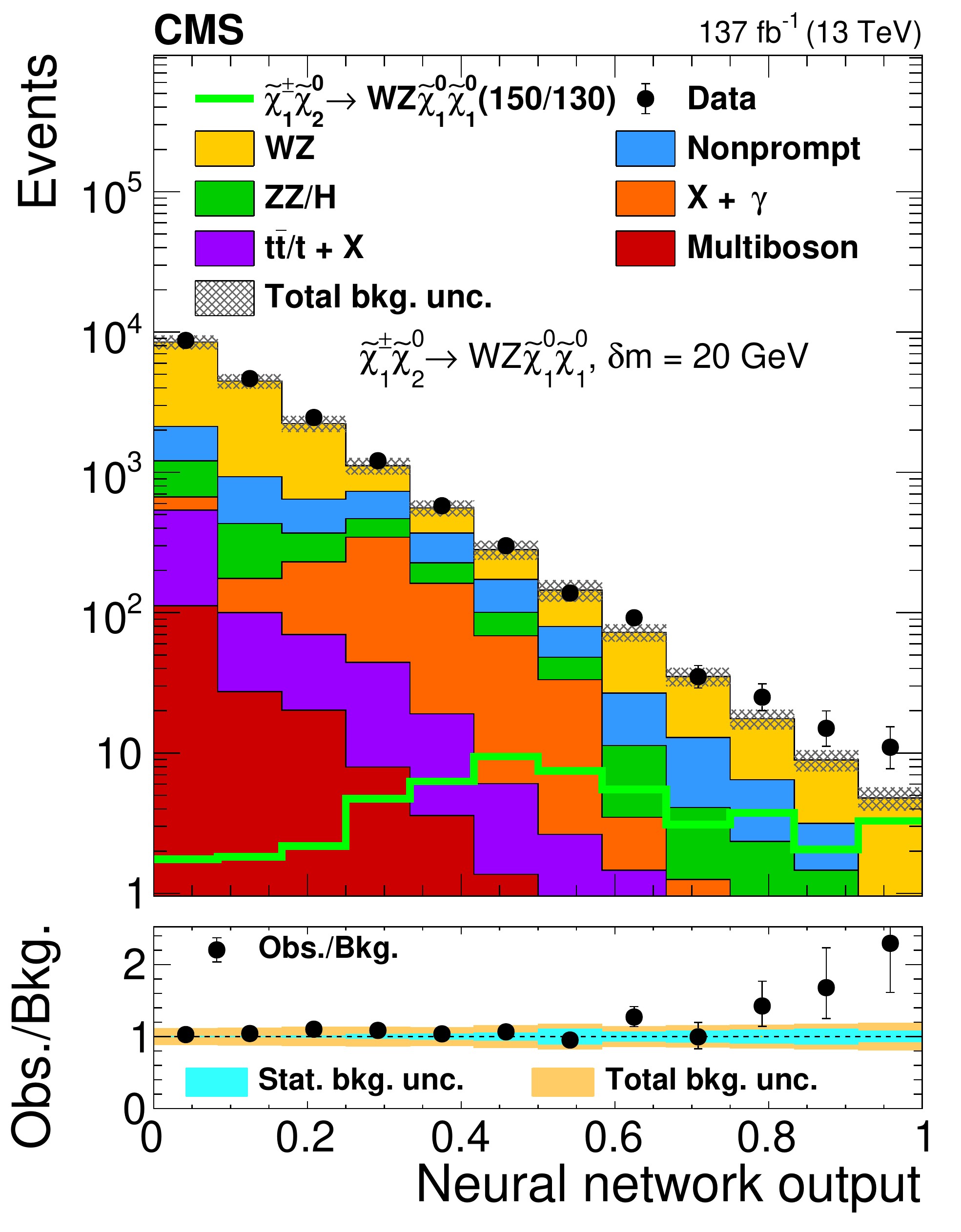}
\includegraphics[width=.33\textwidth]{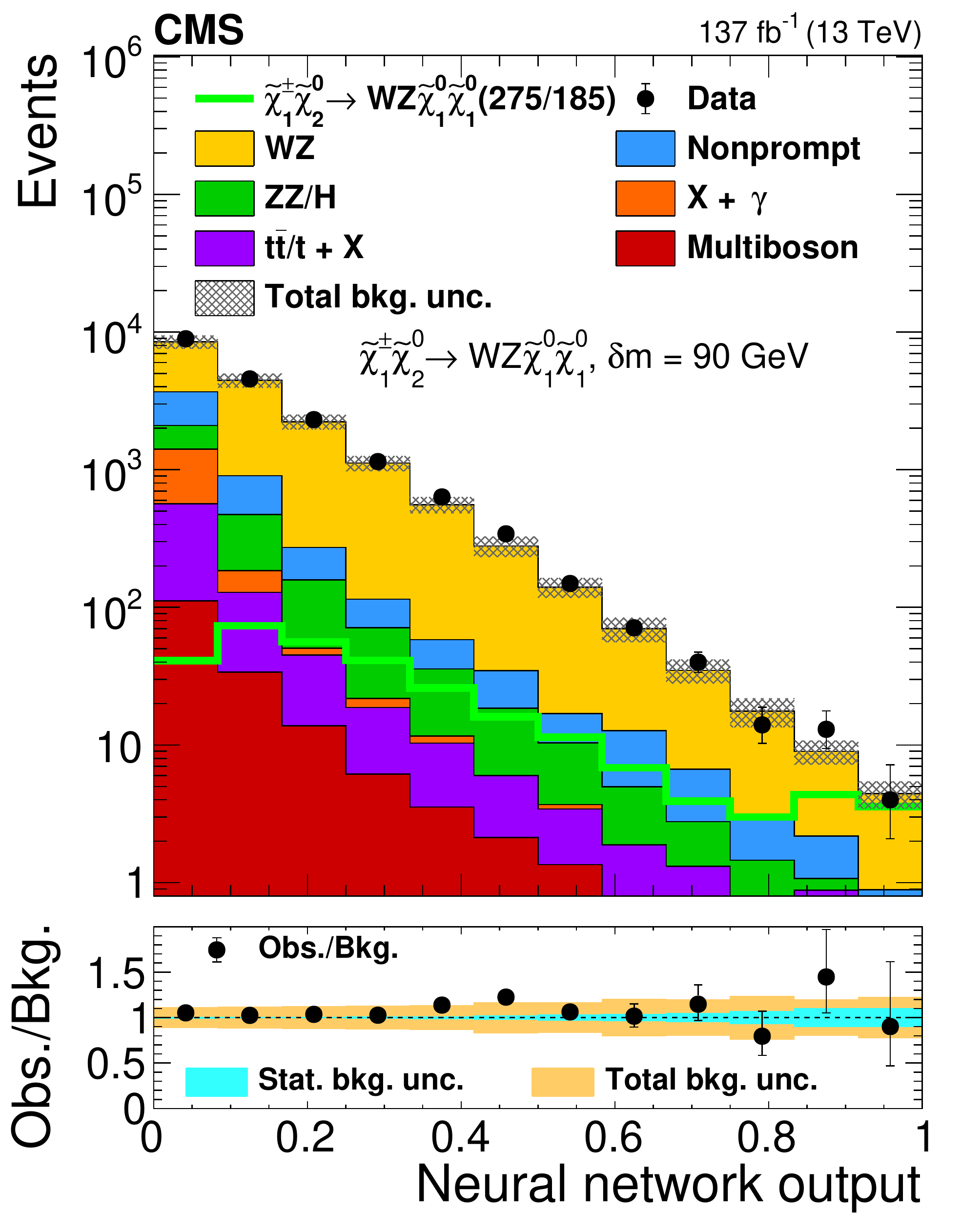}
\includegraphics[width=.33\textwidth]{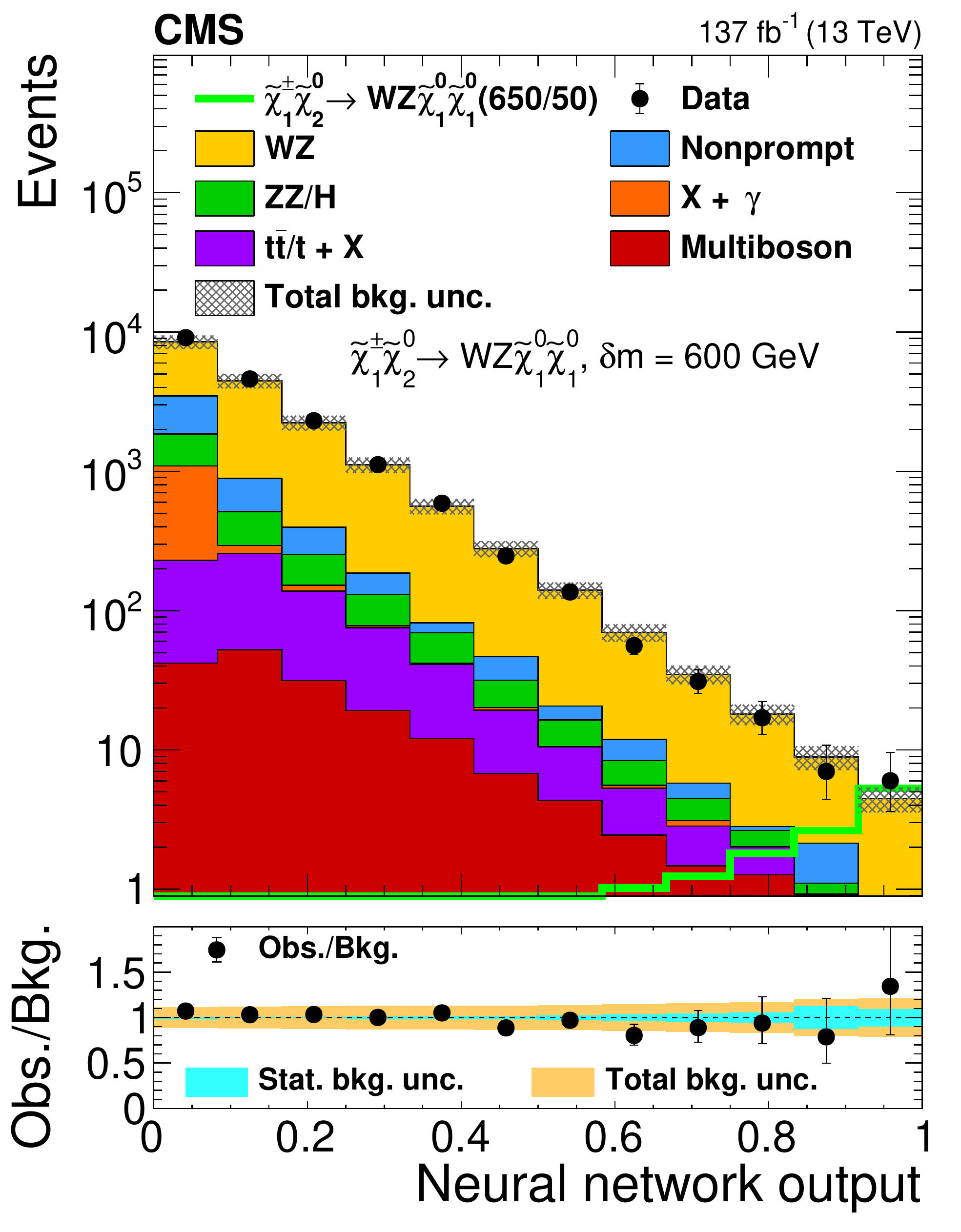}
\caption{Observed and expected yields as functions of the output of the neural network used to search for \xpmxtwo production with \WZ-mediated decays, evaluated at $\dm = 20 \GeV$ (left), $90 \GeV$ (center), and $600 \GeV$ (right). The legends specify the masses of \xtwo and \xone for the shown signal distributions as ($m_{\xtwo}$/$m_{\xone}$). The top panels show only the total uncertainty in the background prediction, while the lower panels show the total and statistical uncertainties separately. The following abbreviations are used in the legends of this and the following figures: ``bkg.'' stands for background, ``unc.'' for uncertainty and ``obs.'' for observed.}
\label{fig:neuralnet_TChiWZ}
\end{figure}

\begin{figure}[htbp!]
\includegraphics[width=.33\textwidth]{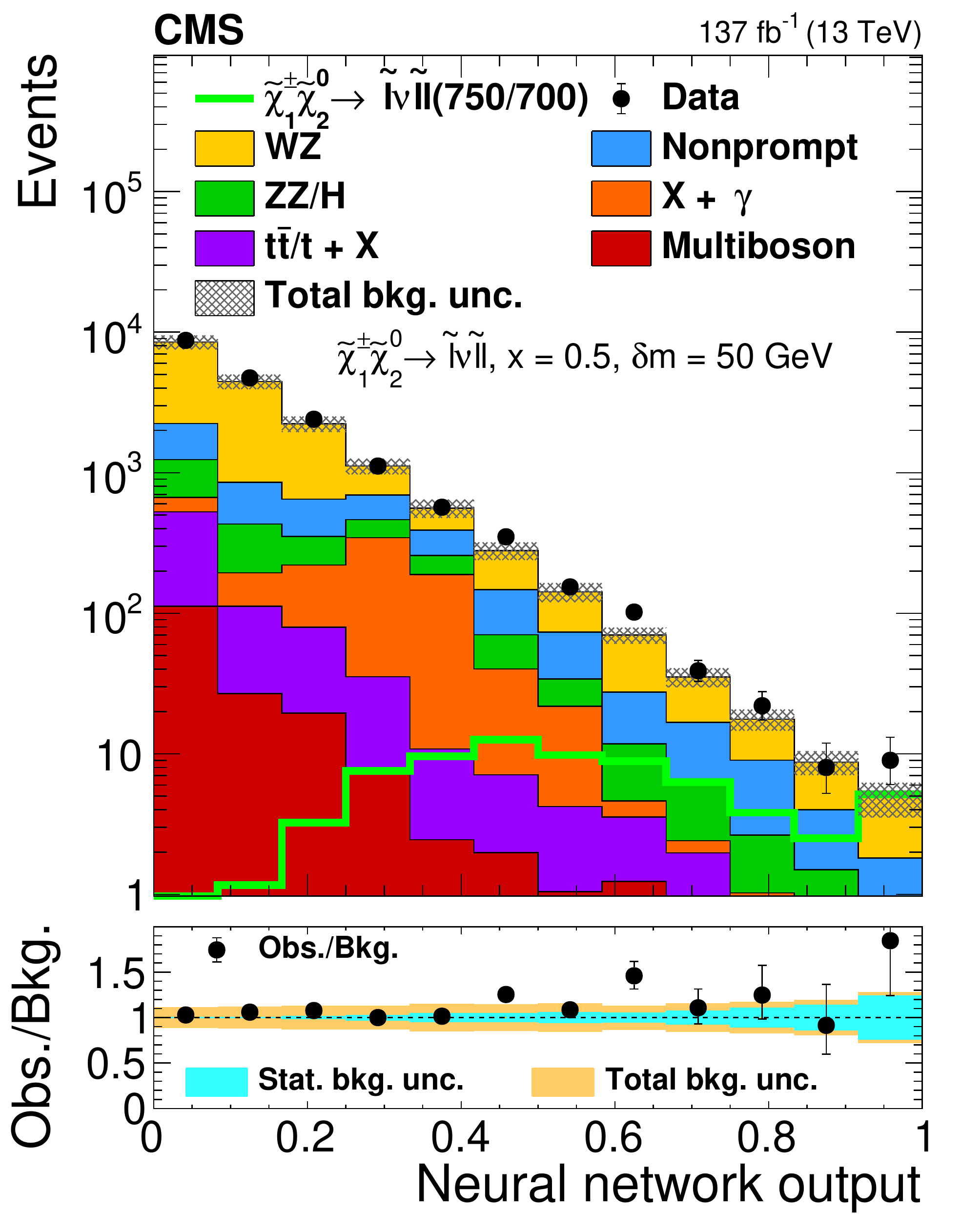}
\includegraphics[width=.33\textwidth]{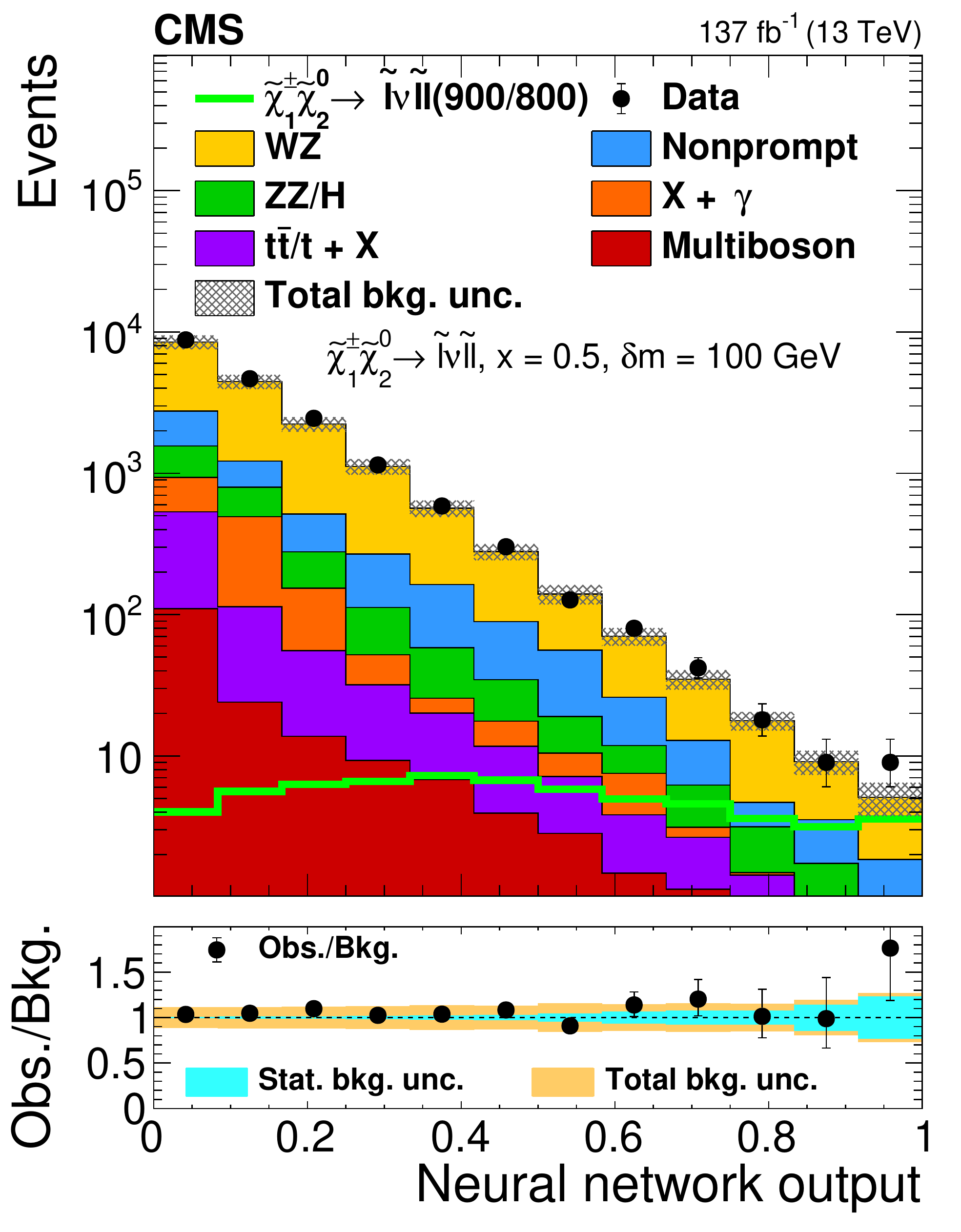}
\includegraphics[width=.33\textwidth]{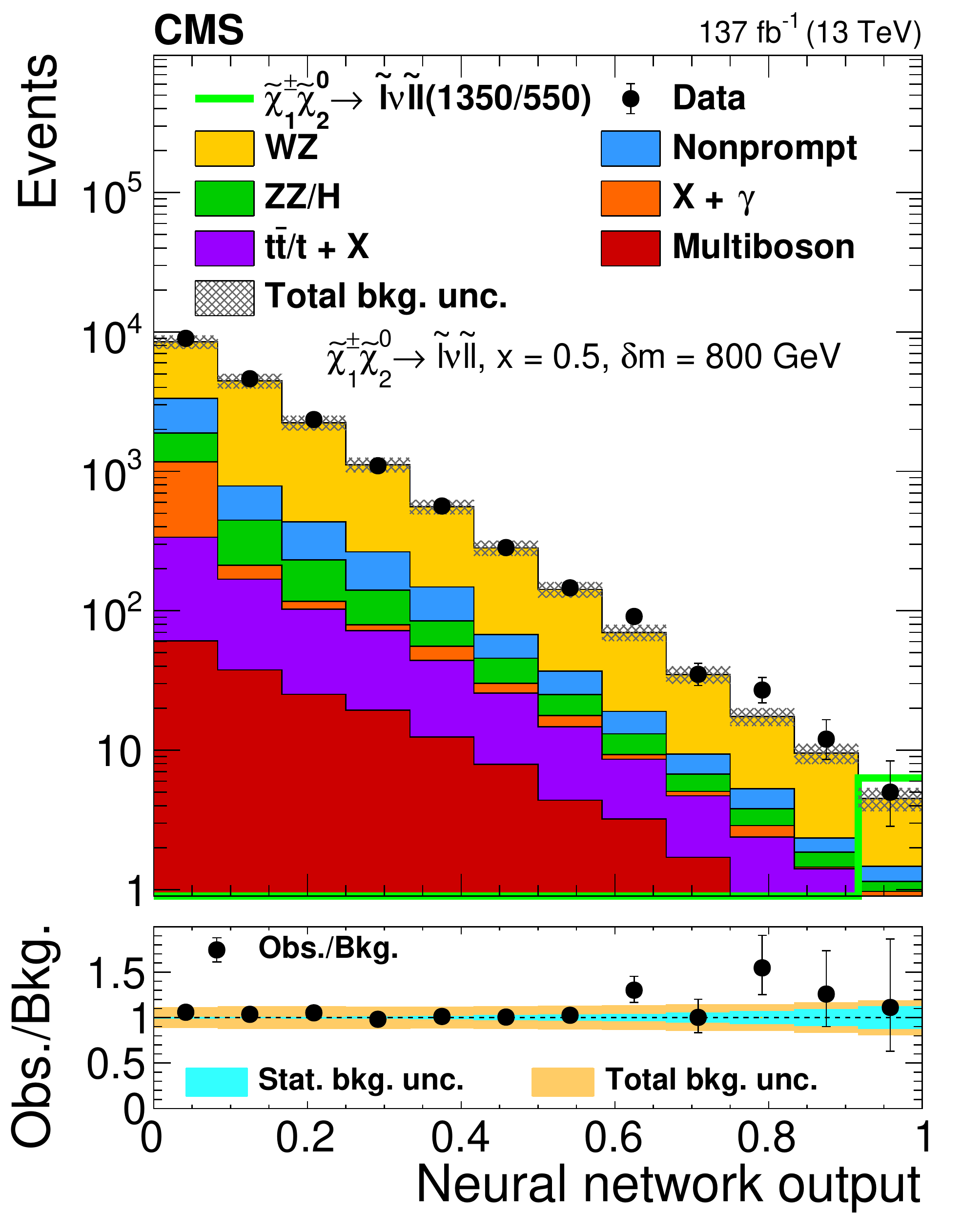}\\
\includegraphics[width=.33\textwidth]{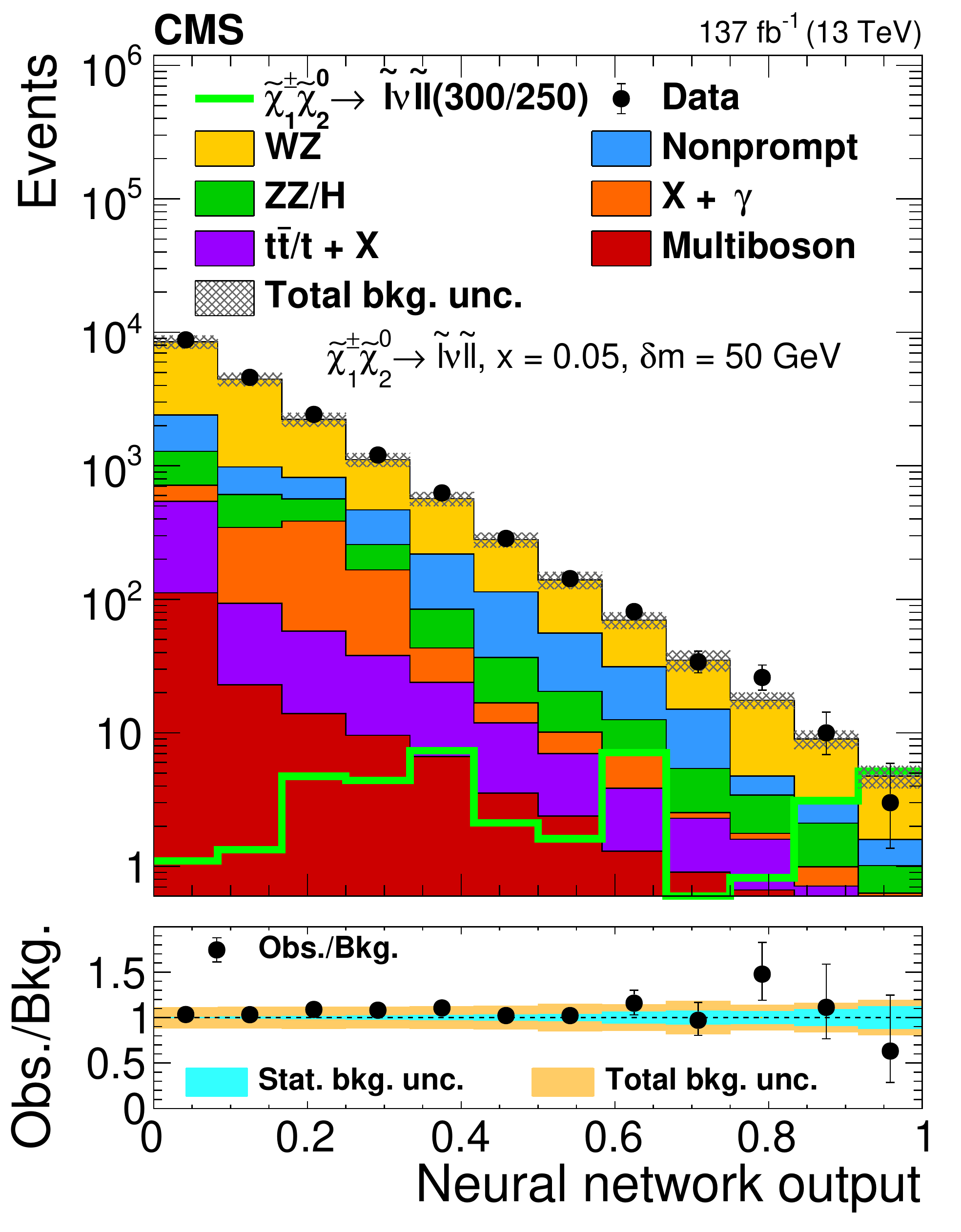}
\includegraphics[width=.33\textwidth]{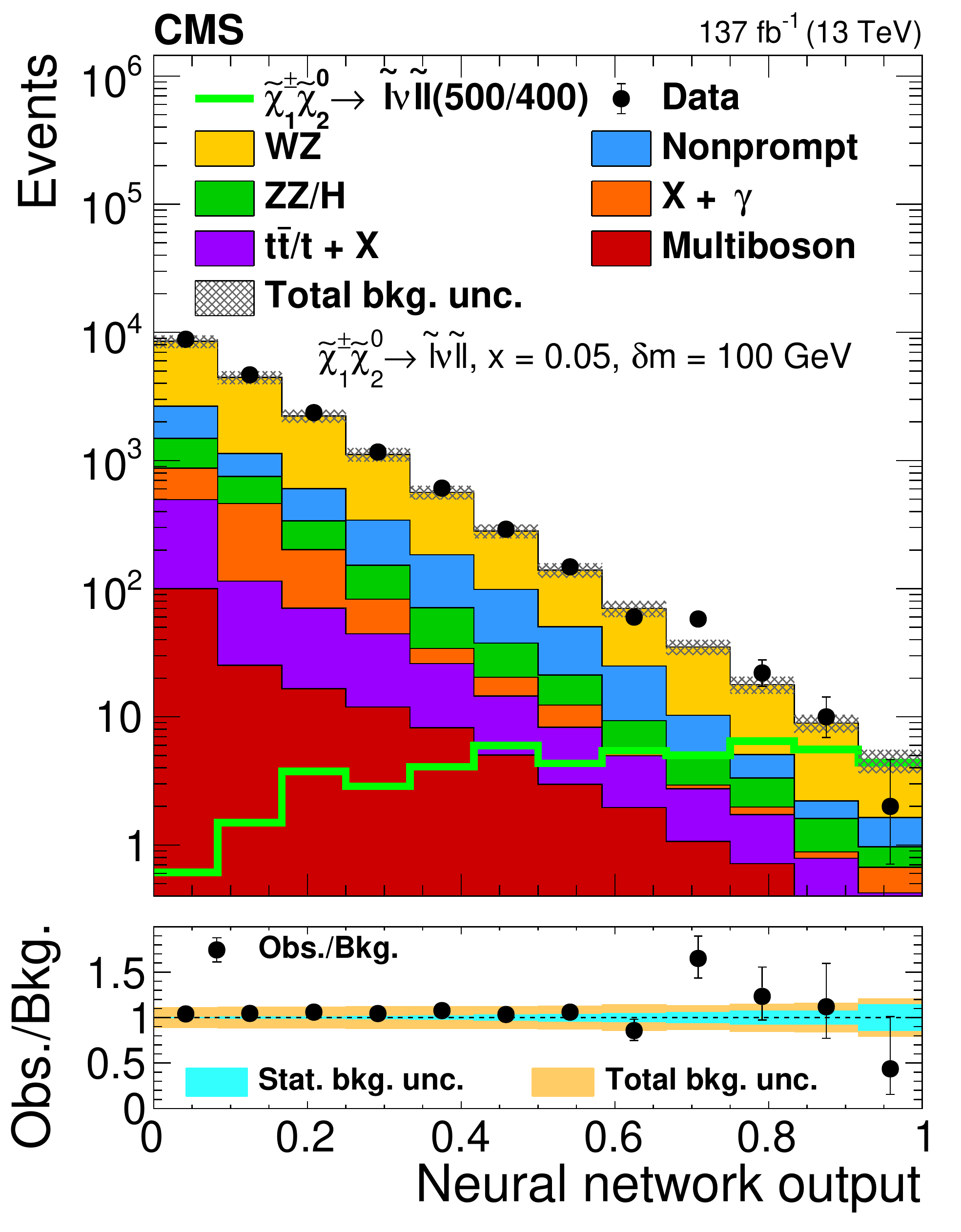}
\includegraphics[width=.33\textwidth]{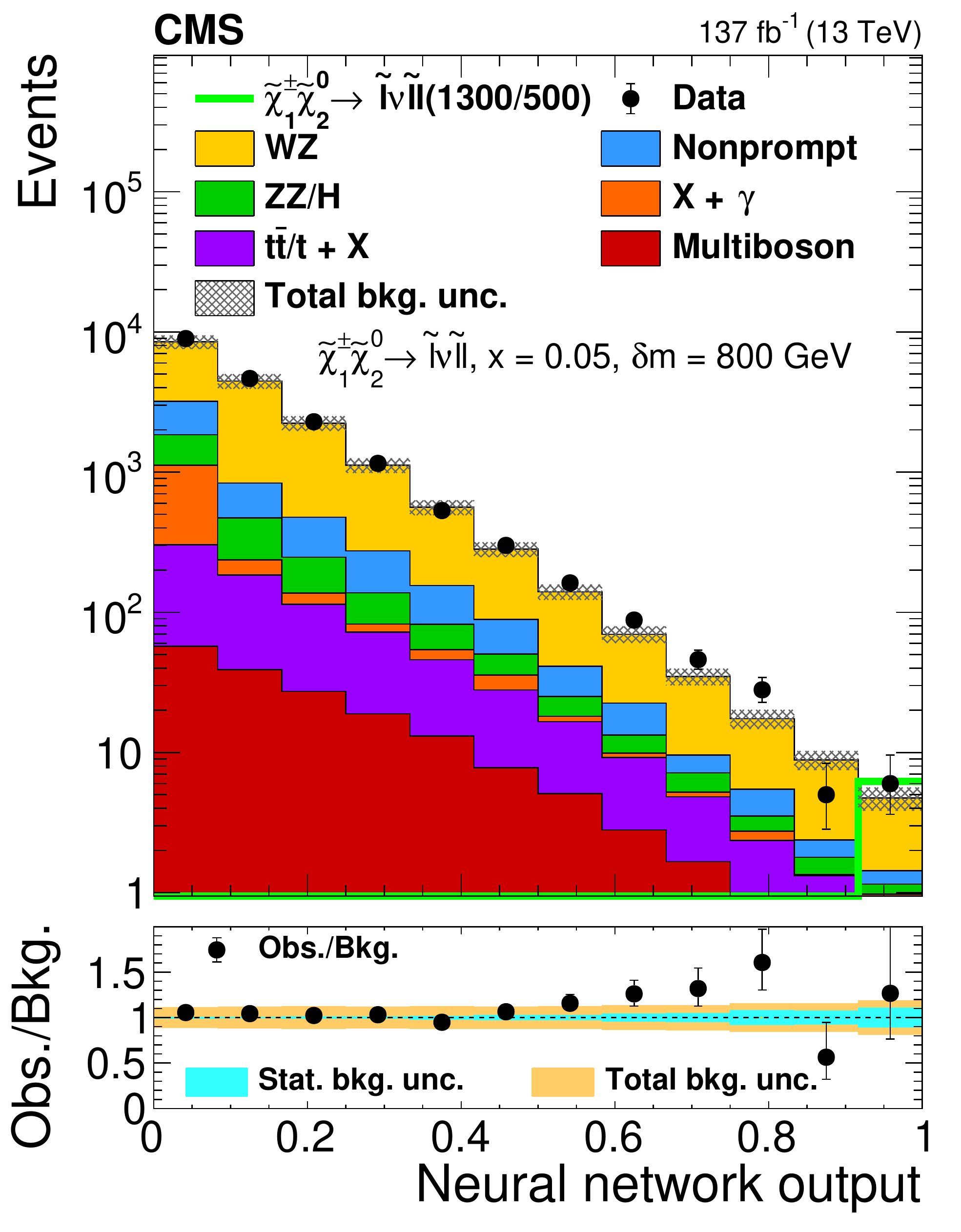}\\
\includegraphics[width=.33\textwidth]{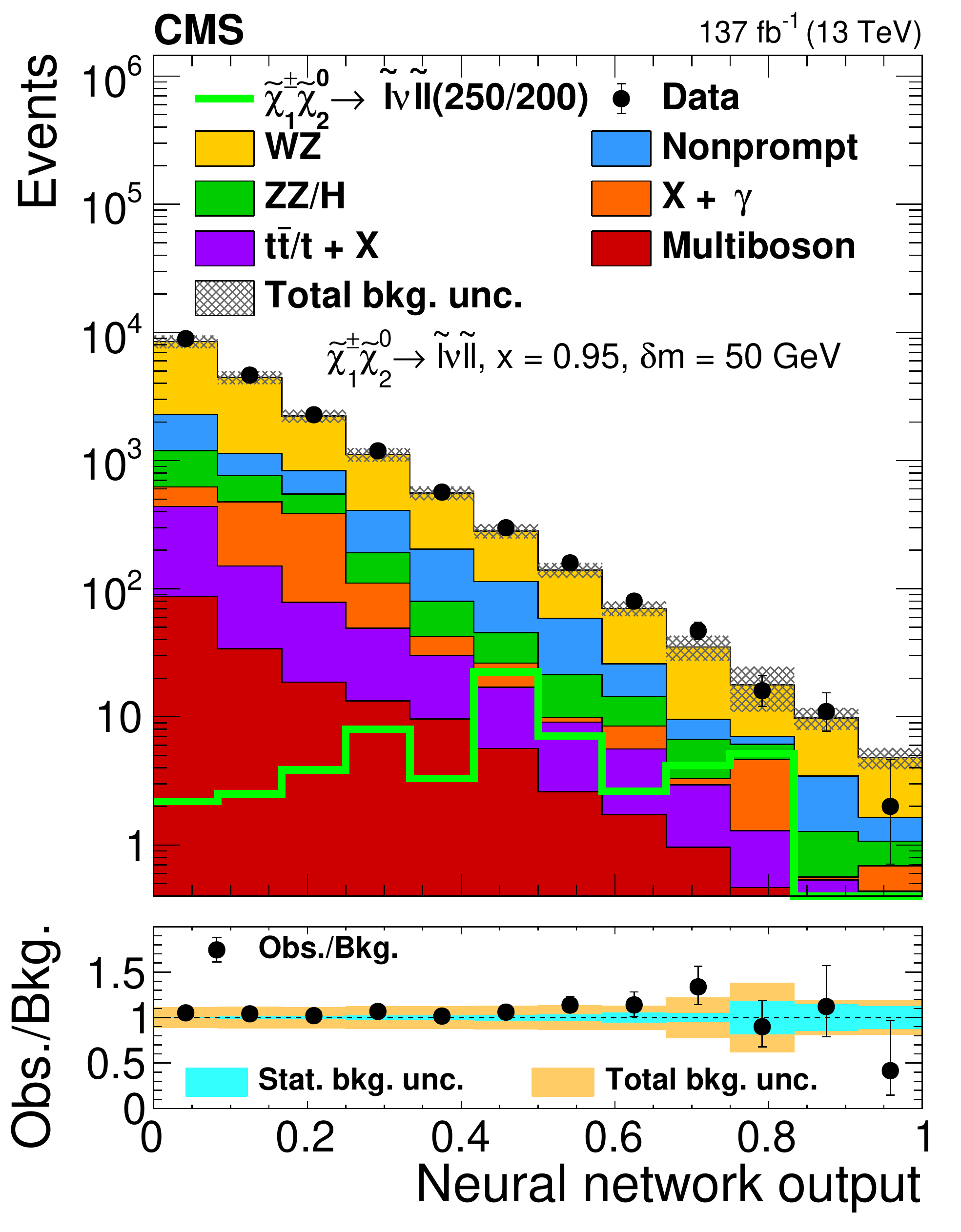}
\includegraphics[width=.33\textwidth]{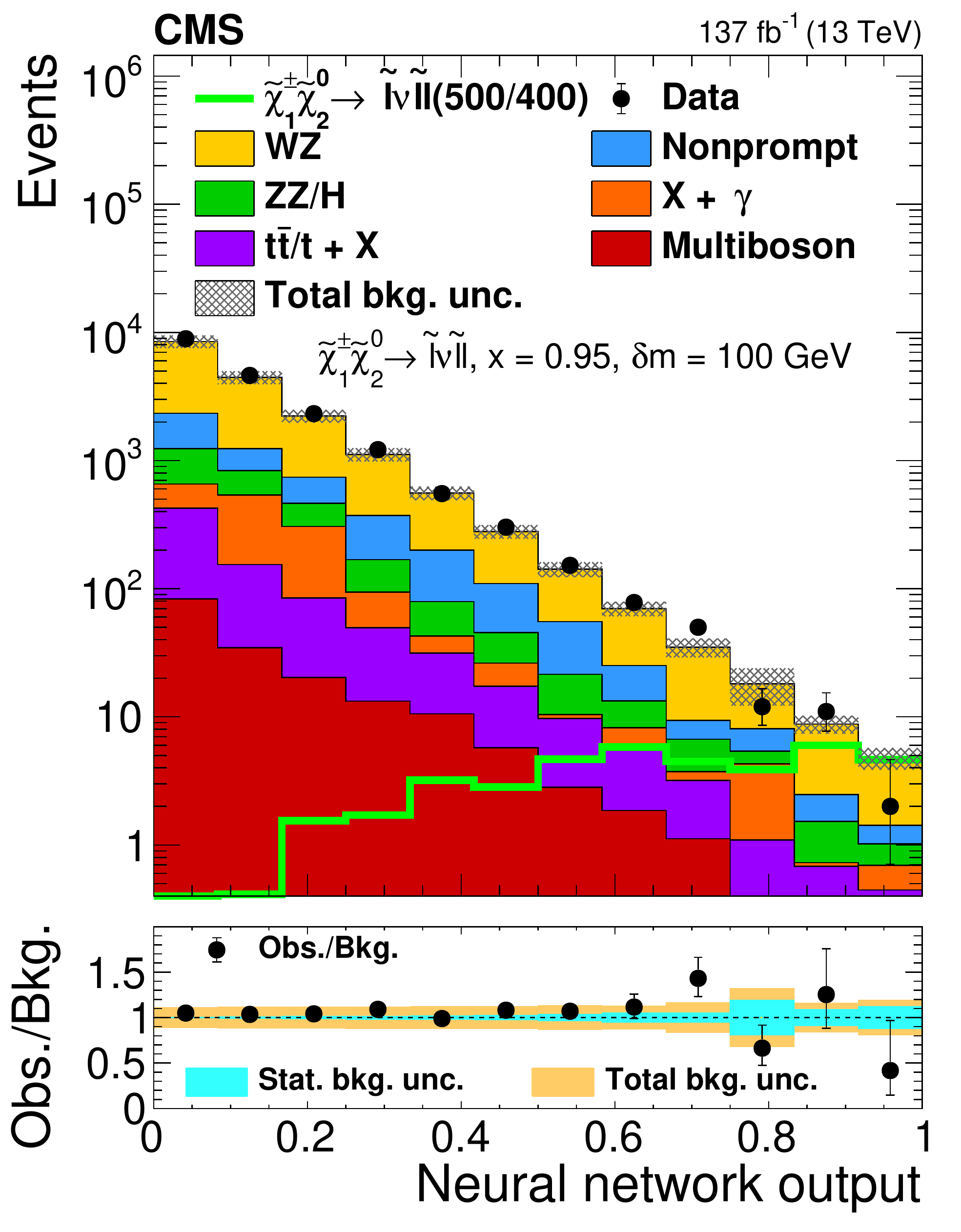}
\includegraphics[width=.33\textwidth]{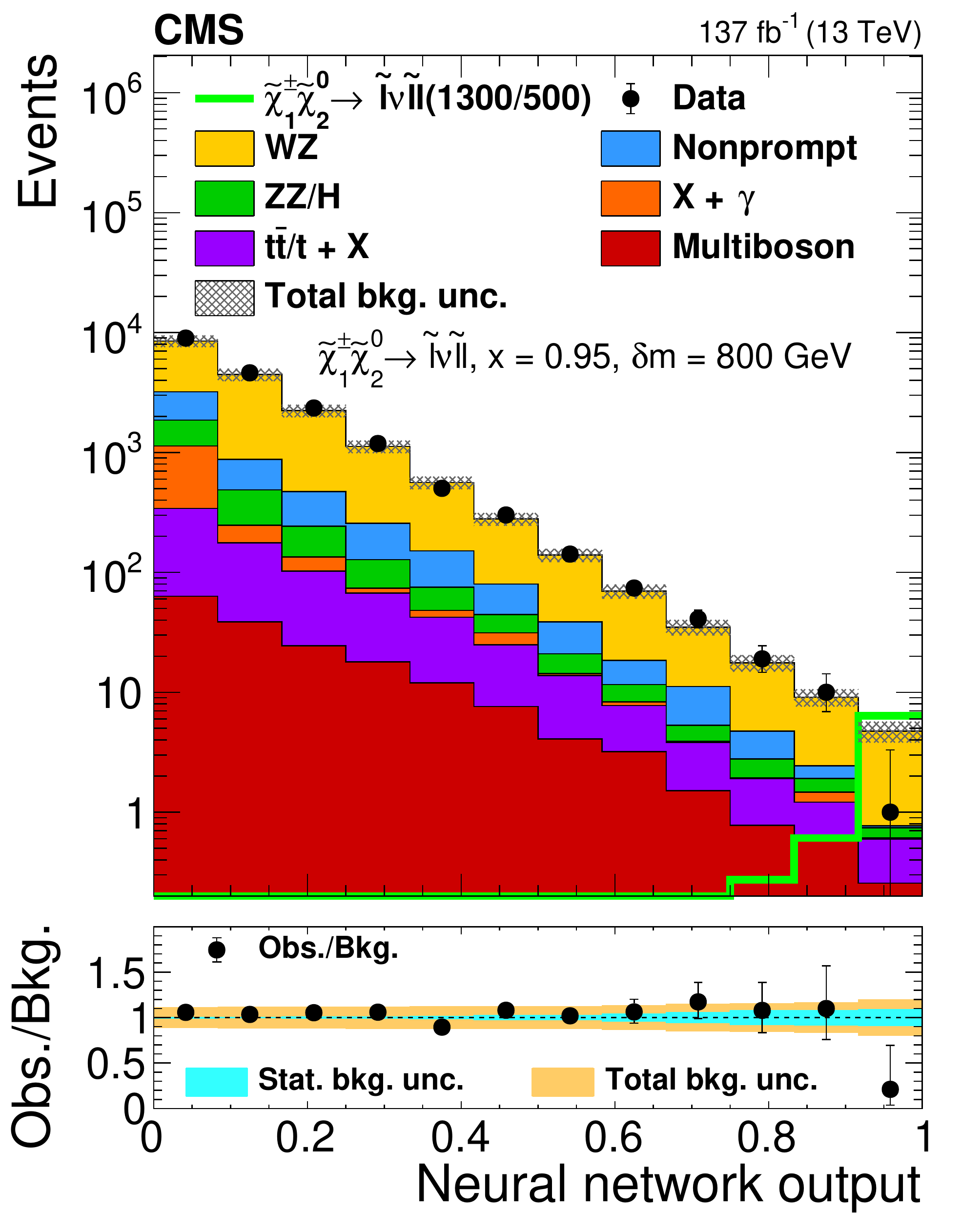}\\
\caption{Observed and expected yields as functions of the output of the neural network used to search for \xpmxtwo production with slepton-mediated decays at $x = 0.5$ (upper), $0.05$ (middle), and $0.95$ (lower), evaluated at $\dm = 50 \GeV$ (left), $100 \GeV$ (center), and $800 \GeV$ (right). The legends specify the masses of \xtwo and \xone for the shown signal distributions as ($m_{\xtwo}$/$m_{\xone}$).}
\label{fig:neuralnet_TChiSlepSnu}
\end{figure}

\begin{figure}[htbp!]
\centering
\includegraphics[width=\textwidth]{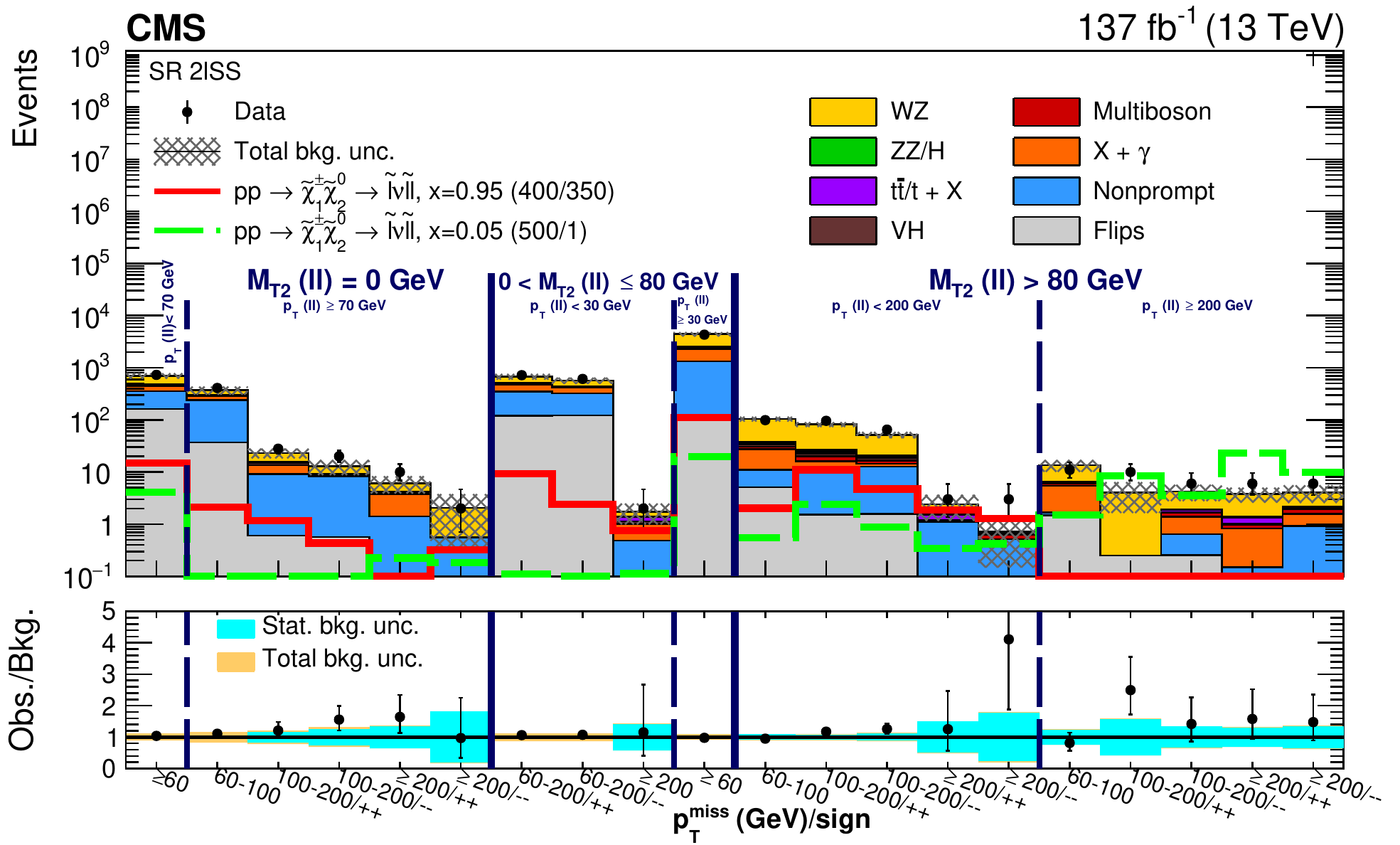}
\caption{Observed and expected yields across the search regions in events with two same-sign light leptons (2$\ell$SS). Several signal models are shown superimposed. They correspond to \xpmxtwo production with slepton-mediated decays in the flavor-democratic hypothesis for a compressed $\dm = 50 \GeV$ (red line) and uncompressed $\dm = 500 \GeV$ (green dashed line) scenario. The legends specify the masses of \xtwo and \xone for the shown signal distributions as ($m_{\xtwo}$/$m_{\xone}$).}
\label{fig:SR2lss}
\end{figure}

\begin{figure}[htbp!]
\centering
\includegraphics[width=\textwidth]{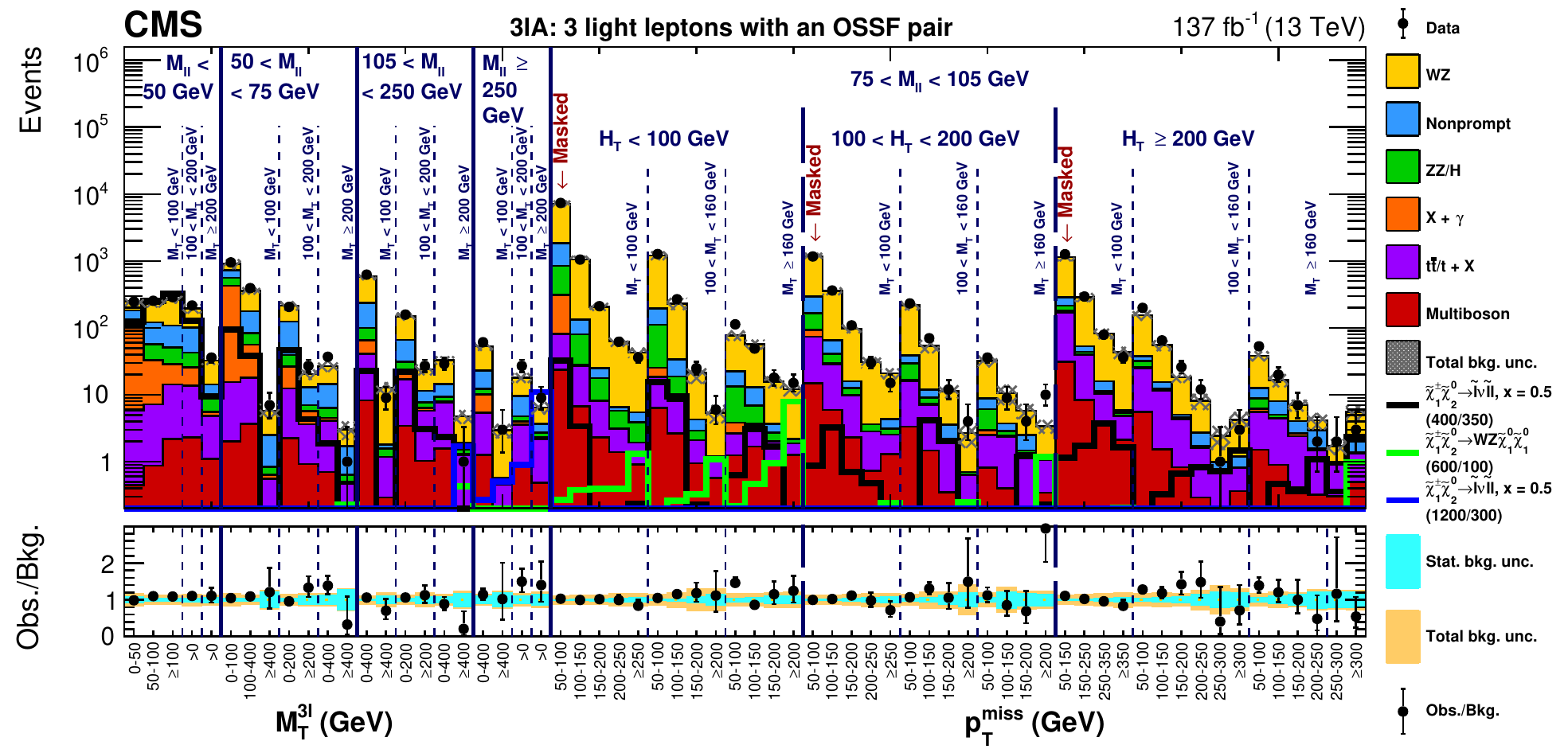}
\caption{Observed and expected yields across the search regions in events with three light leptons, at least two of which form an OSSF pair (3$\ell$A). Several signal models are shown superimposed. They correspond to \xpmxtwo production with slepton-mediated decays in the flavor-democratic hypothesis for a compressed $\dm = 50 \GeV$ (black line) and uncompressed $\dm = 900 \GeV$ (blue line) scenario, and for \WZ-mediated decays in an uncompressed $\dm = 500 \GeV$ scenario (green line). Bins labeled as ``Masked'' are not considered in the interpretation of the results because of overlap with the \WZ control region. The legends specify the masses of \xtwo and \xone for the shown signal distributions as ($m_{\xtwo}$/$m_{\xone}$).}
\label{fig:SR3lA}
\end{figure}

\begin{figure}[htbp!]
\centering
\includegraphics[width=.5\textwidth]{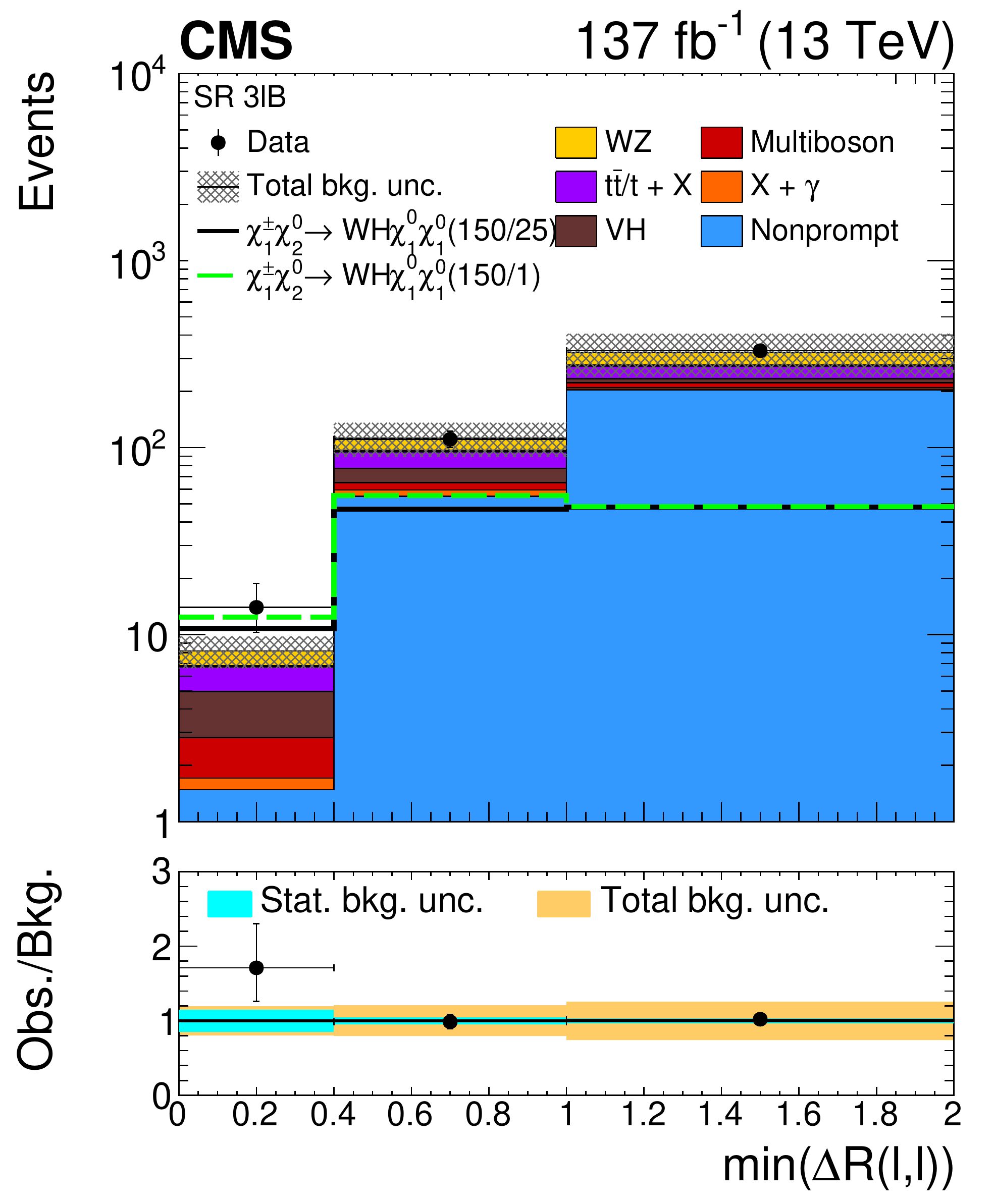}
\caption{Observed and expected yields across the search regions in events with three light leptons, none of which form an OSSF pair (3$\ell$B). Several signal models are shown superimposed. They correspond to \xpmxtwo production with \WH-mediated decays for scenarios corresponding to a \PH boson like mass splitting $\dm = 125 \GeV$ (black line) and a slightly less compressed $\dm = 150 \GeV$ (green dashed line) scenario. The legends specify the masses of \xtwo and \xone for the shown signal distributions as ($m_{\xtwo}$/$m_{\xone}$).}
\label{fig:SR3lB}
\end{figure}

\begin{figure}[htbp!]
\centering
\includegraphics[width=\textwidth]{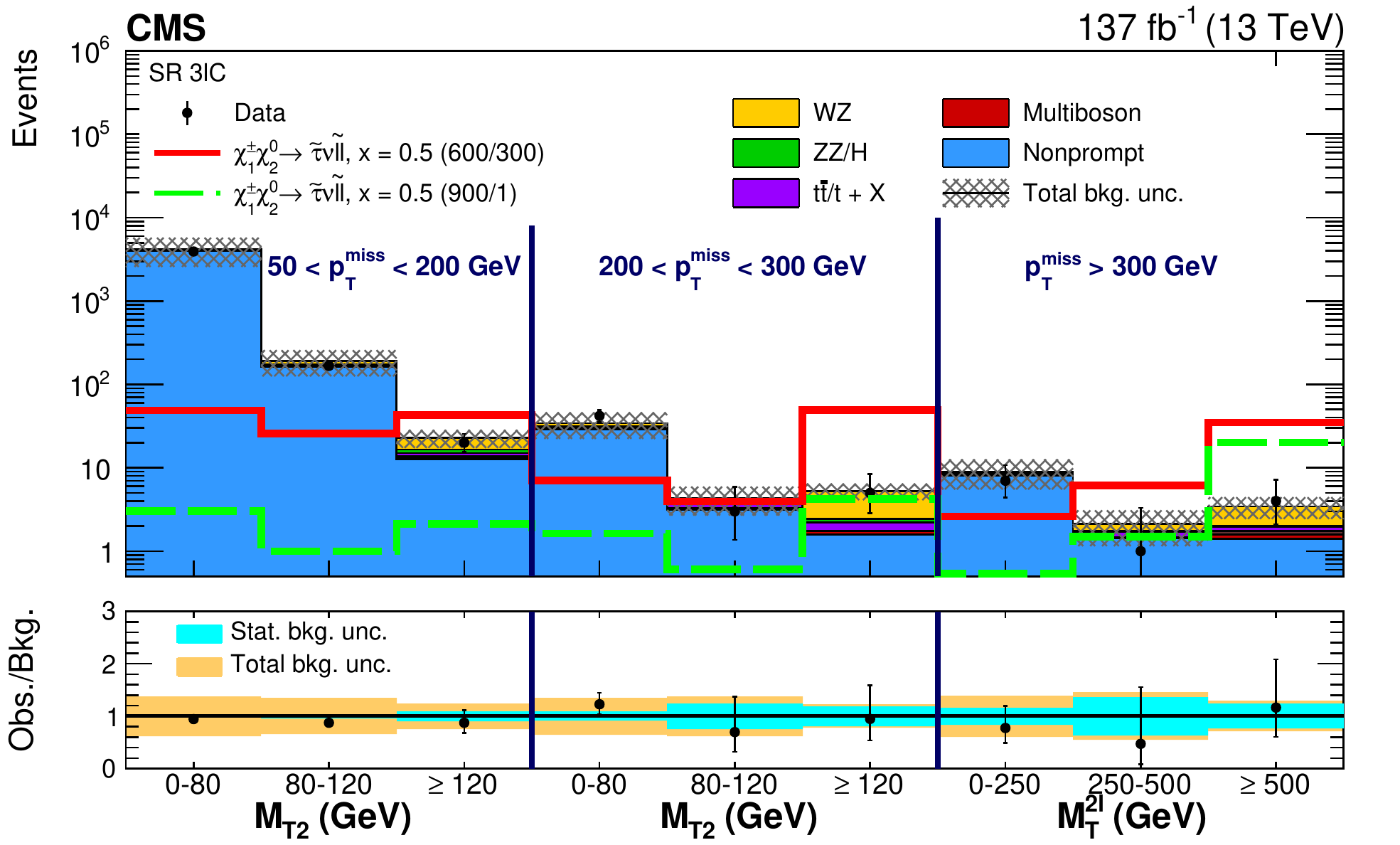}
\caption{Observed and expected yields across the search regions in events with a $\PGmp\PGmm$ or $\Pep\Pem$ pair and an additional \tauh candidate (3$\ell$C). Several signal models are shown superimposed. They correspond to \xpmxtwo production with slepton-mediated decays in the $\tau$ lepton enriched hypothesis for a compressed $\dm = 300 \GeV$ (red line) and uncompressed $\dm = 900 \GeV$ (green dashed line) scenario. The legends specify the masses of \xtwo and \xone for the shown signal distributions as ($m_{\xtwo}$/$m_{\xone}$).}
\label{fig:SR3lC}
\end{figure}

\begin{figure}[htbp!]
\centering
\includegraphics[width=\textwidth]{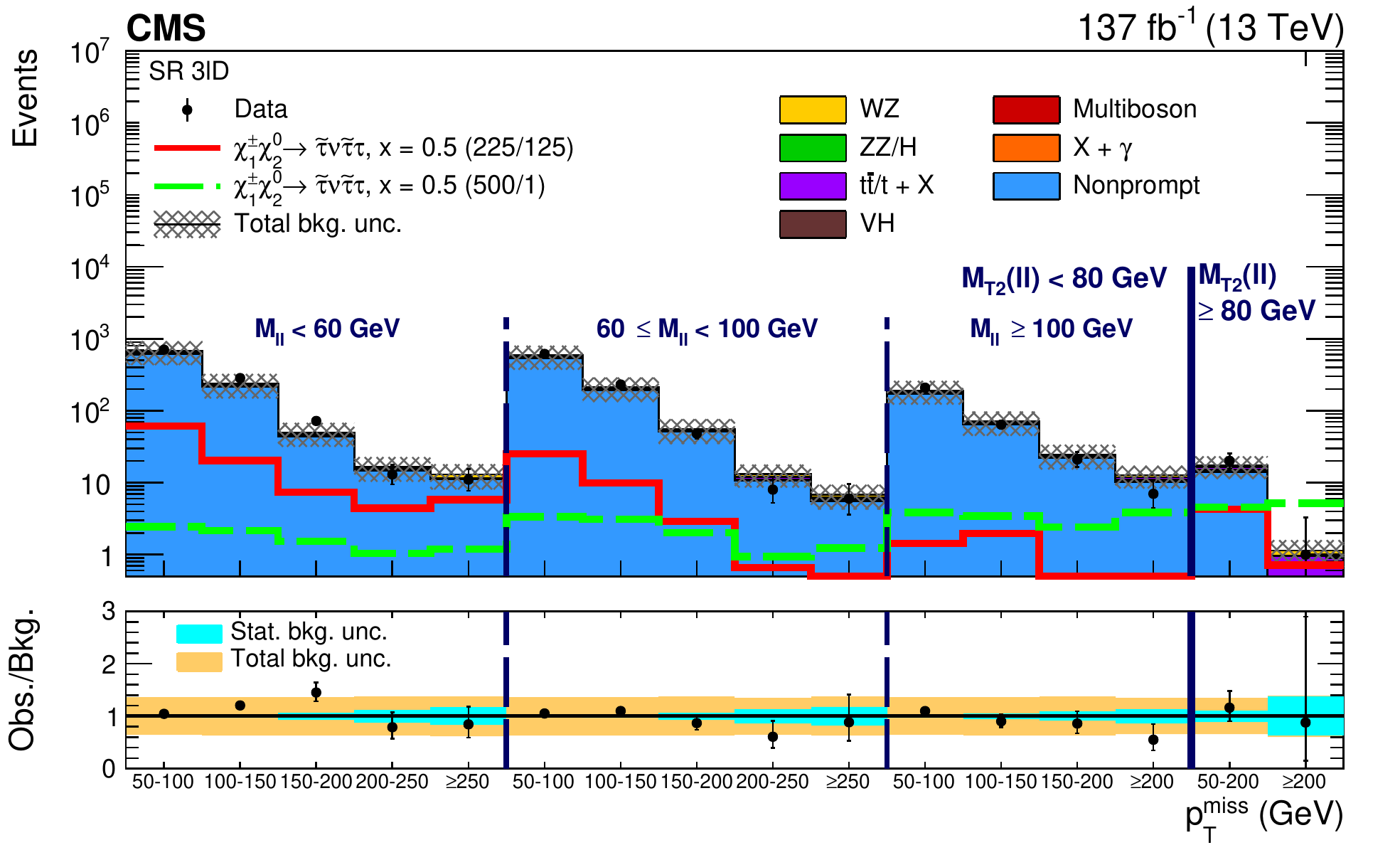}
\caption{Observed and expected yields across the search regions in events with an $\Pe^{\pm}\PGm^{\mp}$ pair and a \tauh candidate (3$\ell$D). Several signal models are shown superimposed. They correspond to \xpmxtwo production with slepton-mediated decays in the $\tau$ lepton dominated hypothesis for a compressed $\dm = 100 \GeV$ (red line) and uncompressed $\dm = 500 \GeV$ (green dashed line) scenario. The legends specify the masses of \xtwo and \xone for the shown signal distributions as ($m_{\xtwo}$/$m_{\xone}$).}
\label{fig:SR3lD}
\end{figure}

\begin{figure}[htbp!]
\centering
\includegraphics[width=\textwidth]{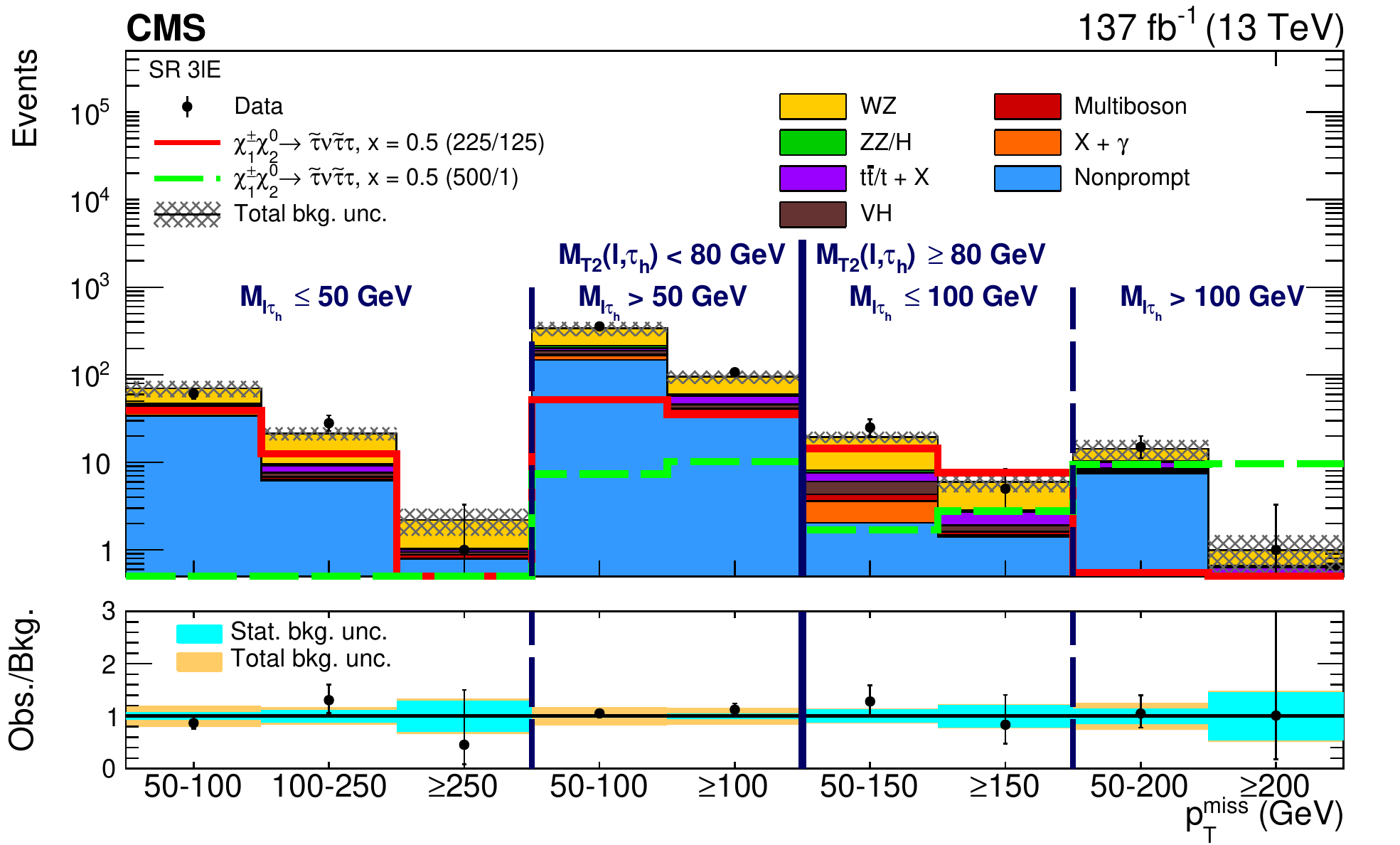}\\
\includegraphics[width=\textwidth]{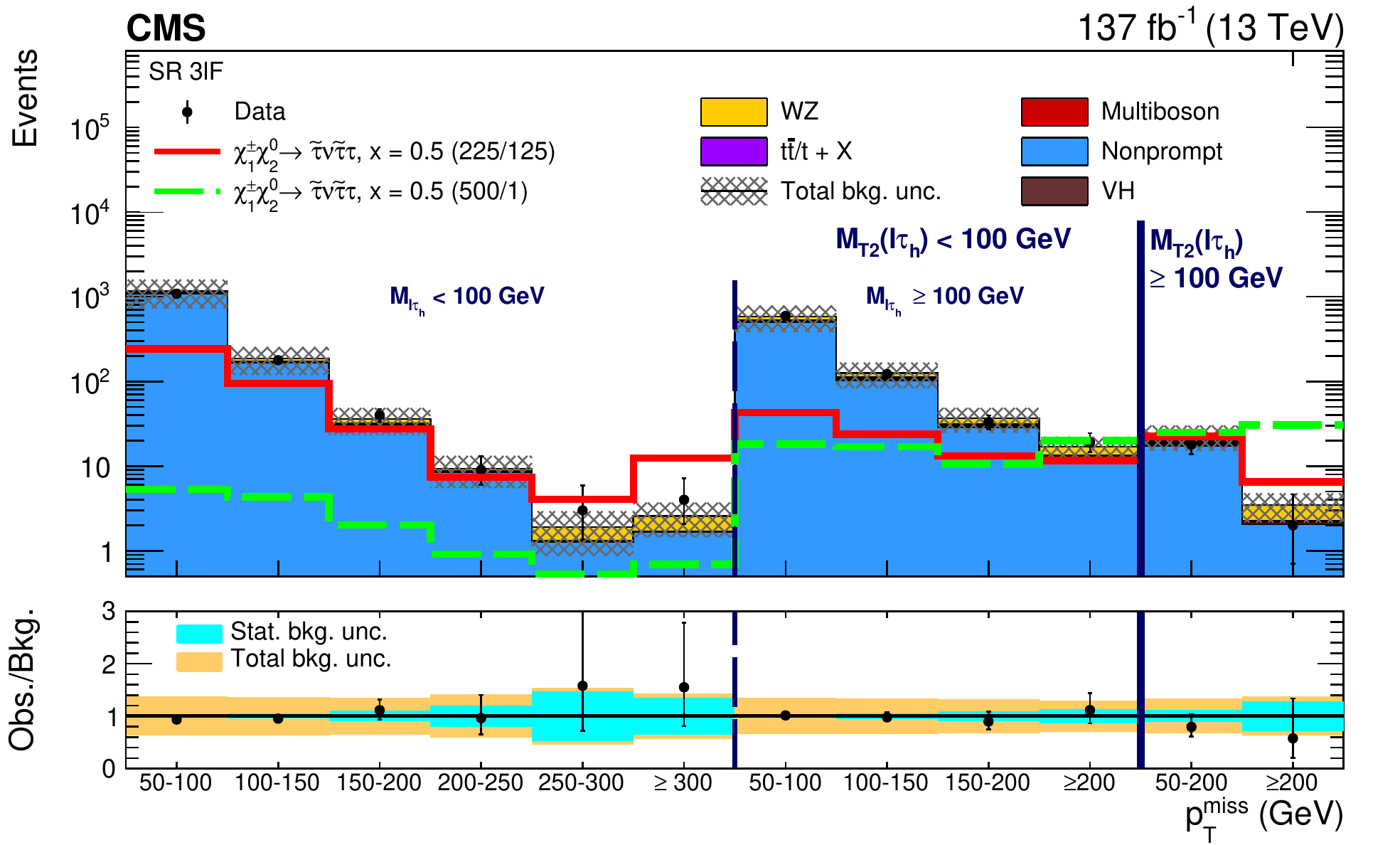}
\caption{Observed and expected yields across the search regions in events (upper plot) with a same-sign light lepton pair and a \tauh candidate (3$\ell$E), and (lower plot) with two \tauh candidates and one light lepton (3$\ell$F). Several signal models are shown superimposed. They correspond to \xpmxtwo production with slepton-mediated decays in the $\tau$ lepton dominated hypothesis for a compressed $\dm = 100 \GeV$ (red line) and uncompressed $\dm = 500 \GeV$ (green dashed line) scenario. The legends specify the masses of \xtwo and \xone for the shown signal distributions as ($m_{\xtwo}$/$m_{\xone}$).}
\label{fig:SR3lEF}
\end{figure}

\begin{figure}[htbp!]
\centering
\includegraphics[width=0.55\textwidth]{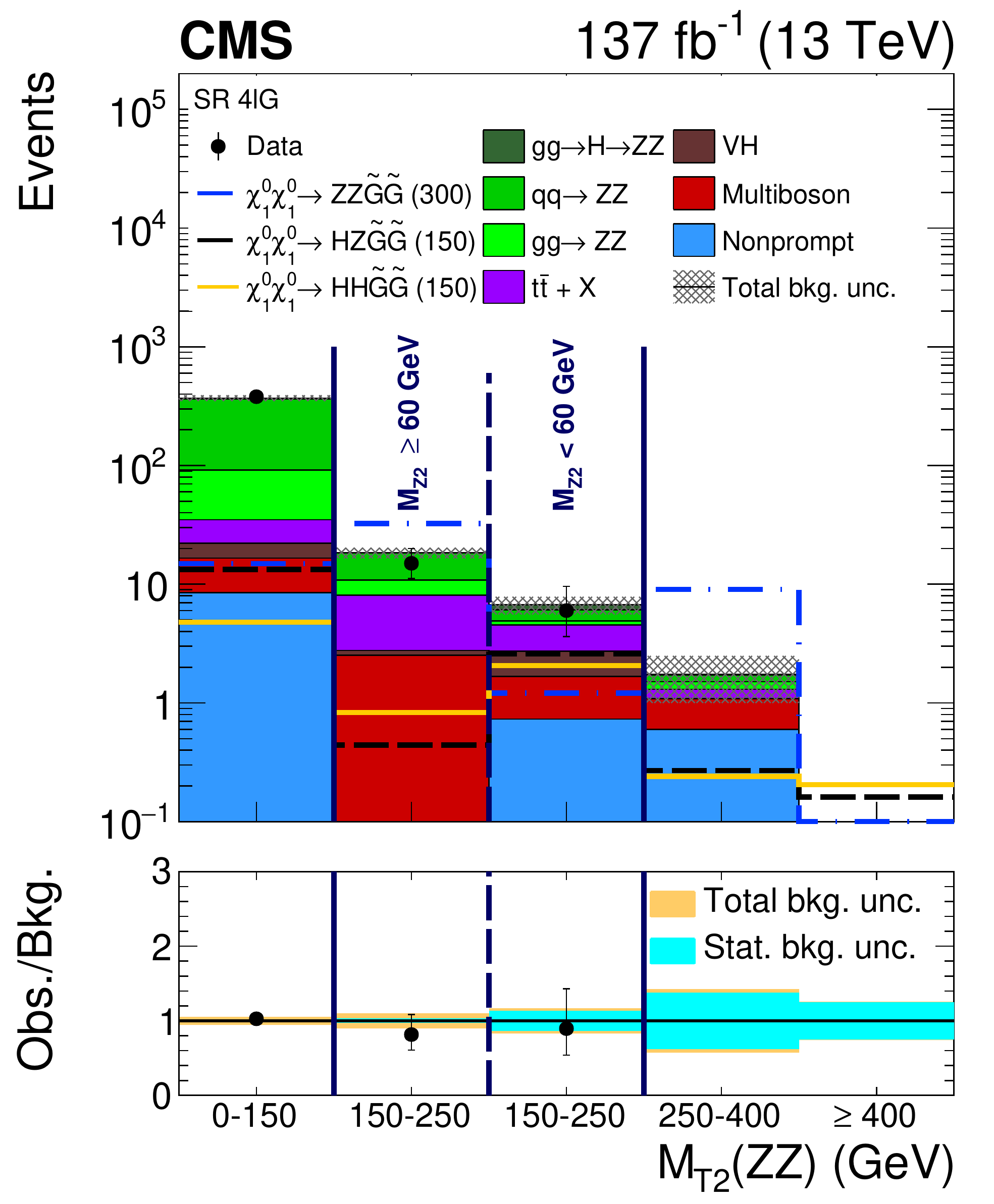}
\caption{Observed and expected yields across the search regions in events with four light leptons, including 2 separate OSSF pairs (4$\ell$G). Several signal models are shown superimposed. They correspond to Higgsino pair production with decays to \ZZ (blue dotted line, Higgsino mass of $300\GeV$), \HZ (black dashed line, Higgsino mass of $150\GeV$), and \HH (dark yellow line, Higgsino mass of $150\GeV$).}
\label{fig:SR4lG}
\end{figure}

\begin{figure}[htbp!]
\centering
\includegraphics[width=.48\textwidth]{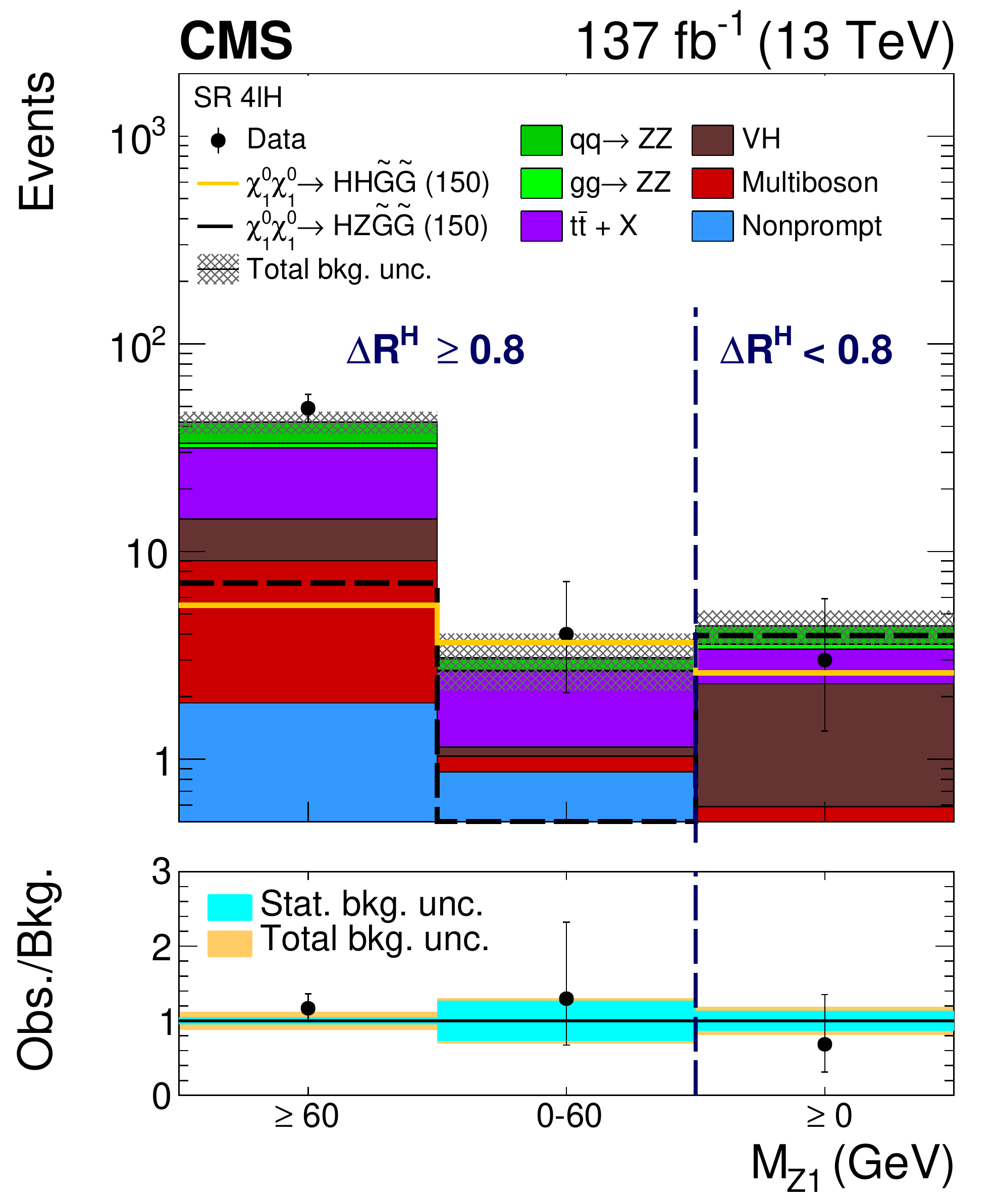}
\includegraphics[width=.48\textwidth]{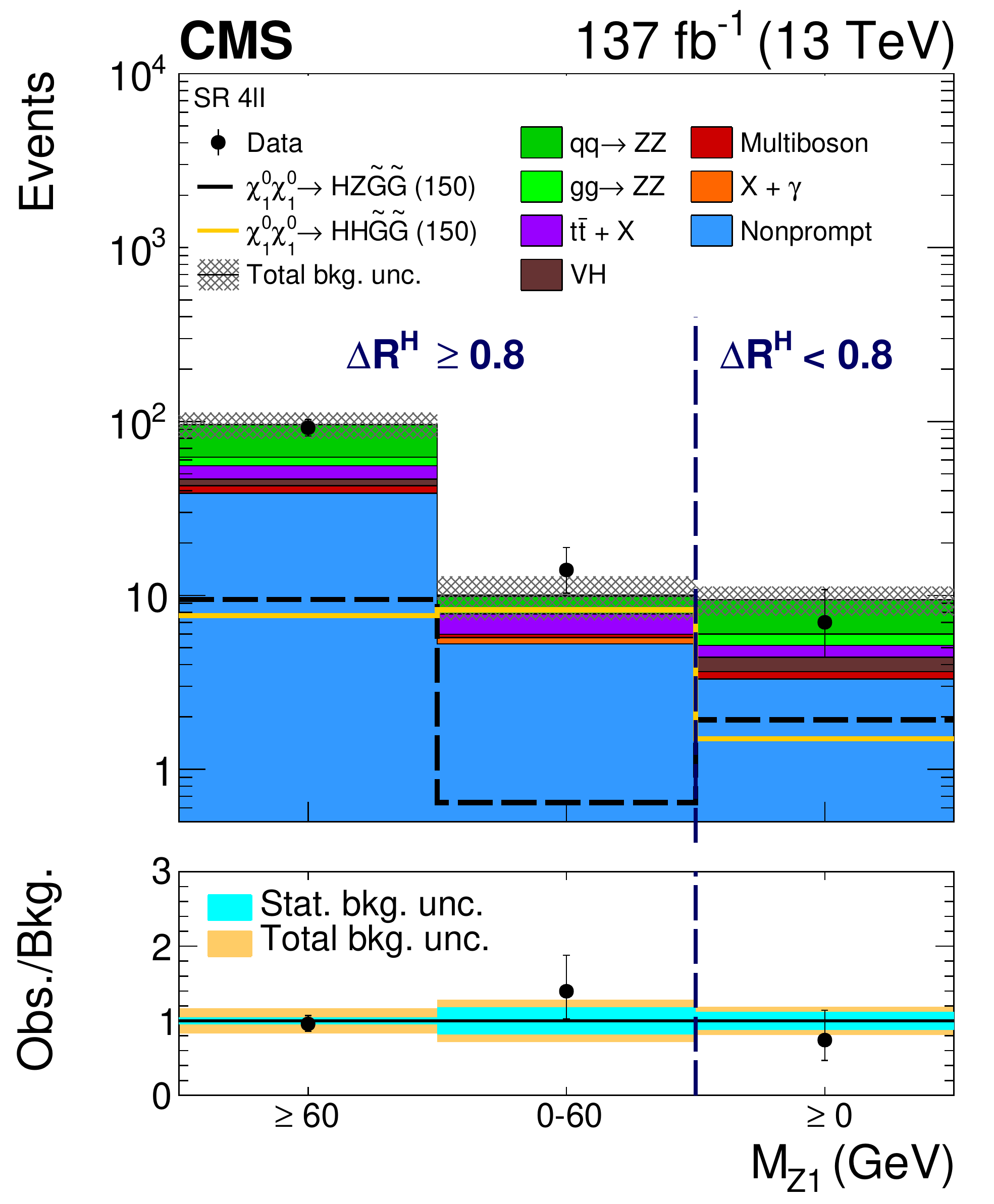}\\
\includegraphics[width=.48\textwidth]{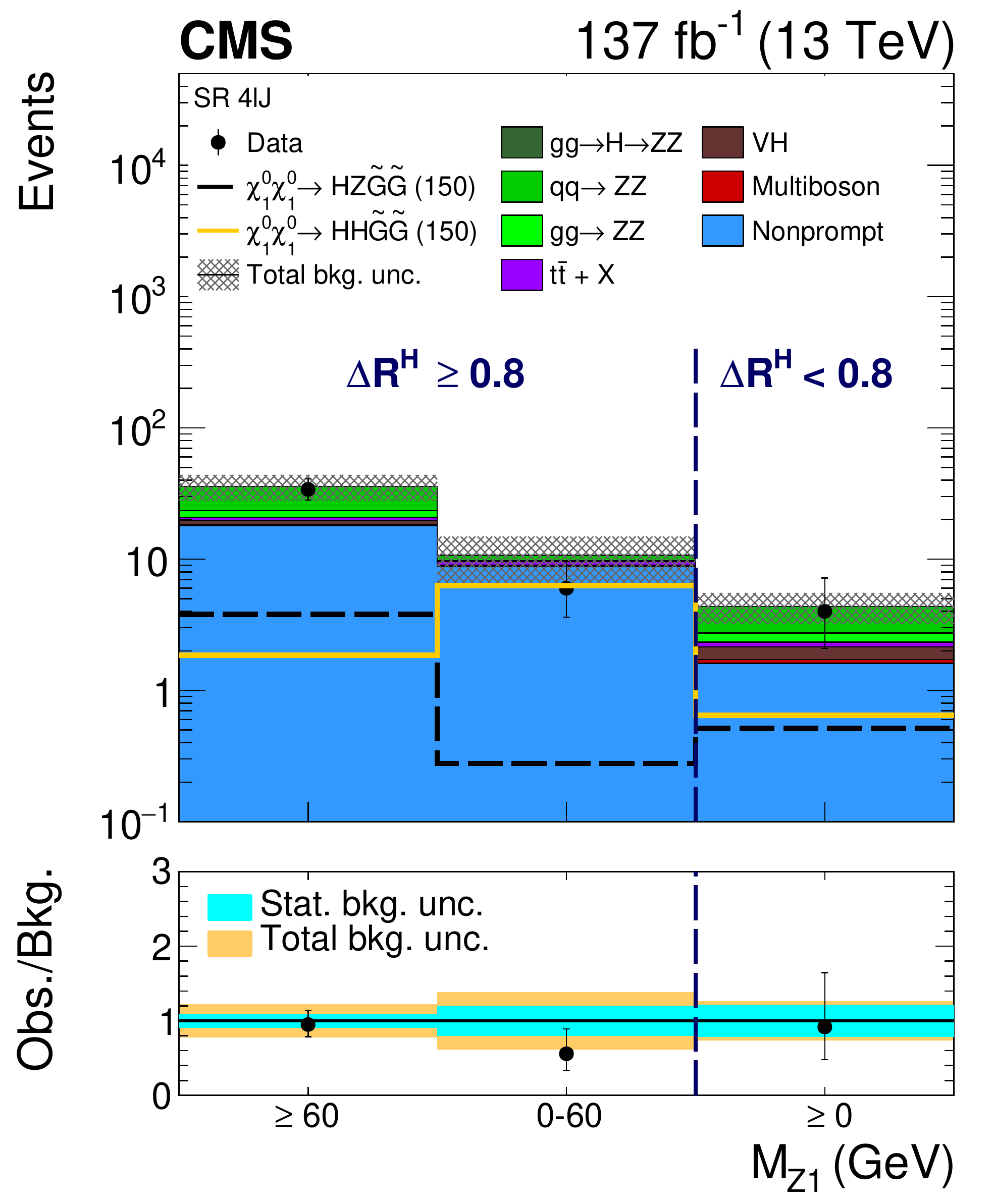}
\includegraphics[width=.48\textwidth]{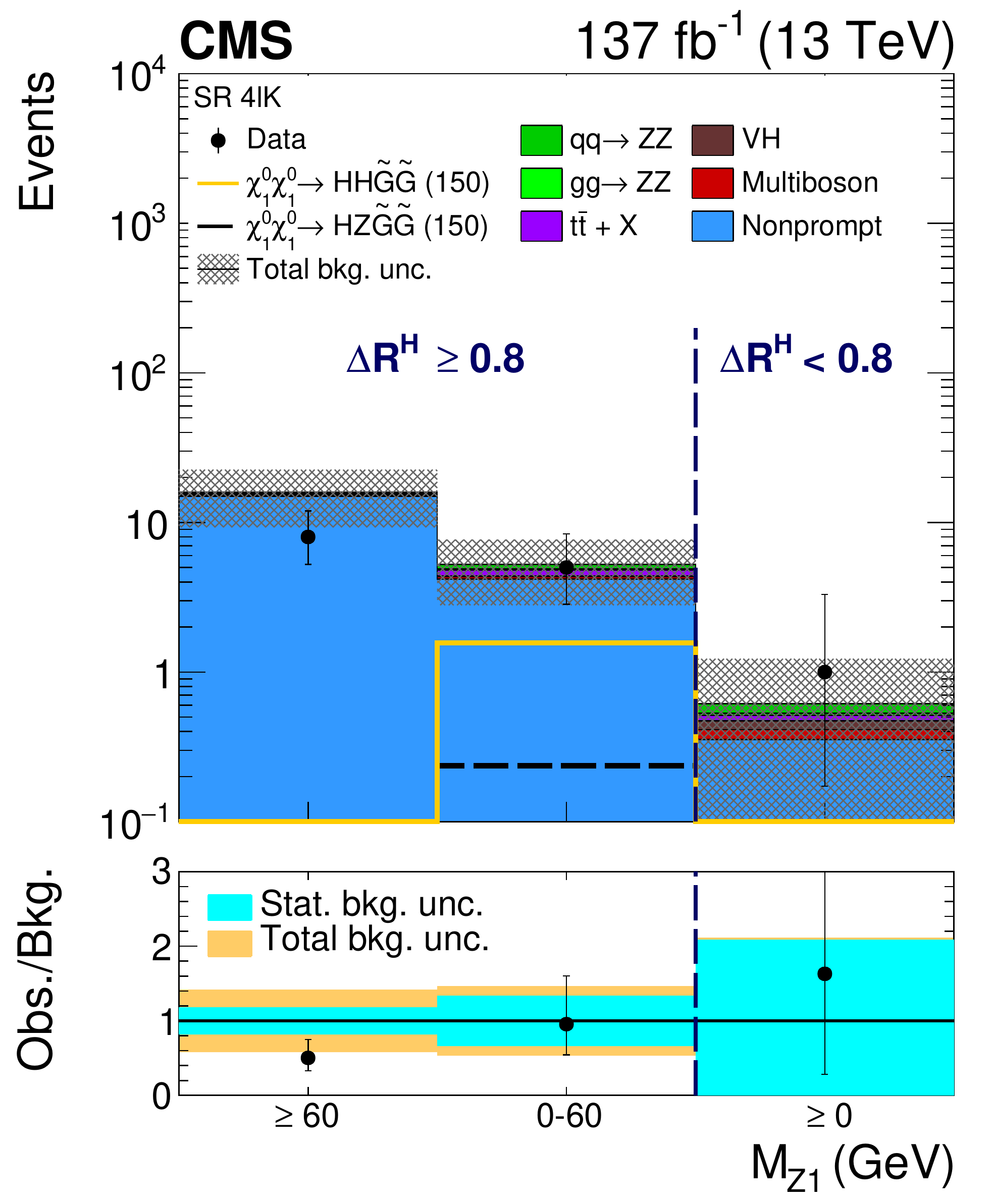}
\caption{Observed and expected yields across the search regions in events with four light leptons not forming two OSSF pairs (4$\ell$H, upper left), events with three light leptons and a \tauh candidate (4$\ell$I, upper right), forming two OSSF pairs (4$\ell$J, lower left), and forming one or less OSSF pairs (4$\ell$K, lower right). Several signal models are shown superimposed. They correspond to Higgsino pair production with decays to \HZ (dashed black line, Higgsino mass of $150\GeV$), and \HH (dark yellow line, Higgsino mass of $150\GeV$).}
\label{fig:SR4lHIJK}
\end{figure}

The yields as a function of the parametric neural network output in 3$\ell$A events are respectively shown in Figs.~\ref{fig:neuralnet_TChiWZ}--\ref{fig:neuralnet_TChiSlepSnu} for the different \xpmxtwo production models considered. For each model the neural network discriminant is shown as evaluated at three distinct \dm hypotheses, representing low, intermediate, and high \dm values. To obtain the final results, the neural network is evaluated at far more \dm parameters, separated by 50\GeV for models with slepton-mediated decays and by 25\GeV in case of \WZ-mediated decays with \dm in excess of 100\GeV. When \dm is below 90\GeV in the former models, the neural network is evaluated in \dm steps of 10\GeV, and in steps of 1\GeV between 90 and 100\GeV. The expected and observed yields as a function of the search regions in each event category are shown in Figs.~\ref{fig:SR2lss}--\ref{fig:SR4lHIJK}.

\begin{figure}[htbp!]
\centering
\includegraphics[width=\textwidth]{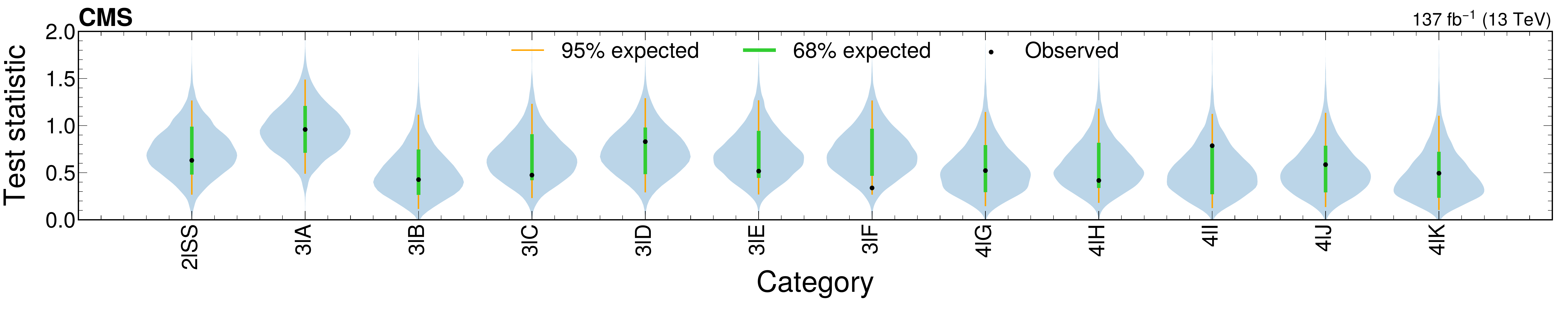}\\
\includegraphics[width=\textwidth]{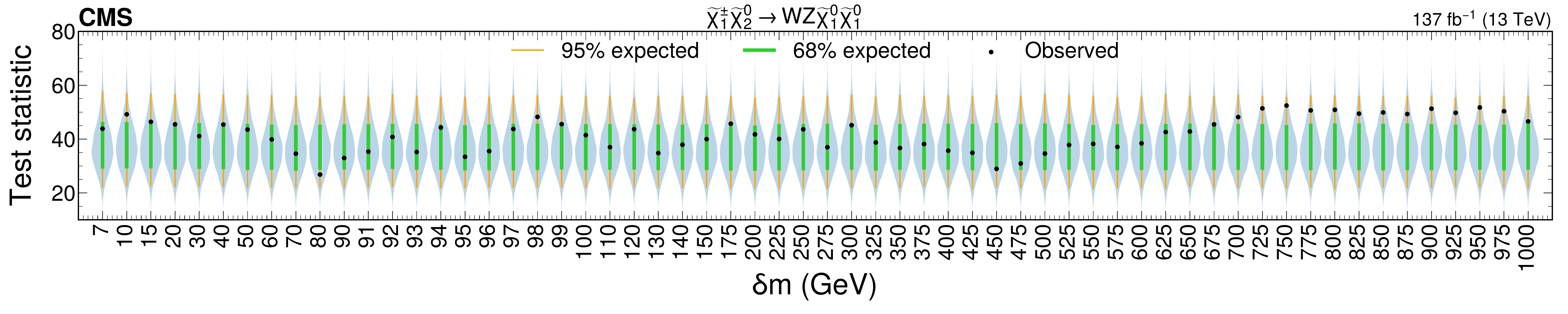}
\caption{Expected test statistic distribution for a background-only fit compared to the observed test statistic value, drawn as black dots, for the search regions in each event category (upper plot) and the neural network targeting \WZ-mediated superpartner decays for each \dm evaluation (lower plot). The gray shaded area represents the (symmetrized) probability density of the expected test statistic distribution, with 68 and 95\% expected ranges respectively drawn in green and orange.}
\label{fig:teststatistic_SR_TChiWZ}
\end{figure}

In all categories, and in both evaluations of 3$\ell$A events, based on the neural networks and on the search regions, the data are found to be consistent with the expectation from the SM backgrounds. The agreement in the search regions is summarized in Fig.~\ref{fig:teststatistic_SR_TChiWZ} (upper plot), where the expected test statistic~\cite{teststatistic} distribution for a background only fit to data is compared to the observed test statistic value. One expects the observed test statistic to lie in a likely region of the expected test statistic distribution in the absence of any unknown physics. A similar plot is shown in Fig.~\ref{fig:teststatistic_SR_TChiWZ} (lower plot) for the neural network targeting \WZ-mediated decays of the chargino-neutralino pair. This is the neural network for which the data are evaluated at the most \dm values, and the agreement is shown for each one of them.

\clearpage

\section{Interpretation} \label{sec:interpretations}

No significant excess of events over the SM-only hypothesis is observed, as shown in Section~\ref{sec:results}. The expected signal and background yields and the observed data are then used to determine 95\% confidence level (CL) upper limits on \xpmxtwo and effective \xonexone production cross sections for the different decay models introduced in Section~\ref{sec:signals}, using the \CLs criterion~\cite{Junk:1999kv,Read_2002}. The asymptotic approximation of the distribution of the profile likelihood test statistic~\cite{Cowan:2010js,ATLAS:2011tau} is used when computing these limits. The systematic uncertainties introduced in Section~\ref{sec:systematics} are included as nuisance parameters with additional constrain terms into the likelihood function. Systematic uncertainties with an effect on overall process normalizations but not distributions are included through log-normal probability density functions while those that have an effect on both normalization and shape of any processes are included via the template morphing technique \cite{Conway:2011in} and represented with Gaussian probability density functions. Uncertainties related to the limited size of the MC samples are introduced into the likelihood following the Barlow--Beeston approach \cite{Barlow:1993dm}.

For each model interpretation, a global fit of the analysis bins is performed, using the events from categories corresponding to the final state of the particular model. The event categories used to interpret each model are listed in Table~\ref{tab:interpretation_categories}. In the interpretations that include the neural network approach, the corresponding distributions used for the interpretation in the 3$\ell$A category in each signal point correspond to the evaluation of the neural network discriminant at the specific \dm value of the chosen point.

\begin{table}
\centering
\topcaption{Summary of the event categories used for the interpretation of the results in terms of different models, and references to the associated figure summarizing the expected and observed 95\% CL upper limits.}

\label{tab:interpretation_categories}
\begin{tabular}{lcc}
Model & Categories used & Figure \\
\hline

\multirow{2}{*}{\xpmxtwo production, flavor-democratic} & 2$\ell$SS, 3$\ell$A (search regions and & \multirow{2}{*}{\ref{fig:limits_tchislepsnu_flavordem}} \\
                                                       & neural network fit separately) & \\

\xpmxtwo production, $\tau$-enriched & 3$\ell$A--3$\ell$F & \ref{fig:limits_tchislepsnu_tauenr}\\

\xpmxtwo production, $\tau$-dominated & 3$\ell$B--3$\ell$F & \ref{fig:limits_tchislepsnu_taudom} \\[\cmsTabSkip]

\multirow{2}{*}{\xpmxtwo production, \WZ-mediated decays} & 2$\ell$SS, 3$\ell$A (search regions and & \multirow{2}{*}{\ref{fig:limits_tchiwz}} \\
                                                       & neural network fit separately) & \\ [\cmsTabSkip]

\xpmxtwo production, \WH-mediated decays & 2$\ell$SS, 3$\ell$A--3$\ell$F & \ref{fig:limits_tchiwh} \\[\cmsTabSkip]

\xonexone production & 3$\ell$A--4$\ell$K & \ref{fig:limits_tchizh}\\
\end{tabular}
\end{table}

The sensitivity of \xpmxtwo production models with slepton-mediated decays is mainly driven by 3$\ell$A events in case of flavor-democratic decays. In case of compressed models, in particular for $x = 0.05$ and $0.95$, 2$\ell$SS events increase the sensitivity significantly, being the leading source of exclusion for mass splittings of $m_{\xtwo} - m_{\xone} < 30 \GeV$. For $\tau$-enriched models, regions 3$\ell$B--3$\ell$F provide the bulk of the sensitivity, with 3$\ell$A and 2$\ell$SS events still adding sensitivity, the latter particularly for compressed models. For $\tau$-dominated models, the sensitivity is driven by 3$\ell$B--3$\ell$F, and only these regions are included in the corresponding fits.

\begin{figure}[ht!]
\centering
  \includegraphics[width=0.48\textwidth]{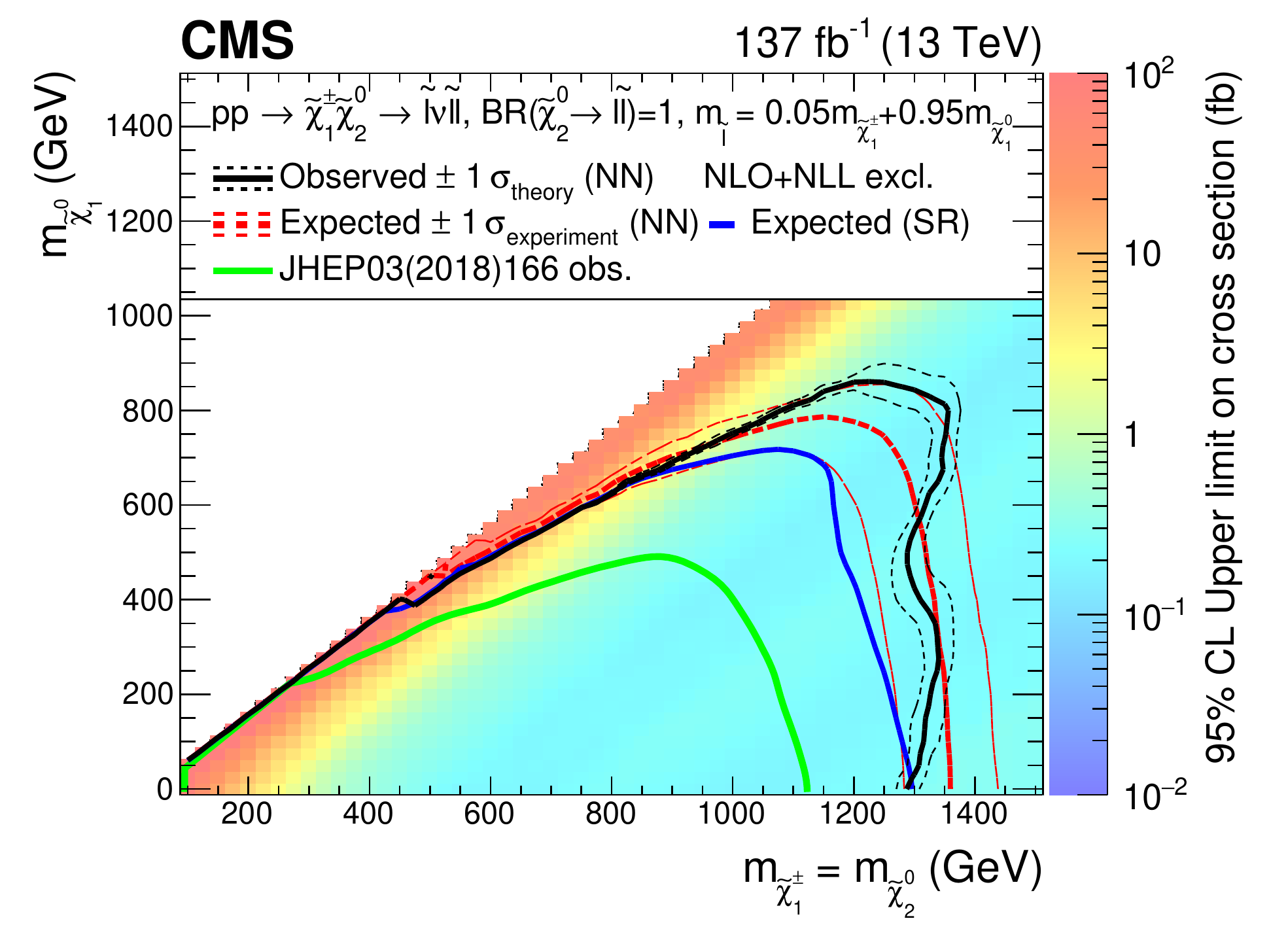}
  \includegraphics[width=0.48\textwidth]{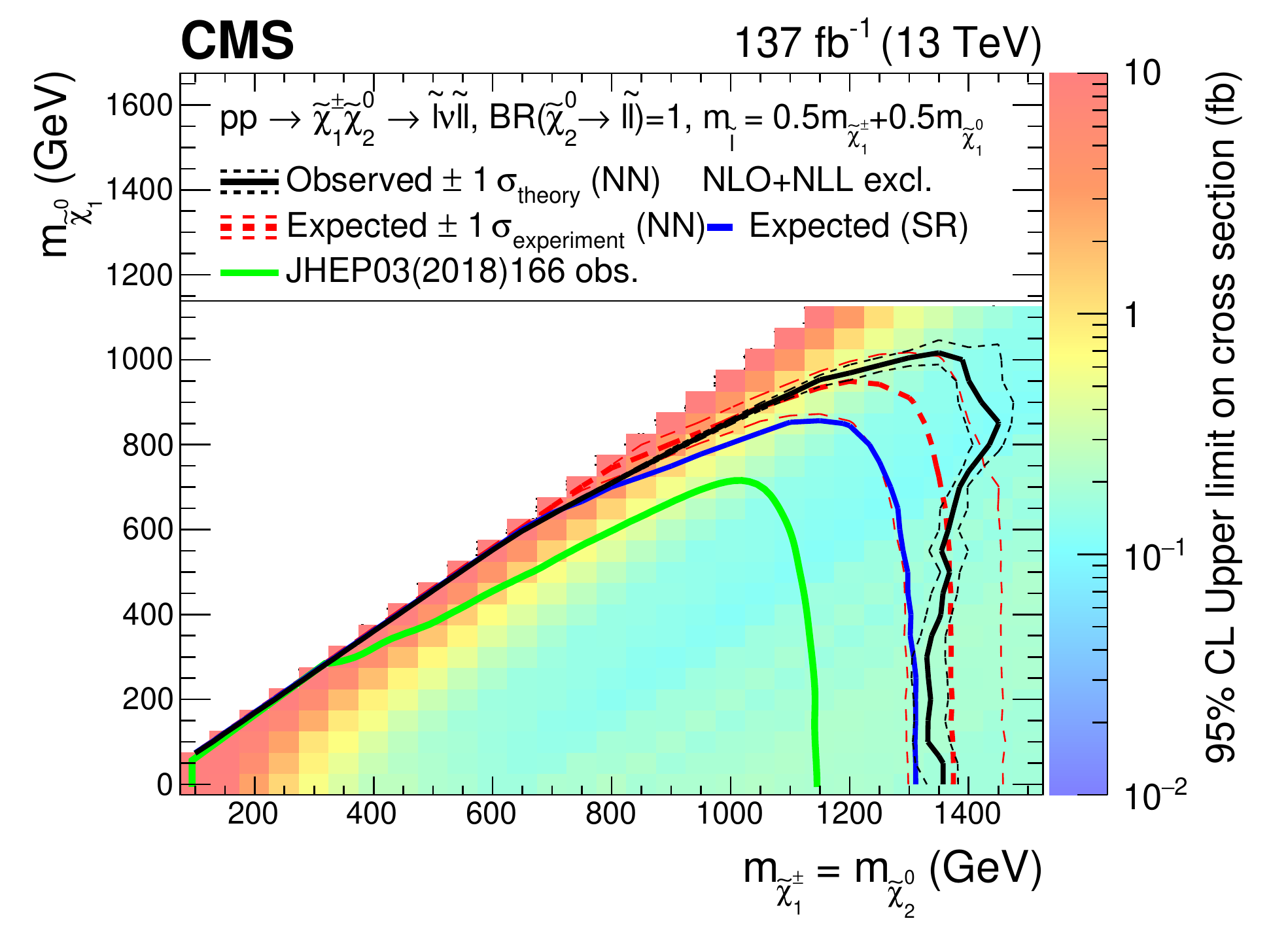}
  \includegraphics[width=0.48\textwidth]{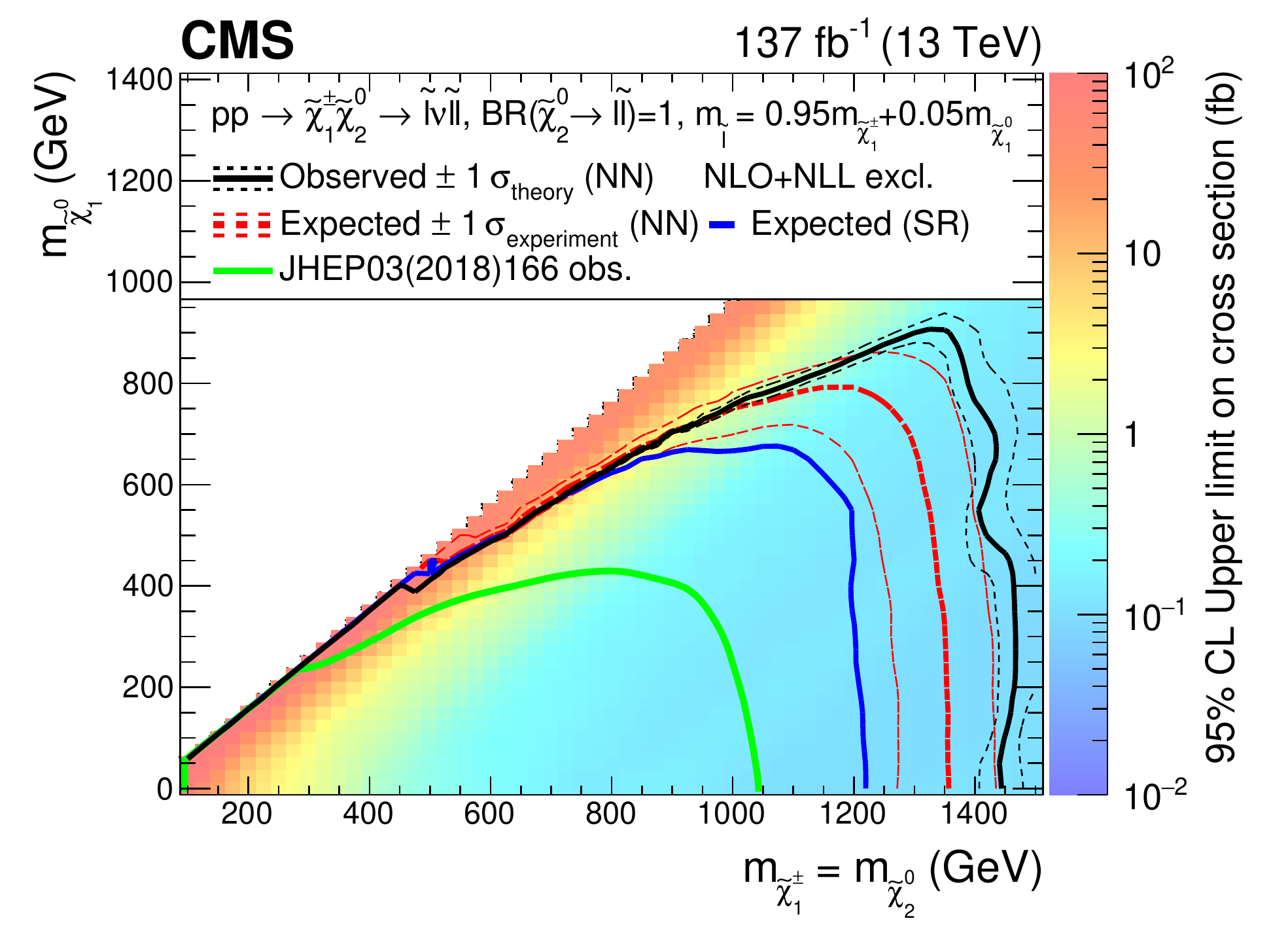}
  \caption{Interpretation of the results for \xpmxtwo production with flavor-democratic slepton-mediated decays, and the parameter governing the mass splittings being $x=0.05$ (upper left), $x=0.5$ (upper right) and $x=0.95$ (lower). The shading in the $m_{\xone}$ versus $m_{\xtwo}$ plane indicates the 95\% CL upper limit on the \xpmxtwo production cross sections. The contours delineate the mass regions excluded at 95\% CL when assuming cross section computed at NLO plus NLL. All masses below the contours are excluded. The observed, observed $\pm 1 \sigma_{\text{theory}}$ ($\pm$1 standard deviation of the theoretical cross sections), median expected, and expected $\pm 1 \sigma_{\text{experiment}}$ bounds obtained with the neural network strategy are shown in black and red. The median expected bound obtained with the search region strategy is shown in blue. The observed limits obtained in the CMS analysis using 2016 data~\cite{SUS-16-039} are shown in green.}
\label{fig:limits_tchislepsnu_flavordem}
\end{figure}

The observed and expected exclusion limits as a function of $m_{\xtwo}$ and $m_{\xone}$ are shown in Fig.~\ref{fig:limits_tchislepsnu_flavordem} for flavor-democratic slepton-mediated decays. For each model the limits are separately computed by fitting 3$\ell$A events as a function of the output of the neural networks or the search regions. The 3$\ell$A search regions and neural networks contain the same events so are never simultaneously fit. In both cases, they are fit together with the 2$\ell$SS search regions as indicated in Table~\ref{tab:interpretation_categories}. When the slepton mass is close to $m_{\xtwo}$ ($x=0.95$), the neural network extends the observed limits by up to about 250\GeV in $m_{\xtwo}$ and 150\GeV in $m_{\xone}$ compared to the search regions. This corresponds to an excluded cross section that is almost of a factor 2 lower when only evaluating 3$\ell$A events, or a relative improvement slightly below 100\%. In models where the mass difference between the sleptons and \xtwo is bigger ($x=0.5$ and $0.05$), the leptons generally have harder \pt spectra and the gains from the neural networks over the search regions are smaller: up to around 100\GeV in both $m_{\xtwo}$ and $m_{\xone}$. This is equivalent to a relative improvement of about 50\% in the excluded cross section. At very low \dm values, the relative improvement varies more and can be greater than 100\% for $x = 0.5$ while typically being smaller than 50\% for the other $x$ values. Chargino masses up to 1450\GeV and LSP masses up to 1000\GeV can be excluded by the neural networks depending on the model parameters. The reach of the previous iteration of the analysis has been improved up to 400\GeV in the chargino masses---a factor of 9 in terms of production cross-section---and up to 500\GeV in the LSP masses.

\begin{figure}[ht!]
\centering
  \includegraphics[width=0.48\textwidth]{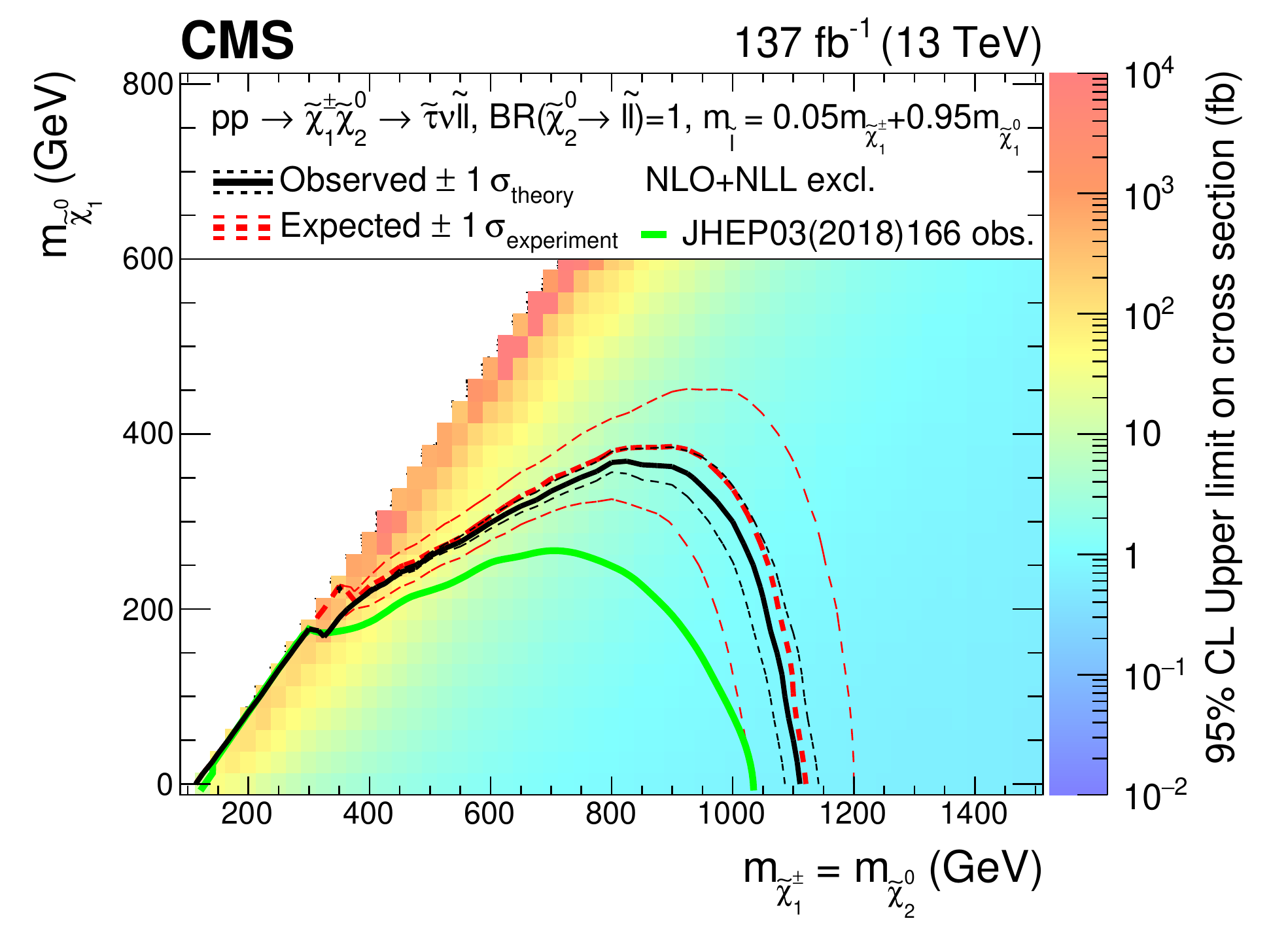}
  \includegraphics[width=0.48\textwidth]{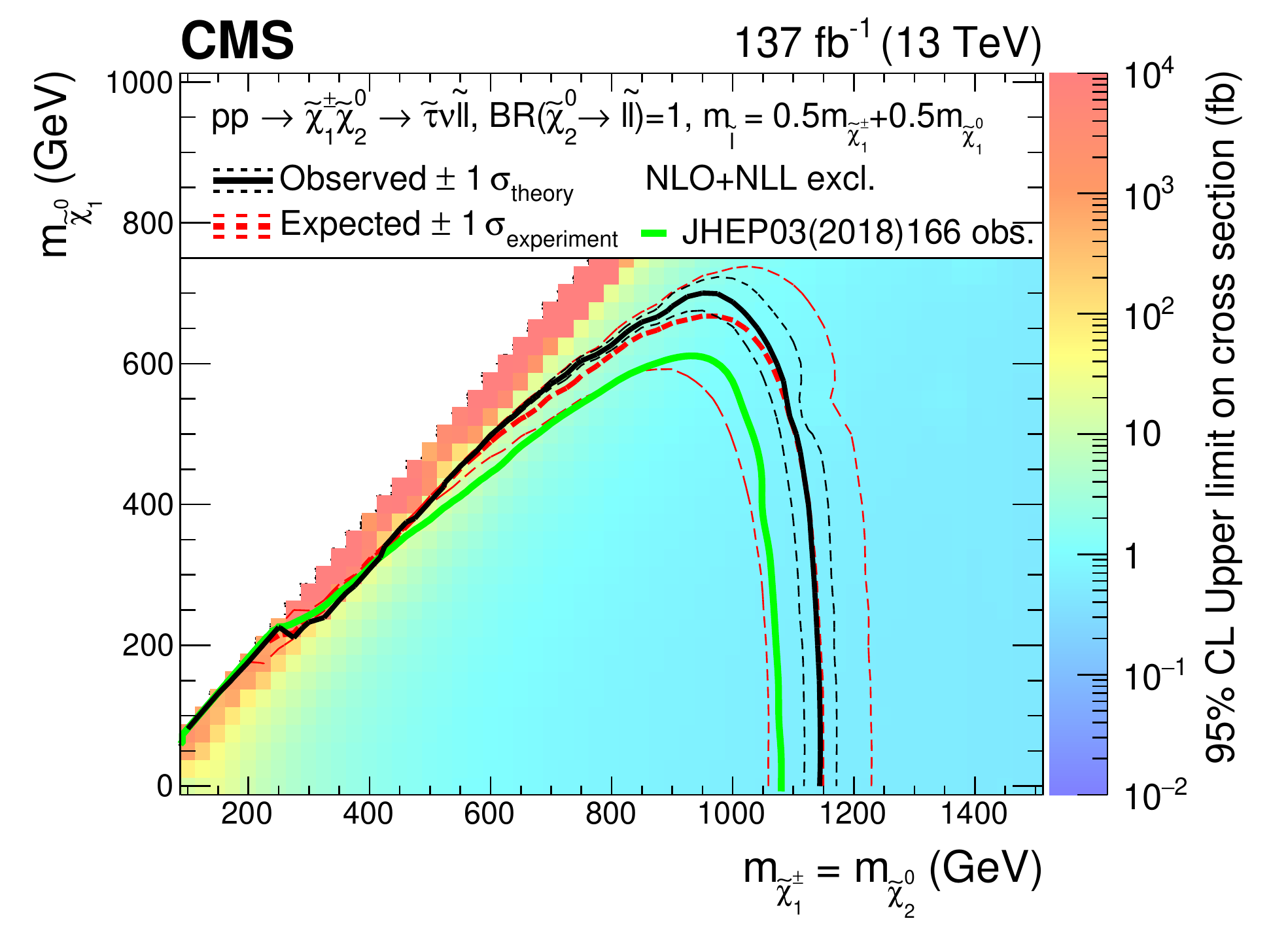}
  \includegraphics[width=0.48\textwidth]{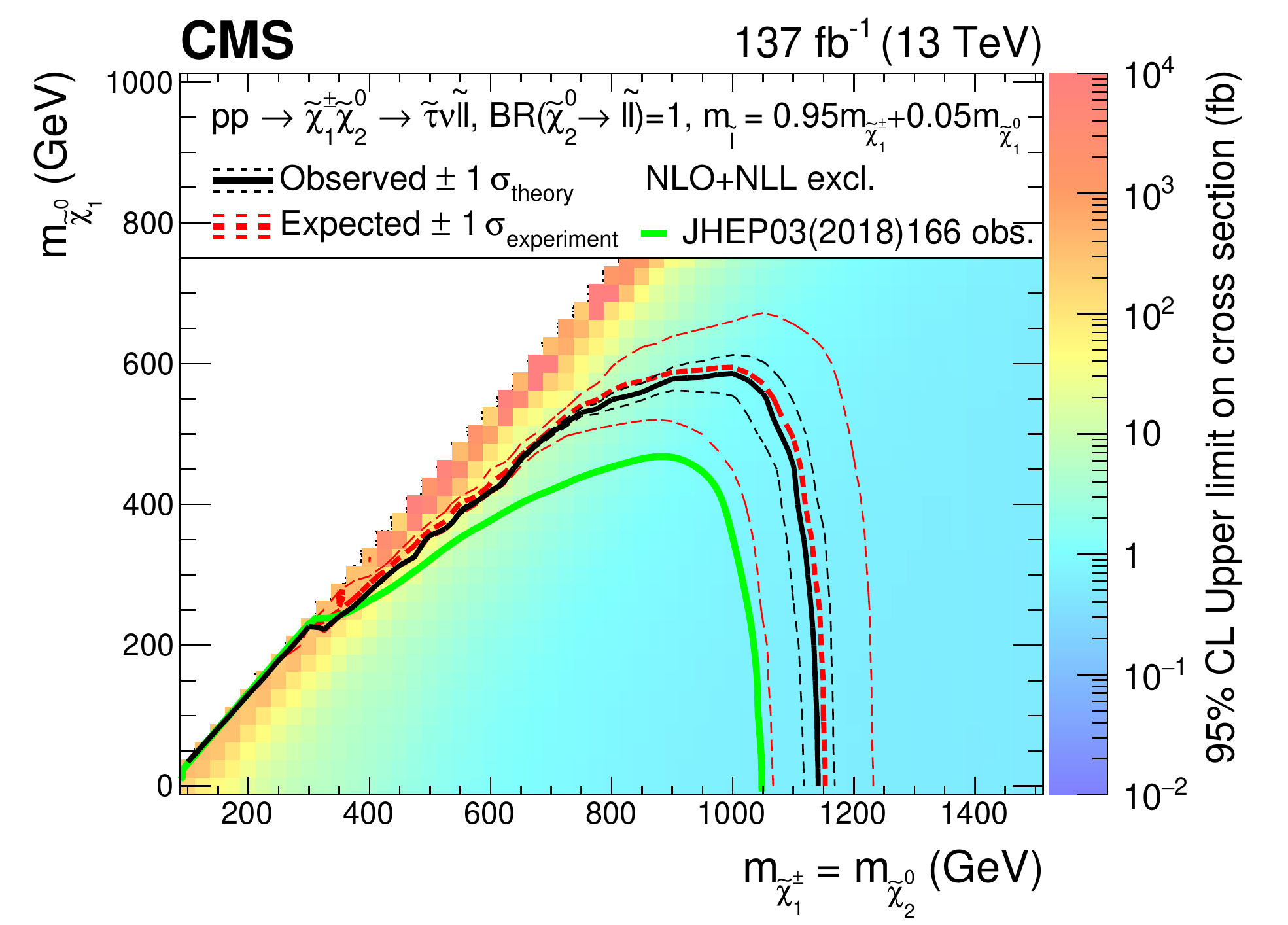}
\caption{Interpretation of the results for \xpmxtwo production with $\tau$-enriched slepton-mediated decays, and the parameter governing the mass splittings being $x=0.05$ (upper left), $x=0.5$ (upper right) and $x=0.95$ (lower). The shading in the $m_{\xone}$ versus $m_{\xtwo}$ plane indicates the 95\% CL upper limits on the \xpmxtwo production cross sections. The contours delineate the mass regions excluded at 95\% CL when assuming cross sections computed at NLO plus NLL. The observed, observed $\pm 1 \sigma_{\text{theory}}$ ($\pm$1 standard deviation of the theoretical cross sections), median expected, and expected $\pm 1 \sigma_{\text{experiment}}$ bounds are shown in black and red. The observed limits obtained in the CMS analysis using 2016 data~\cite{SUS-16-039} are shown in green.}
\label{fig:limits_tchislepsnu_tauenr}
\end{figure}

For $\tau$-enriched models, the exclusion limits are drawn in Fig.~\ref{fig:limits_tchislepsnu_tauenr}. The limits extend up to 1150\GeV in $m_{\xtwo}$ and 700\GeV in $m_{\xone}$ and provide improvements above the reach of previous iterations of the analysis of up to 150\GeV for chargino masses --around a factor 2 in cross section-- and 100\GeV for LSP masses.

\begin{figure}[ht!]
\centering
  \includegraphics[width=0.48\textwidth]{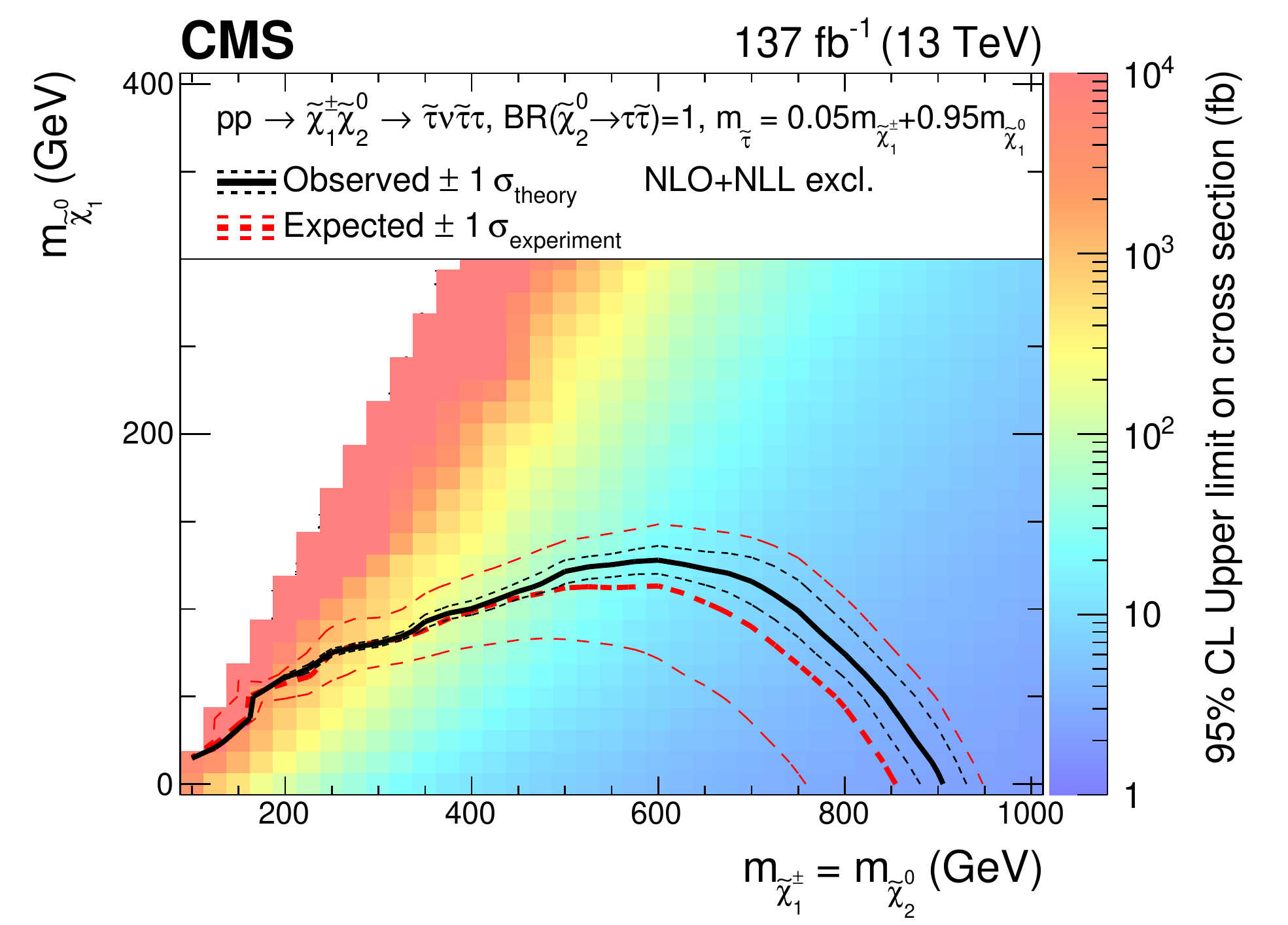}
  \includegraphics[width=0.48\textwidth]{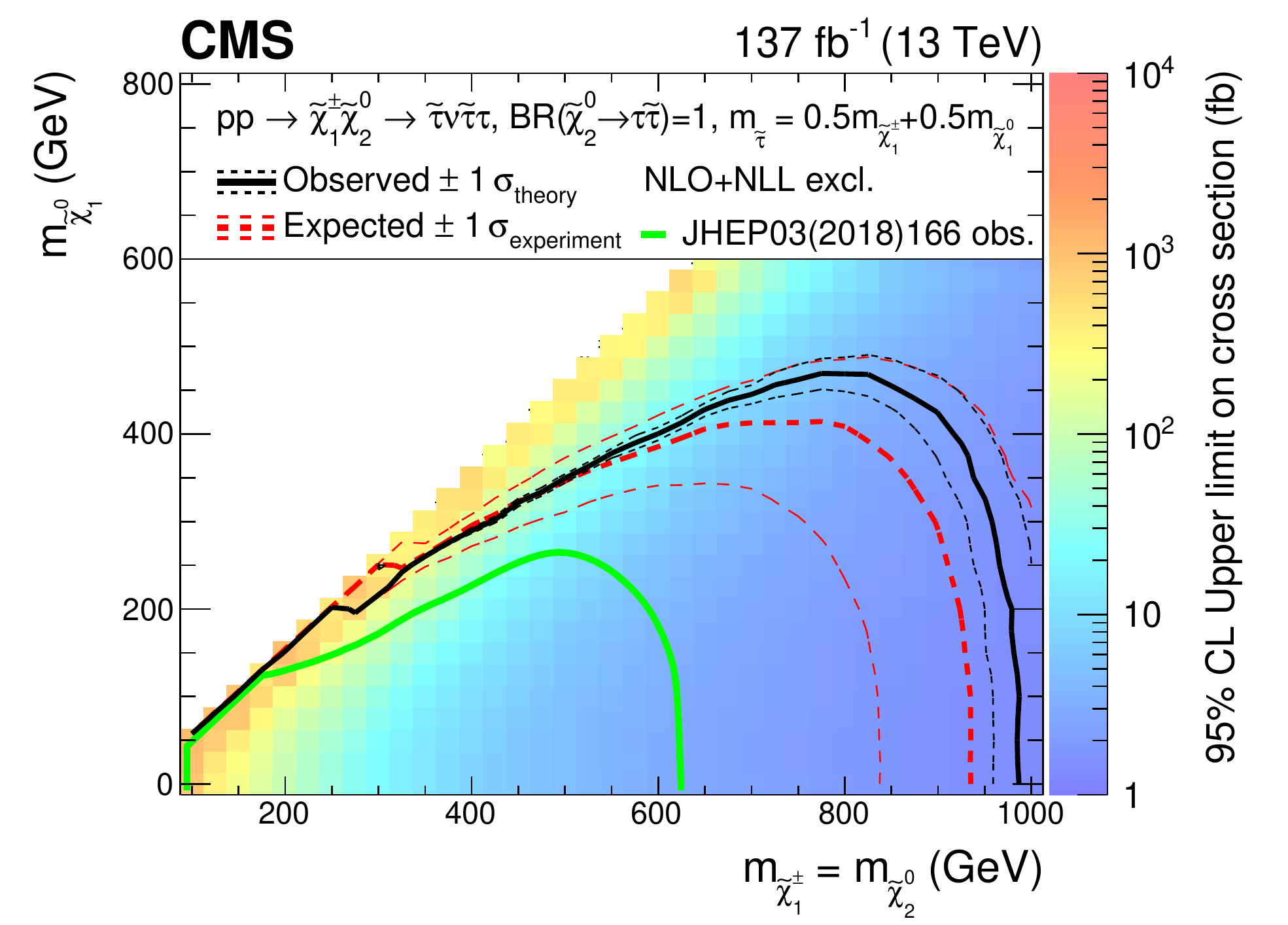}
  \includegraphics[width=0.48\textwidth]{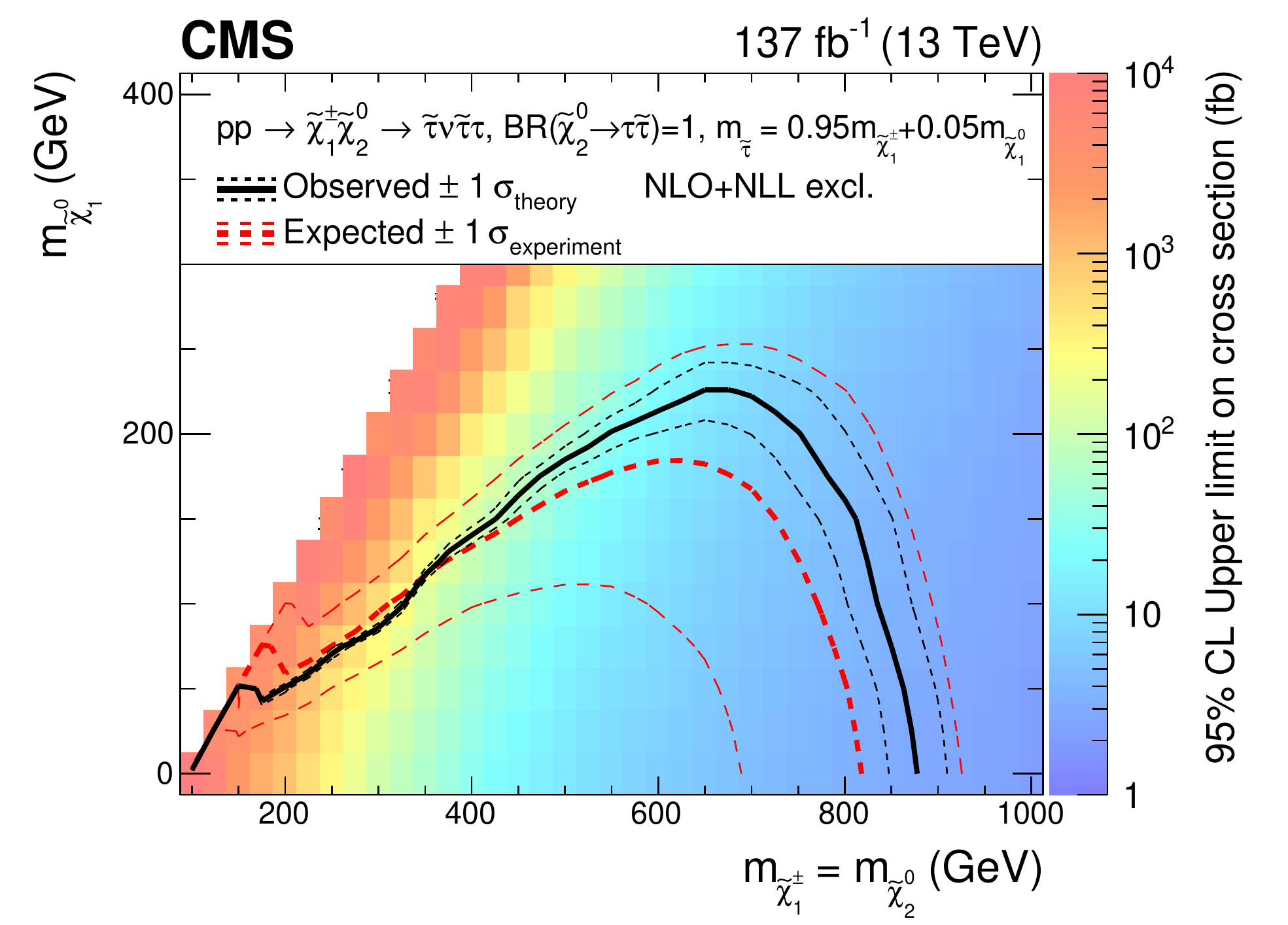}
\caption{Interpretation of the results for \xpmxtwo production with $\tau$-dominated slepton-mediated decays, and the parameter governing the mass splittings being $x=0.05$ (upper left), $x=0.5$ (upper right) and $x=0.95$ (lower). The shading in the $m_{\xone}$ versus $m_{\xtwo}$ plane indicates the 95\% CL upper limits on the \xpmxtwo production cross sections. The contours delineate the mass regions excluded at 95\% CL when assuming cross sections computed at NLO plus NLL. The observed, observed $\pm 1 \sigma_{\text{theory}}$ ($\pm$1 standard deviation of the theoretical cross sections), median expected, and expected $\pm 1 \sigma_{\text{experiment}}$ bounds are shown in black and red. The median expected bound obtained with the search region strategy is shown in blue. The observed limits obtained in the CMS analysis using 2016 data~\cite{SUS-16-039} are shown in green, which only included interpretations in the $x=0.5$ case.}
\label{fig:limits_tchislepsnu_taudom}
\end{figure}

Limits for $\tau$-dominated decay models are shown in Fig.~\ref{fig:limits_tchislepsnu_taudom}, extending up to 970\GeV in $m_{\xtwo}$ and 450\GeV in $m_{\xone}$. For the $\tau$-dominated models the limits for more compressed scenarios are significantly worse because of the relatively large \pt thresholds used in the \tauh selection. Noticeable improvements, owing to the reoptimized search strategy and the updated lepton and $\tau$ identification criteria, lead to an improvement over the results obtained by previous iterations of the analysis of 350\GeV in chargino masses---around a factor 10 in cross section---and up to 200\GeV in the LSP mass.

\begin{figure}
\centering
  \includegraphics[width=0.48\textwidth]{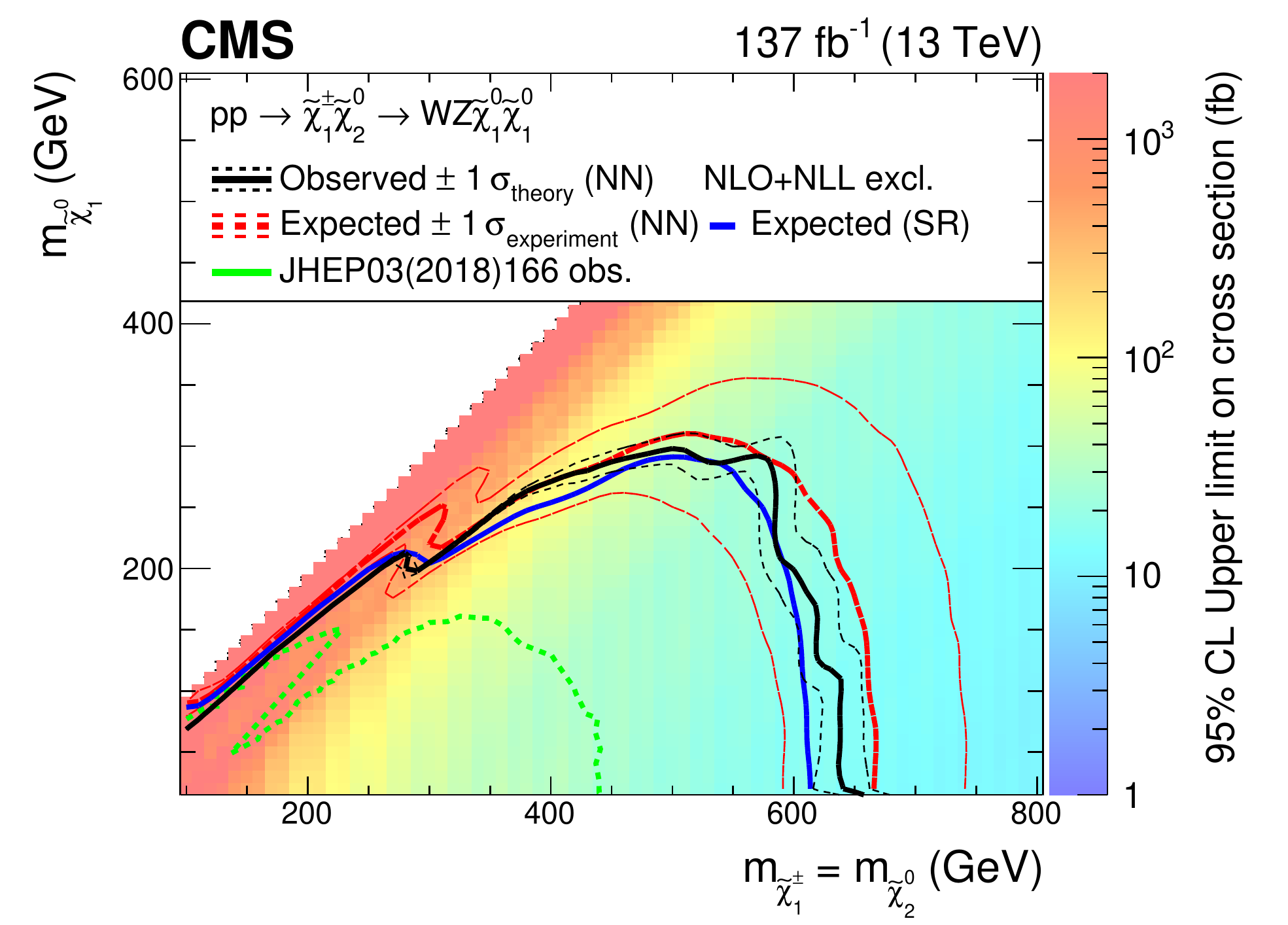}
\caption{Interpretation of the results for \xpmxtwo production with \WZ-mediated decays. The shading in the $m_{\xone}$ versus $m_{\xtwo}$ plane indicates the 95\% CL upper limits on the \xpmxtwo production cross sections. The contours delineate the mass regions excluded at 95\% CL when assuming cross sections computed at NLO plus NLL. The observed, observed $\pm 1 \sigma_{\text{theory}}$ ($\pm$1 standard deviation of the theoretical cross sections), median expected, and expected $\pm 1 \sigma_{\text{experiment}}$ bounds obtained with the neural network strategy are shown in black and red. The median expected bound obtained with the search region strategy is shown in blue. The observed limits obtained in the CMS analysis using 2016 data~\cite{SUS-16-039} are shown in green.}
\label{fig:limits_tchiwz}
\end{figure}

Models of \xpmxtwo production with \WZ-mediated decays are probed using category 3$\ell$A and 2$\ell$SS events, with the former dominating the sensitivity for most mass hypotheses. Similarly to the case of flavor-democratic slepton-mediated decays, the fits are performed twice: once using the neural network, and once using the 3$\ell$A search regions. Other regions provide minimal sensitivity to these models and are thus excluded from the interpretation. The interpretation is done separately, using the parametric neural network and the search region bins. The resulting exclusion limit curves are shown in Fig.~\ref{fig:limits_tchiwz}. The neural network provides maximal sensitivity to the models we are probing, resulting in more stringent exclusion limits by about 130\GeV in $m_{\xtwo}$ and a bit less than 50\GeV in $m_{\xone}$. At most \dm values this corresponds to improvements between 30 and 40\% in the excluded cross section, while at \dm values below 30\GeV the improvement is often larger than 200\%. The current results improve the results of the 2016 analysis by 200\GeV in the excluded chargino masses --around a factor 6 in cross section-- and 150\GeV in the LSP mass. Moreover, the current results provide coverage over the previously nonexcluded gap in the parameter space around mass splittings of $m_{\xtwo} - m_{\xone} = \mz$.

\begin{figure}[htb!]
\centering
  \includegraphics[width=0.48\textwidth]{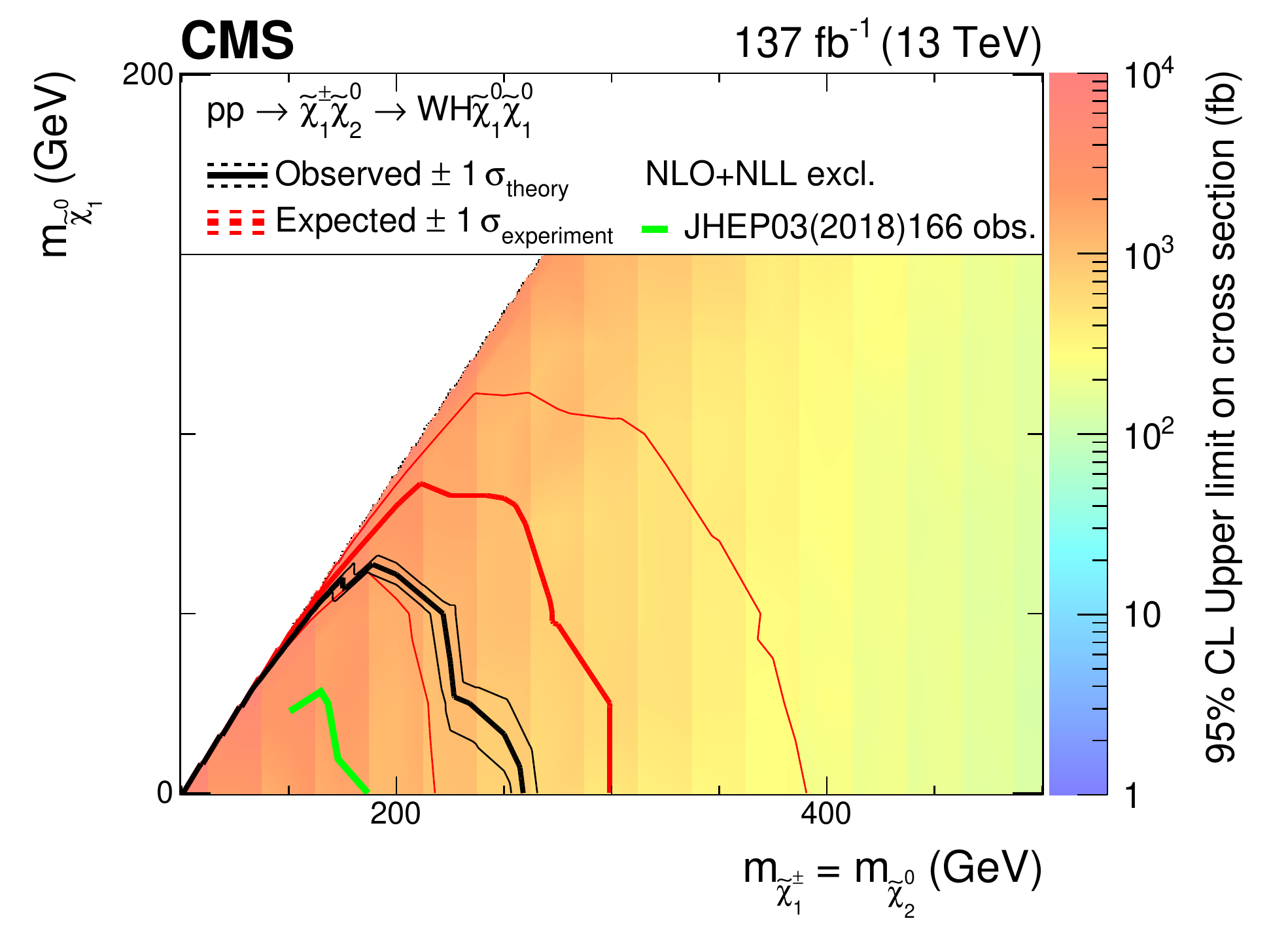}
\caption{Interpretation of the results for \xpmxtwo production with \WH-mediated decays. The shading in the $m_{\xone}$ versus $m_{\xtwo}$ plane indicates the 95\% CL upper limits on the \xpmxtwo production cross sections. The contours delineate the mass regions excluded at 95\% CL when assuming cross sections computed at NLO plus NLL. The observed, observed $\pm 1 \sigma_{\text{theory}}$ ($\pm$1 standard deviation of the theoretical cross sections), median expected, and expected $\pm 1 \sigma_{\text{experiment}}$ bounds are shown in black and red. The observed limits obtained in the CMS analysis using 2016 data~\cite{SUS-16-039} are shown in green.}
\label{fig:limits_tchiwh}
\end{figure}

Because of the diverse set of possible \PH decays, all event categories are used in the interpretation of models with \PH-mediated \xtwo decays. The most important event category in the interpretation of these models is 3$\ell$B, where the search regions are consequently designed to specifically target \PH decays. The resulting limits for this decay hypothesis are shown in Fig.~\ref{fig:limits_tchiwh}, and range up to 300\GeV in $m_{\xtwo}$ and 70\GeV in $m_{\xone}$, excluding regions with cross sections up to four times smaller than those already reached by the previous iteration of the analysis.

\begin{figure}[htbp!]
    \centering
    \includegraphics[width=0.6\textwidth]{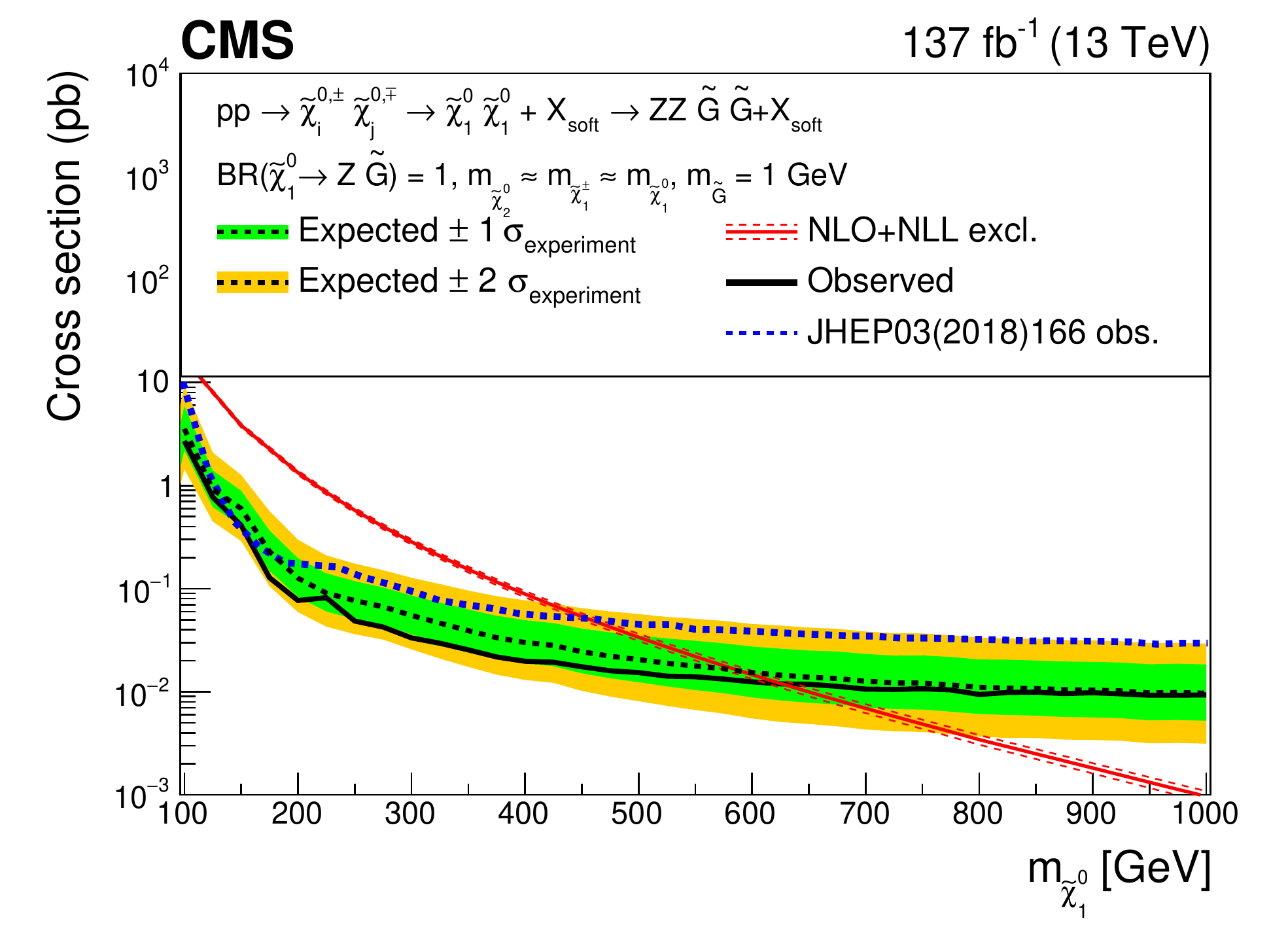} \\
    \includegraphics[width=0.6\textwidth]{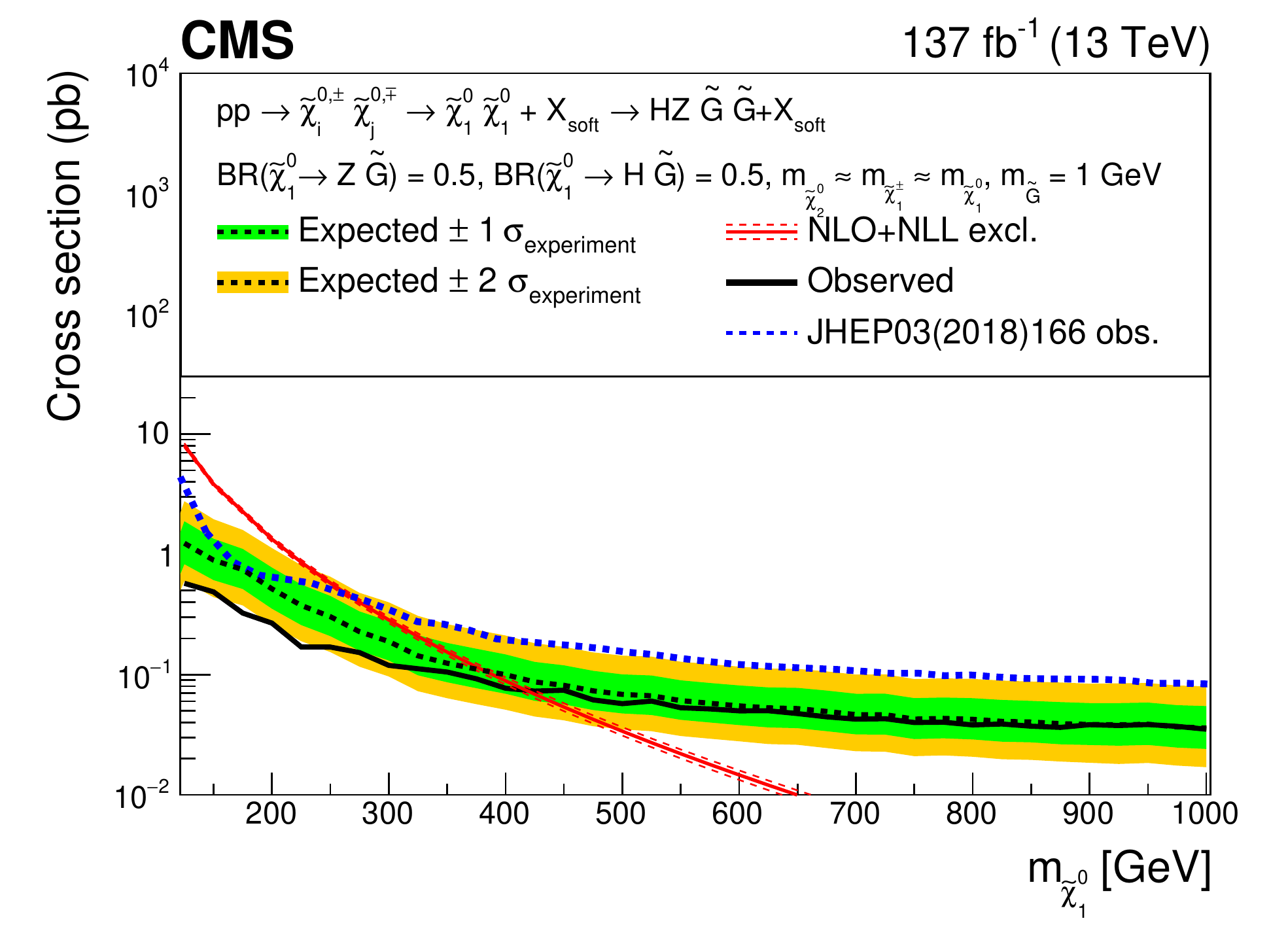} \\
    \includegraphics[width=0.6\textwidth]{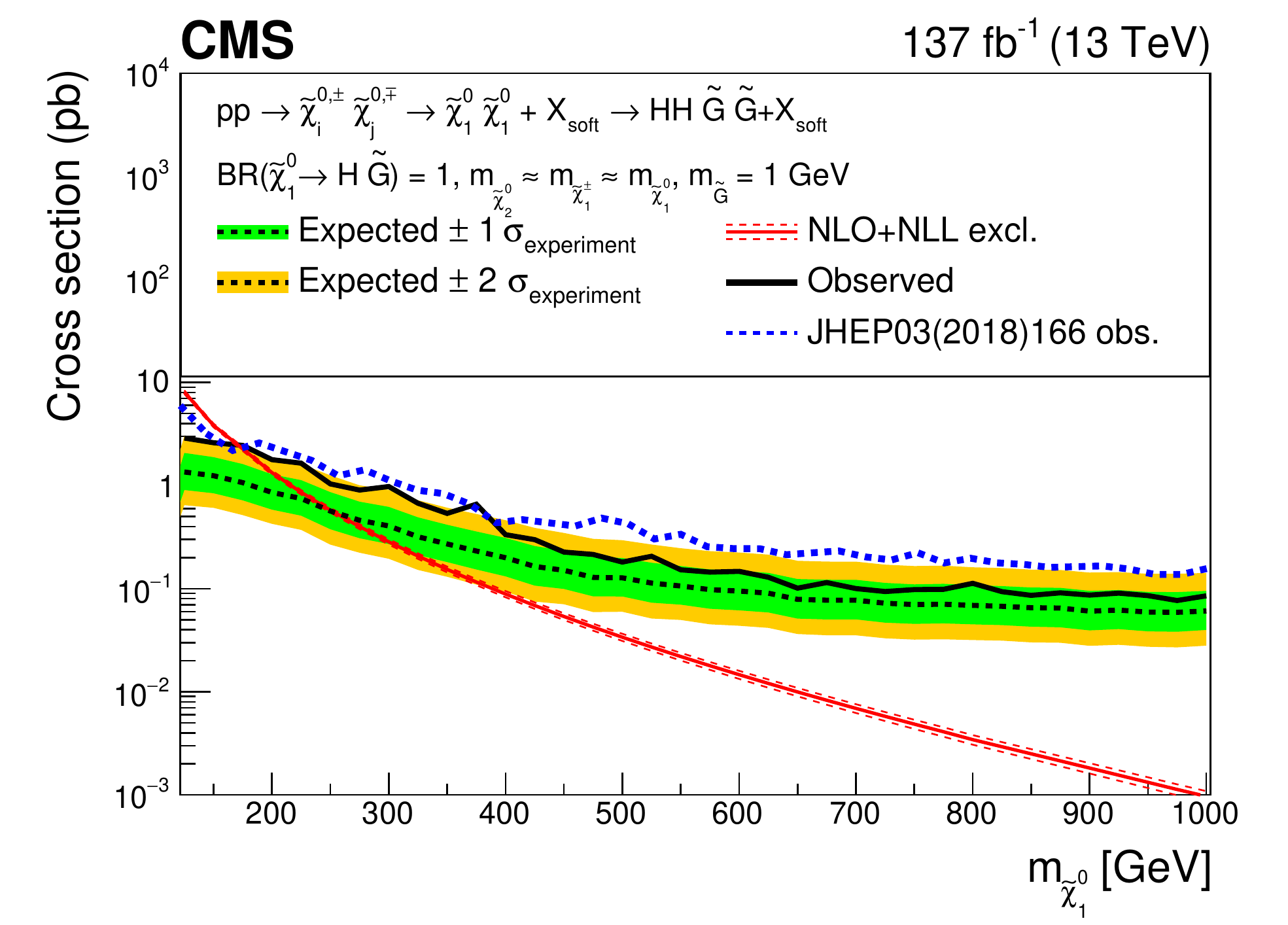}
    \caption{Interpretation of the results for effective \xone pair production, with $\PZ\PZ$-mediated decays (upper), $\PH\PZ$-mediated decays (middle), and $\PH\PH$-mediated decays (lower). The median expected upper limits (black line) are shown along with the $\pm 1 \sigma$ ($0.16$ and $0.84$ quantiles, green) and $\pm 2 \sigma$ ($0.05$ and $0.95$ quantiles, yellow) bands. The predicted production cross sections computed at NLO plus NLL are shown in red and the observed exclusion limits obtained in the CMS analysis using 2016 data~\cite{SUS-16-039} are shown in blue.}
    \label{fig:limits_tchizh}
\end{figure}

The interpretation of \xone pair production models, with subsequent decays via \PH or \PZ bosons uses all event categories. In the case of decays via two \PZ bosons, 4$\ell$G events are the most important contributors to the final exclusion limits. In decays via an \PH and a \PZ boson, four lepton events provide the most sensitivity for low \xone mass hypotheses, while trilepton events become more important at higher \xone masses. When the \xone pair decays via two \PH, trilepton events drive the results. The exclusion limits as a function of $m_{\xone}$ for these models are shown in Fig.~\ref{fig:limits_tchizh}, and extend up to 600\GeV in case of $\PZ\PZ$-mediated decays, up to 400\GeV for decays via $\PH\PZ$, and up to 200\GeV for $\PH\PH$-mediated decays. These improved exclusion limits correspond to excluded cross-sections up to seven times better than those of the previous iteration of the analysis.

\clearpage

\section{Summary} \label{sec:summary}

A search for new physics in events with two leptons of the same sign, or with three or more leptons with up to two hadronically decaying $\tau$ leptons, is presented. A data set of proton-proton collisions with $\sqrt{s}=13\TeV$ collected with the CMS detector at the LHC, corresponding to an integrated luminosity of 137\fbinv, is analyzed. Events are categorized according to the number of leptons, their signs, and flavors. Events in each category are further binned using a plethora of kinematic quantities to maximize the sensitivity of the search to an extensive set of hypotheses of supersymmetric particle production via the electroweak interaction. In events with three light leptons, of which two have opposite sign and same flavor, parametric neural networks are used to significantly enhance the sensitivity of the search to several signal hypotheses.

No significant deviation from the standard model expectation is observed in any of the event categories. The results are interpreted in terms of a number of simplified models of superpartner production. Models of chargino-neutralino pair production with the neutralino forming the lightest supersymmetric particle (LSP), as well as models of effective neutralino pair production with a nearly massless gravitino as the LSP are considered. The signal topologies depend on the masses of the leptonic superpartners and the mixing of the gauge eigenstates.

If left-handed sleptons lighter than the chargino existed, the chargino-neutralino pair might undergo slepton-mediated decays resulting in final states with three leptons. The results of the analysis lead to a lower limit in the chargino mass up to 1450\GeV when using a parametric neural network. Searches in events with three light leptons including an opposite-sign, same-flavor pair provide sensitivity to these models. Events with two same-sign leptons further enhance the sensitivity in experimentally challenging scenarios with small mass differences between the chargino and the LSP.

If sleptons were right-handed, the chargino, or both the chargino and the neutralino, might decay almost exclusively to $\tau$ leptons. In the former scenario, a chargino mass up to 1150\GeV is excluded, while a mass up to 970\GeV is excluded in the latter.

If sleptons were sufficiently heavy, charginos and neutralinos would undergo direct decay to the LSP via the emission of \PW, \PZ, or Higgs bosons. For decays of the chargino-neutralino pair via a \PW and a \PZ boson, values of the chargino mass up to 650\GeV are excluded through the use of a parametric neural network. In case of a neutralino decay via the emission of a Higgs boson, charginos with a mass below 300\GeV are excluded for nearly massless LSPs.

In models of effective neutralino production we assume the neutralinos decay to almost massless gravitino LSPs via \PZ and Higgs bosons. This leads to excluded values of the neutralino mass up to 600\GeV.

The obtained results currently provide the most stringent limits for chargino-neutralino production with mass splittings close to the \PZ boson mass, nearly closing the gap in the exclusion plane found in this region of the parameter space. The exclusions obtained for the slepton-mediated decays are as well the most stringent results currently for all the considered branching fraction hypotheses. In the case of the flavor-democratic decay scenario, the obtained exclusion limits of up to 1450\GeV are the overall highest exclusion values obtained for the production of electroweak superpartners.

The analysis techniques have been considerably refined compared to the earlier version of this search that used 35.9 \fbinv of proton-proton collision data at $\sqrt{s}=13\TeV$~\cite{SUS-16-039}. The integrated luminosity has increased by just short of a factor four, which in the absence of novel analysis techniques would result in improvements in the excluded cross sections by a factor of roughly two. In most search categories the improvements in analysis techniques result in significantly larger improvements to the results than the increased data volume, \eg the limits on the excluded superpartner production cross sections are improved by up to a factor ten, and more than a factor five for most model hypotheses.

\begin{acknowledgments}

We dedicate this paper to the memory of our friend and colleague Luc Pape whose seminal contributions to the CMS experiment and to searches for new physics, including supersymmetry, made this work possible.

We congratulate our colleagues in the CERN accelerator departments for the excellent performance of the LHC and thank the technical and administrative staffs at CERN and at other CMS institutes for their contributions to the success of the CMS effort. In addition, we gratefully acknowledge the computing centers and personnel of the Worldwide LHC Computing Grid and other centers for delivering so effectively the computing infrastructure essential to our analyses. Finally, we acknowledge the enduring support for the construction and operation of the LHC, the CMS detector, and the supporting computing infrastructure provided by the following funding agencies: BMBWF and FWF (Austria); FNRS and FWO (Belgium); CNPq, CAPES, FAPERJ, FAPERGS, and FAPESP (Brazil); MES (Bulgaria); CERN; CAS, MoST, and NSFC (China); MINCIENCIAS (Colombia); MSES and CSF (Croatia); RIF (Cyprus); SENESCYT (Ecuador); MoER, ERC PUT and ERDF (Estonia); Academy of Finland, MEC, and HIP (Finland); CEA and CNRS/IN2P3 (France); BMBF, DFG, and HGF (Germany); GSRT (Greece); NKFIA (Hungary); DAE and DST (India); IPM (Iran); SFI (Ireland); INFN (Italy); MSIP and NRF (Republic of Korea); MES (Latvia); LAS (Lithuania); MOE and UM (Malaysia); BUAP, CINVESTAV, CONACYT, LNS, SEP, and UASLP-FAI (Mexico); MOS (Montenegro); MBIE (New Zealand); PAEC (Pakistan); MSHE and NSC (Poland); FCT (Portugal); JINR (Dubna); MON, RosAtom, RAS, RFBR, and NRC KI (Russia); MESTD (Serbia); SEIDI, CPAN, PCTI, and FEDER (Spain); MOSTR (Sri Lanka); Swiss Funding Agencies (Switzerland); MST (Taipei); ThEPCenter, IPST, STAR, and NSTDA (Thailand); TUBITAK and TAEK (Turkey); NASU (Ukraine); STFC (United Kingdom); DOE and NSF (USA).

\hyphenation{Rachada-pisek} Individuals have received support from the Marie-Curie program and the European Research Council and Horizon 2020 Grant, contract Nos.\ 675440, 724704, 752730, 765710 and 824093 (European Union); the Leventis Foundation; the Alfred P.\ Sloan Foundation; the Alexander von Humboldt Foundation; the Belgian Federal Science Policy Office; the Fonds pour la Formation \`a la Recherche dans l'Industrie et dans l'Agriculture (FRIA-Belgium); the Agentschap voor Innovatie door Wetenschap en Technologie (IWT-Belgium); the F.R.S.-FNRS and FWO (Belgium) under the ``Excellence of Science -- EOS" -- be.h project n.\ 30820817; the Beijing Municipal Science \& Technology Commission, No. Z191100007219010; the Ministry of Education, Youth and Sports (MEYS) of the Czech Republic; the Deutsche Forschungsgemeinschaft (DFG), under Germany's Excellence Strategy -- EXC 2121 ``Quantum Universe" -- 390833306, and under project number 400140256 - GRK2497; the Lend\"ulet (``Momentum") Program and the J\'anos Bolyai Research Scholarship of the Hungarian Academy of Sciences, the New National Excellence Program \'UNKP, the NKFIA research grants 123842, 123959, 124845, 124850, 125105, 128713, 128786, and 129058 (Hungary); the Council of Science and Industrial Research, India; the Latvian Council of Science; the Ministry of Science and Higher Education and the National Science Center, contracts Opus 2014/15/B/ST2/03998 and 2015/19/B/ST2/02861 (Poland); the National Priorities Research Program by Qatar National Research Fund; the Ministry of Science and Higher Education, project no. 0723-2020-0041 (Russia); the Programa Estatal de Fomento de la Investigaci{\'o}n Cient{\'i}fica y T{\'e}cnica de Excelencia Mar\'{\i}a de Maeztu, grant MDM-2015-0509 and the Programa Severo Ochoa del Principado de Asturias; the Thalis and Aristeia programs cofinanced by EU-ESF and the Greek NSRF; the Rachadapisek Sompot Fund for Postdoctoral Fellowship, Chulalongkorn University and the Chulalongkorn Academic into Its 2nd Century Project Advancement Project (Thailand); the Kavli Foundation; the Nvidia Corporation; the SuperMicro Corporation; the Welch Foundation, contract C-1845; and the Weston Havens Foundation (USA).
\end{acknowledgments}

\bibliography{auto_generated}
\cleardoublepage \appendix\section{The CMS Collaboration \label{app:collab}}\begin{sloppypar}\hyphenpenalty=5000\widowpenalty=500\clubpenalty=5000\vskip\cmsinstskip
\textbf{Yerevan Physics Institute, Yerevan, Armenia}\\*[0pt]
A.~Tumasyan
\vskip\cmsinstskip
\textbf{Institut f\"{u}r Hochenergiephysik, Wien, Austria}\\*[0pt]
W.~Adam, J.W.~Andrejkovic, T.~Bergauer, S.~Chatterjee, M.~Dragicevic, A.~Escalante~Del~Valle, R.~Fr\"{u}hwirth\cmsAuthorMark{1}, M.~Jeitler\cmsAuthorMark{1}, N.~Krammer, L.~Lechner, D.~Liko, I.~Mikulec, P.~Paulitsch, F.M.~Pitters, J.~Schieck\cmsAuthorMark{1}, R.~Sch\"{o}fbeck, M.~Spanring, S.~Templ, W.~Waltenberger, C.-E.~Wulz\cmsAuthorMark{1}
% ...
\vskip\cmsinstskip
% ...
1:  Also at TU Wien, Wien, Austria\\
% ...
\end{sloppypar}
\end{document}